\documentclass[a4paper,11pt]{article}
\pdfoutput=1 

\usepackage{jheppub} 

\usepackage[T1]{fontenc} 

\title{\boldmath  Two-Measure Electroweak Standard Model.
\\
 Some aspects of cosmological evolution 
\\
and vacuum stability}

\author[a,b,1]{Alexander B. Kaganovich,\note{Corresponding author.}}

\affiliation[a]{Physics Department, Ben Gurion University of the Negev,\\Beer Sheva, Israel}
\affiliation[b]{Sami Shamoon College of Engineering, \\Beer Sheva, Israel}

\emailAdd{alexk@bgu.ac.il}

\abstract{
In the FLRW universe, the homogeneous scalar field $\phi(t)$ obtained by cosmological averaging of the local Higgs field $H(x)$ 
is considered as a classical field for which the electroweak SM quantization  procedure  is meaningless.
 When applying the Two-Measure theory (TMT)
 to study cosmology, the ratio $\zeta$
of the measure densities of two volume elements is a scalar function, which has the following properties: 
1) $\zeta$ enters into all equations of motion; 2) through the constraint, $\zeta$ is defined as a function of $\phi(t)$;
 3) during cosmological evolution, $\zeta(\phi)$ changes from $\zeta\approx 0$ at the inflationary stage to $\zeta= 1$ 
at the cosmological stage of approaching vacuum . Each stage of the evolution of the classical cosmological background 
is determined by the set \{$\phi(t)$, curvature, $\zeta(\phi(t))$\}. The Two-Measure electroweak SM (TMSM) is realized
 in the context of cosmology as a set of cosmologically modified copies of the GWS model, such that each of the copies 
exists as a local quantum field theory defined on the classical cosmological background at the appropriate stage of its evolution. 
This basic idea is studied in detail for two stages of the cosmological background evolution: for slow-roll inflation 
and for the stage of approaching vacuum. Mainly due to the presence of $\zeta(\phi(t))$ in all equations of motion, 
all TMSM coupling constants turn out to be a kind of running  (classical) TMT-effective parameters. During the evolution of 
the cosmological background, changing these parameters yields new results that are important for both particle physics and inflation:
  1) the classical running TMT-effective Higgs self-coupling increases from $\lambda\sim 10^{-11}$
 (which ensures consistency with the Planck's CMB data at $\xi=\frac{1}{6}$)
 to $\lambda\sim 0.1$ at the stage close to the vacuum; 
2) the mass term in the TMT-effective Higgs potential changes sign from positive to negative,
which provides SSB in the standard way of GWS theory; 3) the classical running constants of the gauge and Yukawa couplings change by several orders of magnitude;
4) The GWS theory is reproduced in such a way that the observed  hierarchy of fermion masses is obtained quite naturally. We show that, thanks to these classical-level results, taking into account quantum corrections 
 in the one-loop approximation preserves the slow-roll inflation regime and does not violate the vacuum stability.
}

\begin{document} 
\maketitle
\flushbottom

\section{Introduction}
\label{sec:intro}

The study of possible ways of extrapolating the Standard Model (SM) to the Planck scale has been the subject of numerous papers, see for example 
ref.\cite{Buttazzo} and references therein. When gravity is taken into account, a central problem arising at the interface between particle field theory and cosmology is establishing a link between physics at the low energies typical for the SM collider experiments and the inflationary energy scale. The standard approach to solving the latter problem can be roughly described as follows.
 The tree-level  electroweak SM defined in Minkowski space can also be described  in  general covariant form  to allow gravity to be included in the model.
This combination of particle physics and gravity, which we will call SM+Gravity for brevity, is the basis for how the Higgs sector of the electroweak SM is typically used to study cosmology at energies well above the electroweak energy scale.
But the tree-level Higgs sector (like other SM fields) in this picture is actually the same as in the SM in Minkowski space. Adding a non-minimal  coupling
of the Higgs field to the scalar curvature allows inflation to be successfully described, but the predicted inflationary parameters agree with CMB measurements only if the nonminimal coupling constant $\xi$ is unnaturally huge: depending on the type of model, metric  \cite{MetricHI-1}-\cite{MetricHI-10} or Palatini
\cite{Palatini-H-1}-\cite{Palatini-H-8}, the coupling constant varies in the range $10^4\lesssim\xi \lesssim 10^8$.
Nevertheless, apart from some effects due to nonminimal coupling (see, for example, review \cite{MetricHI-7}), in the SM+Gravity approach to Higgs inflation models at the tree level, the problem of finding a reasonable connection between low-energy particle physics and physics at the inflation energy scale remains unsolved. One might expect that such a connection could manifest itself in the form of running coupling constants as solutions of the renormalization group (RG) equations, that is, only when quantum effects are taken into account. However, the presence of gravity makes these Higgs inflation models perturbatively non-renormalizable.
Moreover, in both the metric Higgs inflation models \cite{MetricHI-1}-\cite{MetricHI-7} and the Palatini formulation \cite{Palatini-H-1}-\cite{Palatini-H-8} the transition in the original action to the Einstein frame leads to an essentially non-polynomial Lagrangian, which is also the reason for non-renormalizability.
Therefore, to calculate the energy dependences of the SM coupling constants, and in particular the value of the Higgs field selfcoupling constant $\lambda$ on the inflationary scale, it becomes necessary to use the UV cutoff, higher-dimensional operators and construct effective field theory models. Within this approach, for example, it was shown \cite{Palatini-H-5} that the corresponding modifications of the potential shape can be consistent with the measured CMB data only if there is a very clear correlation between the constraint on the value of  $\xi$ and the bounds on the low-energy value of the top Yukawa coupling.
Nevertheless, given that the reliability of such methods  is difficult to control, a more field-theoretically based approach to solving the problem would be desirable. Thus,  the need to use an unnaturally huge value of $\xi$, the non-renormalizability and vacuum stability problems in existing Higgs inflation models do not allow us to be confident that we are indeed dealing with the proper application of particle physics to the cosmology of the early Universe.

\vspace{0.3cm}

{\bf Motivation.}

This paper is a logical continuation of the paper \cite{1-st paper}, and the main goal of these papers is to show that the radical changes that Two-Measure theory (TMT) introduces to the way of combining gravity with particle physics (compared to SM+Gravity) allows us to successfully overcome the above-mentioned problems.
Indeed, gravity is built into TMT in such a way that in all TMT models the equations derived from  the principle of least action describe a self-consistent system of gravity and matter, and this system of equations, represented in the Einstein frame, is valid at any energy available to classical field theory. One of the most important consequences of this feature of TMT is that, in the context of cosmology, the Two-Measure Standard Model (TMSM) we study describes what the tree-level SM looks like at various stages of cosmological evolution, which is equivalent to a description at any allowed energies.
The radical difference between TMSM and SM+Gravity is most clearly manifested in the fact that in TMSM the description of matter in Minkowski space can 
{\bf only} be obtained as a result of the limiting transition from curved space-time, that is, the limiting transition to a state close to vacuum. Moreover, the latter turns out to be possible only as a result of fine tuning when choosing the integration constant ${\mathcal M}$, the value of which significantly affects {\bf all} equations of motion. This means that the problem of the cosmological constant (CC) and the method of ensuring its desired value also turn out to be built into the TMT in a fundamentally different way than in SM+Gravity: ensuring, for example, a zero value of the CC with a special choice of the integration constant 
${\mathcal M}$ leads to a corresponding fixation of the TMT-effective parameter values in the equations of all particle fields.
 This is indeed radically different  from SM+Gravity, where it is usually assumed that somehow the zero value of the vacuum energy is already ensured, and so the CC in the Lagrangian is set to zero "by hand". This  means that the Higgs-inflation models \cite{MetricHI-1}-\cite{Palatini-H-8} are formulated within a theory that is not self-sufficient in such a fundamental matter.

From a structural point of view, comparing SM+Gravity models with TMSM, we can say that SM+Gravity belongs to the “conventional” type of models, in the sense that in the original action there is only  the usual volume element $\sqrt{-g}d^4x $.
In models of the Two-Measure Theory, in the  integral of the primordial action, along with terms with the standard volume measure $dV_g=\sqrt{-g}d^4x$, there are terms with an alternative, metric independent volume measure $dV_{\Upsilon}$
 constructed as the following 4-form using 4 scalar functions $\varphi_a$ ($a =1 ,..,4$)
\begin{eqnarray}
dV_{\Upsilon}=\Upsilon d^4x\equiv
\varepsilon^{\mu\nu\gamma\beta}\varepsilon_{abcd}\partial_{\mu}\varphi_{a}
\partial_{\nu}\varphi_{b}\partial_{\gamma}\varphi_{c}
\partial_{\beta}\varphi_{d}d^4x
=4!d\varphi_1\wedge d\varphi_2\wedge d\varphi_3\wedge d\varphi_4.
\label{Phi}
\end{eqnarray}
Here $\Upsilon$ is a scalar density, that is under general coordinate transformations with positive Jacobian it has the same transformation law as $\sqrt{-g}$.
A specific {\em dynamic} feature of TMT is that the ratio of volume measures 
\begin{equation}
\zeta=\frac{\Upsilon}{\sqrt{-g}}
\label{zeta definition}
\end{equation}
 appears in all equations of motion, and all coupling constants become $\phi$-dependent due to $\zeta(\phi)$. On the other hand, the function $\zeta(\phi)$ is determined by a {\em constraint}, which is a self-consistency condition  of the equations obtained by varying the metric $g_{\mu\nu}$ and the functions $
\varphi_a$. The constraint and the function $\zeta(\phi)$ it defines  play a key role in obtaining all the new results presented both in the  paper \cite{1-st paper} and in this paper\footnote{To denote the TMT effect, which manifests itself in the appearance of $\zeta(\phi)$ in all equations of motion and the $\zeta(\phi)$-dependence of all coupling constants, we will add “TMT-effective” to the names of the corresponding quantities. For example, "TMT-effective parameters",
 “TMT-effective energy-momentum tensor” and “TMT-effective potential” will be used for parameters, the energy-momentum tensor and potential that appear after performing all steps of the TMT procedure and passing to the Einstein frame in the fields equations.
Instead of this system of equations, it is more convenient to operate with an “effective” action, variation of which leads to these equations. For the corresponding action and Lagrangian in ref.\cite{1-st paper} and in this paper we use the terms “TMT-effective  action” and  “TMT-effective Lagrangian”.
Note that this use of the term "effective" refers to the classical description and, for example, is in no way connected with the usual use of the term "effective potential" calculated taking into account quantum corrections.}. To illustrate this, the following two examples can be given, based on the results of the paper \cite{1-st paper}.
During the cosmological evolution after the end of inflation, a change in $\zeta(\phi)$ leads to a change in the sign of the TMT-effective Higgs mass term in the TMT-effective Lagrangian.
This  effect provides the TMSM answer  to the mystery of the Higgs potential structure and leads to spontaneous symmetry breaking (SSB) in a standard way. 
The second example is related to the  model parameter $\lambda$ of the Higgs self-coupling,  equal to $\lambda\approx 10^{-11}$, which allowed in ref. \cite{1-st paper} to realize a slow-roll inflation  in agreement  with the CMB data, while the constant of non-minimal coupling to the scalar curvature is small, e.g. $\xi=1/6$.
But at the stage of the cosmological evolution approaching the vacuum, due to the $\zeta(\phi)$ dependence, the TMT-effective  Higgs self-coupling
  turns out to be $\lambda_{(near\,vac)}\sim 0.1$ required by the GWS theory. It is important to note that in this effect the role of nonminimal coupling turns out to be minor. These examples demonstrate that in the TMSM formulated as a model of classical field theory, the TMT-effective mass and the Higgs self-coupling behave as a kind of {\em classical running parameters}.
Thus, in TMSM, to achieve very non-trivial results in the entire admissible energy spectrum (from the energies of the electroweak SM to inflationary scales), it is sufficient to use {\em the classical TMT-effective action}. This suggests that the standard approach to quantization and finding the effective potential used in conventional Higgs inflation models may be irrelevant in TMSM.

\vspace{1cm}

{\bf Main idea.}

In light of the above, a more radical idea suggests itself:
 since the homogeneous scalar field $\phi(t)$ used in the Higgs inflation model is the result of cosmological averaging of the local Higgs field $H(x)$, {\em $\phi(t)$ should be regarded as a classical field defined  globally} \footnote{Of course, local cosmological perturbations are possible, but here we are discussing quantization in the context of the SM.}. {\em Such definition of the field $\phi(t)$ makes it meaningless to include $\phi(t)$ in the SM quantization procedure. Instead, the field $\phi(t)$ together with the function $\zeta$ and the curvature describe the classical cosmological background on which the electroweak SM exists as a local quantum field theory.} It should be emphasized once again that the presence of the  function $\zeta$  and the role it plays in TMSM are the main reason that leads to the need for such fundamental theoretical changes compared to conventional models. In this paper, we explore how this idea can be implemented, and we will call the approach developed on the basis of this idea the {\em “Concept of the cosmological realization of TMSM”}. Following this concept, in this paper we show that in addition to the above examples, a similar effect of the appearance of a kind of {\em classical running coupling constants} occurs for all electroweak TMSM  coupling constants.

For a clearer understanding of the main idea of the concept of the cosmological realization of the TMSM, it is worth paying attention to the fact that {\em the Higgs field plays a dual role in the theory being developed}. Since the Higgs field fills the entire space of the Universe, it can be averaged. At each stage of the evolution of the Universe (of course, in a simplified model without matter), this cosmologically averaged field $\phi(t)$ together with the curvature and the scalar $\zeta$ forms the cosmological background. But when studying the SM at a certain stage of cosmological evolution, the Higgs field and other SM fields should be considered as local quantum fields, which we must study on the corresponding cosmological background.

\vspace{0.3cm}

{\bf Main results and organization of the paper.}

In sections \ref{sec:bosonic TMSM} and \ref{Fermions in the primordial TMSM action} we define the TMT primordial action\footnote{In "conventional" alternative gravity theories, the original action differs from the Einstein-Hilbert action only in the form of the Lagrangian, for example, by the presence of a non-minimal coupling with curvature. The transformation to the Einstein  frame, which in conventional theories is performed in the original action, simply changes the set of variables used to describe the theory, but does not change the theory itself. In TMT, the main difference is that the volume measure contains $\Upsilon$, although a modification of the Lagrangian is also possible. Therefore, the original TMT action differs from the Einstein-Hilbert action in the form of the Lagrangian {\em density}, and to bring the TMT action to the Einstein frame, the conformal transformation must exclude from the gravitational term of the action not only the non-minimal coupling, but also $\Upsilon$. In a theory obtained in this way, the constraint determining $\zeta$ cannot arise, that is, by acting in this way we would be dealing with a different theory. It is therefore incorrect to use the term "Jordan frame" in the usual sense for the set of variables in the original action of  TMT. To emphasize this distinction, we will use the terms "primordial variables", "primordial model parameters",  "primordial Lagrangian" and "primordial action" instead of "variables in the original frame",  
"original model parameters", "original Lagrangian" and "original action".} 
for the bosonic and fermion sectors of the $SU(2)\times U(1)$ gauge-invariant electroweak TMSM, which also contains the
 TMT-modifcation of the gravitational action in the Palatini formulation. According the prescription of TMT, all gravity and matter terms of the primordial action contain additional factors in the form of linear combinations $(b_i\sqrt{-g}\pm \Upsilon)$, where $b_i$ is a set of dimensionless model
 parameters. Another fundamentally important novelty is that, unlike conventional models,
 the TMT structure implies the need to take into account the possibility of two different
 types of vacuum-like terms. A summary of the main results of ref.\cite{1-st paper} is also given in section \ref{sec:bosonic TMSM}.

In section \ref{Higgs-grav background} we explore how the classical Higgs+gravity sector of the TMSM can serve as a cosmological background on which a local particle fields theory  must be built. In the FLRW universe, this background is determined by a homogeneous field $\phi(t)$, a curvature, and a scalar function $\zeta$. For such a description of the cosmological background we use the term "cosmological frame" (CF). The presence of $\zeta$ radically distinguishes the description of the cosmological background and its evolution in the TMSM from conventional models. We analyze in particular detail  two stages of evolution of the cosmological background: inflationary stage in the slow-roll regime and  the stage of approaching the vacuum.

Following the concept of the cosmological realization of TMSM, in section \ref{Towards} we study  general features of the description of a local particle field theory on an arbitrary cosmological background. If we consider the Glashow-Weinberg-Salam (GWS) theory as a pattern, then at each specific stage of the evolution of the cosmological background, the particle field theory is realized as a {\em cosmologically modified copy of the electroweak SM}. To describe the latter, we have to operate with a set of variables for which we use the term "local particle physics frame" (LPPF).

In section \ref{TMSM in LPP near vacuum} we show that the  realization of the TMSM in LPPF at the stage of evolution of the cosmological background approaching vacuum  reproduces the tree-level GWS theory. The attractiveness of this  realization is due to the  new possibilities existing in the TMSM, which allow us to obtain the following fundamentally new results that may have great significance both in cosmology and in the GWS theory.

\begin{itemize}

\item

The above mentioned effect of sign reversal of the TMT-effective Higgs mass term in the TMT-effective Lagrangian,  giving the TMSM answer  to the mystery of the Higgs potential structure.

\item

The above mentioned effect of the change of the TMT-effective parameter of the Higgs self-coupling  from its primordial value $\lambda\approx 10^{-11}$ at the inflationary stage (which is in agreement whith the CMB data)
 to the value  $\lambda_{(near\,vac)}\sim 0.1$ required by the GWS theory.

\item

Near the vacuum, the TMT effective gauge coupling constants take the values required by the GWS theory, while the values of the corresponding primordial model parameters (in the primordial TMT-action) are of the order of $g\sim g'\sim 10^{-3}$.
 This effect turns out to be very important in studying the influence of 1-loop quantum corrections on slow-roll inflation.

\item

Possibility to implement fermion mass generation in a standard way due to  SSB, when the Yukawa coupling constants in the  primordial TMT- action are chosen to be universal (the same) for all charged leptons and similar for all up-quarks. Despite the choice of the primordial universal Yukawa coupling constants 
$\approx 10^{-6}$ for leptons and $\approx 10^{-5}$ for up-quarks, the values of the TMT-effective Yukawa coupling constants near the vacuum of the GWS theory give the correct values of the fermion masses. Thus, TMSM is able to resolve the problem of the fermion mass hierarchy. 

\end{itemize} 

In section \ref{TMSM in LPP on back slow-rol} we study the cosmologically modified copy of the electroweak SM  in the LPPF 
 when the cosmological background is in the slow-roll inflation stage. For brevity, we use the name "up-copy" to denote this copy of the electroweak SM, and we also add the prefix 'up' to the names of physical quantities and effects to distinguish them from the corresponding names in the GWS theory.
The study is carried out in the aprroximation where we neglect relative corrections of the order $\lesssim 3\cdot 10^{-2}$. Using the notations $\phi_b(t)$ for the classical slow-rolling  background Higgs field and $R_b$ for the background scalar curvature, we obtain that $\xi R_b<0$ plays the role of the squared mass in the classical TMT-effective Higgs potential. The latter has a minimum at $\tilde{v}=\sqrt{6}M_P$, which within the above-specified accuracy turns out to be independent of the background field $\phi_b(t)$. The up-SSB generates masses of the up-Higgs boson $m_{\tilde{h}}\approx 2.9\cdot 10^{13} \, GeV$
and the up-gauge bosons $M_{\tilde{W}} \approx 7.8\cdot 10^{15}\, GeV$, \, $M_{\tilde{Z}}\approx 8.8\cdot 10^{15}\, GeV$. 

Section \ref{The essence of the cosmological concept} is devoted to additional discussion of some important quantitative results concerning
  the concept of cosmological realization of TMSM. 

In section \ref{Quantization} the features of the quantization of the up-copy in the LPPF are studied in detail and the 1-loop effective potential is calculated. It is shown that, in contrast to the conventional theory, the radiative corrections $\propto g^4\sim g'^4$ have a weak effect on the effective quartic Higgs self-interaction. On the other hand, due to the extreme smallness of the primordial universal Yukawa coupling constants, it turns out that the contribution of the gauge fields significantly dominates the fermion contribution to the 1-loop effective potential. Therefore, the up-minimum of the one-loop effective potential is absolute, and the  vacuum of the up-copy of the electroweak SM is stable (although its description is possible only in the adiabatic approximation). Finally we show that 
taking into account quantum corrections in the 1-loop approximation does not change the fact that inflation quickly turns into a slow-roll regime.

In  section \ref{Discussion} we analyze some key aspects of TMSM and discuss possible directions for its further development. Three appendices provide some important technical calculations and discussions.

\section{Bosonic  sector of the electroweak TMSM at the tree level}
\label{sec:bosonic TMSM}

\subsection{Primordial action of the bosonic sector}
\label{primord action}

The bosonic sector of the electroweak TMSM includes the gravitational, Higgs and gauge fields. Gravity is described in the Palatini formulation and therefore the metric tensor and affine connection are treated as independent variables in the principle of least action. The scalars $\varphi_a(x)$, from which the density $\Upsilon$ of the alternative volume element is constructed, are additional degrees of freedom; their inclusion in the principle of least action is of key importance for the emergence of fundamentally new results that distinguish TMSM from SM+GR. The primordial action is chosen as follows:
\begin{equation}
S_{gr,bos}=S_{gr}+S_{nonmin}+ S_{H}+S_{g}+S_{vac},
\label{S-gr-H-VB}
\end{equation}
where 
\begin{equation}
S_{gr}=-\frac{M_P^2}{2}\int d^4x(\sqrt{-g}+\Upsilon)R(\Gamma,g),
\label{S-gr}
\end{equation}
is the TMT-modification of the gravitational action in the Palatini formulation. 
Here $M_P$  is the reduced Planck mass;  \,  $\Gamma$ stands for affine connection; $R(\Gamma,g)=g^{\mu\nu}R_{\mu\nu}(\Gamma)$,
$R_{\mu\nu}(\Gamma)=R^{\lambda}_{\mu\nu\lambda}(\Gamma)$ and
$R^{\lambda}_{\mu\nu\sigma}(\Gamma)\equiv \Gamma^{\lambda}_{\mu\nu
,\sigma}+ \Gamma^{\lambda}_{\gamma\sigma}\Gamma^{\gamma}_{\mu\nu}-
(\nu\leftrightarrow\sigma)$;
The TMT-modification of the nonminimal coupling of the Higgs isodoublet 
\begin{equation}
H= \left(\begin{matrix} \phi^{+} \\ \phi^{0} \end{matrix}\right) 
\label{Higgs  doublet}
\end{equation}
to the scalar curvature is described by
\begin{equation}
S_{nonmin}=-\xi\int d^4x(\sqrt{-g}+\Upsilon)R(\Gamma,g)|H|^2,
\label{S-nonmin}
\end{equation}
where $|H|^2=H^{\dag}H$. 
In our notations, in a theory with only the volume element $\sqrt{-g}d^4x$, the model of massless scalar field non-minimally coupled  to scalar curvature
would be conformally invariant if the parameter $\xi$ were equal to $\xi=-\frac{1}{6}$.

The  TMT- primordial action $S_H$ for the Higgs field $H$  has a standard $SU(2)\times U(1)$ gauge invariant structure
\begin{equation}
S_{H}=\int d^4x \left[(b_k\sqrt{-g}-\Upsilon) g^{\alpha\beta}\left( \mathit{D}_{\alpha}H\right)^{\dag} \mathit{D}_{\beta}H
 -(b_p\sqrt{-g}+\Upsilon)\lambda |H|^4-(b_p\sqrt{-g}-\Upsilon)m^2 |H|^2\right],
\label{S-H 12}
\end{equation}
where additional factors associated with the existence of two types of volume elements are selected in a special way. In paper \cite{1-st paper}, this choice played a decisive role in the possibility of realizing the idea of Higgs inflation at small $\xi$.
 The parameters $b_k>0$, $b_p>0$,
$\lambda>0$ and $m^2>0$ are chosen to be positive.
The standard definition for the operator 
\begin{equation}
  \mathit{D}_{\mu}=\partial_{\mu}
-ig\mathbf{\hat{T} A}_{\mu}-i\frac{g'}{2}\hat{Y}B_{\mu},
\label{Nabla SM}
\end{equation}
is used with the  gauge coupling constants $g$ and $g'$.
 Here, as usually, $\mathbf{\hat{T}}$ stands for the three generators of the $SU(2)$ group and $\hat{Y}$ is the generator of the U(1) group.

 With standard notations $\mathbf{F}_{\mu\nu}=\partial_{\mu}\mathbf{A}_\nu -\partial_{\nu}\mathbf{A}_\mu +g\mathbf{A}_\mu\times \mathbf{A}_\nu$ and $B_{\mu\nu}=\partial_{\mu}B_\nu -\partial_{\nu}B_\mu$  for the field strengths of the isovector $\mathbf{A}_{\mu}$ and isoscalar $B_{\mu}$, the gauge fields primordial action has the standard $SU(2)\times U(1)$ gauge invariant structure
\begin{eqnarray}
S_{g}=\int d^4x(b_k\sqrt{-g}-\Upsilon)\left[-\frac{1}{4} g^{\mu\alpha}g^{\nu\beta}\mathbf{F}_{\mu\nu}\mathbf{F}_{\alpha\beta}-\frac{1}{4} g^{\mu\alpha}g^{\nu\beta}B_{\mu\nu}B_{\alpha\beta}\right],
 \label{S gauge}
\end{eqnarray}
where the density of the volume measure is chosen to be the same as in the kinetic term of $S_{H}$.

Contribution of the  vacuum-like terms to the  primordial action is defined by\footnote{For the justification of the term with $V_2$, see ref. \cite{1-st paper}} 
\begin{equation}
S_{vac}=\int d^4x\left(-\sqrt{-g}V_1-\frac{\Upsilon^2}{\sqrt{-g}}V_2\right).
\label{S_vac}
\end{equation}
It is convenient to introduce a parametrization $|V_2|=(qM_P)^4$.
In model of ref. \cite{1-st paper} and in this paper we use the choice 
\begin{equation}
V_1=V_2=-(qM_P)^4\approx -(10^{16}GeV)^4, \,\, \text{which means that} \,\,  q^4\approx 3\cdot10^{-10}.
\label{V1=V2=10 16}
\end{equation}

\subsection{Higgs+gravity sector. 
Summary of the main results of ref.\cite{1-st paper}}
\label{summary of 1-st paper}

In the paper \cite{1-st paper}, which was mainly devoted to the application of the model to the study of inflation and in which we worked in the CF, all contributions of gauge fields were omitted.
Therefore, the TMT primordial action of the Higgs+gravity sector reduces from the action (\ref{S-gr-H-VB}) to the following
\begin{eqnarray}
&&S_{H+gr}=\int d^4x\Biggl[(\sqrt{-g}+\Upsilon)\left( -\frac{M_P^2}{2}\right)\left(1+\xi\frac{2|H|^2}{M_P^2}\right)R(\Gamma,g)+\Bigl(\sqrt{-g}+\frac{\Upsilon^2}{\sqrt{-g}}\Bigr)q^4M_P^4
\nonumber
\\
&+&(b_k\sqrt{-g}-\Upsilon) g^{\alpha\beta}\left(\partial_{\alpha}H\right)^{\dag} \partial_{\beta}H
 -(b_p\sqrt{-g}+\Upsilon)\lambda |H|^4-(b_p\sqrt{-g}-\Upsilon)m^2 |H|^2\Biggr].
\label{S-Higgs+gr class}
\end{eqnarray}

Large scale homogeneity and isotropy of our Universe allows to study its evolution as a whole neglecting  inhomogeneities on smaller scales. This means that using the approximation of a homogeneous and isotropic universe is based on the procedure of "cosmological averaging" over an appropriate scales.   If a homogeneous and isotropic spatially flat universe is described by the 
Friedmann-Robertson-Walker (FRW) metric
\begin{equation}
ds^2=dt^2-a^2(t)d\vec{x}\,^2,
\label{FRW metric}
\end{equation}
then the value of the classical scalar field $\phi(t)$ which drives the cosmological evolution at the moment $t$ of the cosmological time, is a result of {\em the cosmological averaging of an inhomogeneous field $\phi(x)$}. Applying this idea to the Higgs isodoublet $H(x)$  we have
\begin{equation}
\langle H(x)\rangle_{cosm. \, aver.}= \left(\begin{matrix} 0 \\ \frac{1}{\sqrt{2}}\phi(t)\end{matrix}\right)
\label{cosm aver Higgs  doublet}
\end{equation}
where  the  field $\phi(t)/\sqrt{2} $ is the only nonzero component of   the Higgs isodoublet $H(x)$ obtained as a result of the cosmological averaging or, using the terminology of refs. \cite{MetricHI-8}-\cite{MetricHI-10},  the cosmological state. In this paper we adopt the idea that, by the method of obtaining $\phi(t)$, it is a classical field, the quantization of which is meaningless in the context of  the electroweak SM quantization  procedure.

Following the prescription of the TMT procedure, we first obtain equations by varying the primordial action (\ref{S-gr-H-VB}) with respect to $\varphi_a$, $ \Gamma^{\lambda}_{\mu\nu}$, $g_{\mu\nu}$ and $\phi$. It is very importat  that as a result of varying the primordial action with respect to $\varphi_a$ an arbitrary integration constant $\mathcal M$ appears. The equations are then rewritten in the Einstein frame using the Weyl transformation
\begin{equation}
\tilde{g}_{\mu\nu}=\bigl(1+\zeta\bigr)\Omega g_{\mu\nu},
 \label{gmunuEin}
\end{equation}
where 
\begin{equation}
\Omega=1+\xi\frac{\phi^2}{M_P^2}
\label{Omega}
\end{equation}
and $\zeta$, which was defined in eq.(\ref{zeta definition}), turns out to be present {\em in all equations of motion}. The value of the 
 integration constant $\mathcal M$ is chosen so that the TMT-effective vacuum energy density $\Lambda$ satisfies the natural constraint $0\leq\Lambda\lesssim (electroweak \, scale)^4$ and it  is achieved if 
\begin{equation}
 {\mathcal M}=2(qM_P)^4(1+\delta), \qquad {\text where} \qquad |\delta|\ll 1,
\label{M 2sqrt}
\end{equation}
Then the self-consistency condition of the resulting system of equations determines $\zeta$ as the following function of $\phi$ and $X_{\phi}$
\begin{equation}
\zeta(\phi, X_{\phi})=\frac{2(qM_P)^4\zeta_v(1+\zeta_v)-(2b_p-1)\frac{\lambda}{4}\phi^4-(2b_p+1)\frac{m^2}{2}\phi^2+(1+b_k)\Omega X_{\phi}}{2(qM_P)^4(1+\zeta_v)+\frac{\lambda}{4}\phi^4-\frac{m^2}{2}\phi^2-(1+b_k)\Omega X_{\phi}},
\label{1+ zeta fine tun phi M}
\end{equation}
where we have neglected $\delta$ compared to 1 and the following definition has been used:
\begin{equation}
X_{\phi}=\frac{1}{2}\tilde{g}^{\mu\nu}\partial_{\mu}\phi\partial_{\nu}\phi.
\label{Xphi}
\end{equation}
To refer to the equality (\ref{1+ zeta fine tun phi M}) we use the term 'constraint'\footnote{It is important to note that in Palatini's formalism $\zeta$ is not a dynamical variable and its value is determined by the constraint (\ref{1+ zeta fine tun phi M})}.
A very important novelty in the TMSM, formulated in the Introductory section \ref{sec:intro} and studied in detail in sections 
\ref{Towards}, \ref{TMSM in LPP near vacuum}, \ref{TMSM in LPP on back slow-rol} and \ref{The essence of the cosmological concept} of this paper, is the need to distinguish between the results of the description in the cosmological frame (CF) and in the local particle physics frame (LPFF).
In particular, it turns out that the VEV of the Higgs field in LPPF differs from  the value  $\sigma$ of the field $\phi$ at the minimum of its potential  in CF
\begin{equation}
\sigma = \langle\phi\rangle_{cf}.
\label{def of sigma}
\end{equation}  
When the field $\phi$ is  in the vacuum state $\sigma$ the value of the scalar function $\zeta(\sigma)=\zeta_v$ is determined by the constraint (\ref{1+ zeta fine tun phi M}) as follows
\begin{equation}
\zeta_v=1+ \mathcal{O}\left(\frac{\lambda \sigma^4}{(qM_P)^4}\right).
\label{zeta v}
\end{equation} 

The gravitational equations in Einstein frame have the canonical GR form
\begin{equation}
R_{\mu\nu}(\tilde{g})-\frac{1}{2}\tilde{g}_{\mu\nu}R(\tilde{g})=\frac{1}{M_P^2}T_{\mu\nu}^{(eff)}
\label{grav eq Ein}
\end{equation}
with the same Newton constant as in the primordial action.
Here $R_{\mu\nu}(\tilde{g})$ and $R(\tilde{g})$ are the Ricci tensor and the scalar curvature of the metric $\tilde{g}_{\mu\nu}$, respectively.
The $\zeta$-dependent form of the TMT-effective energy-momentum tensor $T_{\mu\nu}^{(eff)}$ is as follows 
\begin{equation}
T_{\mu\nu}^{(eff)}(\phi;\zeta)=\frac{1}{\Omega}\Bigl(\frac{b_k-\zeta}{1+\zeta}\phi_{,\mu}\phi_{,\nu}+\tilde{g}_{\mu\nu}X_{\phi}\Bigr)
+\tilde{g}_{\mu\nu}U_{eff}^{(tree)}(\phi;\zeta)
\label{Tmn phi zeta}
\end{equation}
\begin{equation}
U_{eff}^{(tree)}(\phi;\zeta)=\frac{ 1}{(1+\zeta)^2\Omega^2}\Bigl[ q^4M_P^4(\zeta_v^2-\zeta^2)+(1-b_p)\frac{\lambda}{4}\phi^4
-(1+b_p)\frac{m^2}{2}\phi^2\Bigr]
\label{Ueff with zeta}
\end{equation}
After substituting $\zeta(\phi, X_{\phi})$ into $T_{\mu\nu}^{(eff)}(\phi;\zeta)$ its expression  reduces to 
\begin{equation}
T_{\mu\nu}^{(eff)}=\frac{K_1(\phi)}{\Omega}\cdot\big(\phi_{,\mu}\phi_{,\nu}
-\tilde{g}_{\mu\nu}X_{\phi}\big)-K_2(\phi)\cdot \frac{X_{\phi}}{M_P^4}\Big(\phi_{,\mu}\phi_{,\nu}
-\frac{1}{2}\tilde{g}_{\mu\nu}X_{\phi}\Big)
+\tilde{g}_{\mu\nu}U_{eff}^{(tree)}(\phi),
 \label{Tmn 0 phi K1 K2}
\end{equation}
where the TMT effective potential $U_{eff}^{(tree)}(\phi)$ and
 functions $K_1(\phi)$ and $K_2(\phi)$
are defined as follows:
\begin{equation}
U_{eff}^{(tree)}(\phi)=\frac{q^4\Bigl[(\zeta_v+b_p)\lambda\phi^4-2(\zeta_v-b_p)m^2\phi^2 \Bigr]+\frac{1}{M_P^4}\bigl(\frac{\lambda}{4}\phi^4
-\frac{m^2}{2}\phi^2\bigr)^2}{4\Omega^2\left[q^4(1+\zeta_v)^2+(1-b_p)\frac{\lambda}{4}\frac{\phi^4}{M_P^4}-(1+b_p)\frac{m^2\phi^2}{2M_P^4}\right]},
\label{V eff}
\end{equation}

\begin{equation}
K_1(\phi)=\frac{2q^4(1+\zeta_v)(b_k-\zeta_v)+(2b_p+b_k-1)\frac{\lambda}{4}\frac{\phi^4}{M_P^4}+(2b_p+1-b_k)\frac{m^2\phi^2}{2M_P^4}}
{2\left[q^4(1+\zeta_v)^2+(1-b_p)\frac{\lambda}{4}\frac{\phi^4}{M_P^4}-(1+b_p)\frac{m^2\phi^2}{2M_P^4}\right]},
\label{K1 phi}
\end{equation}

\begin{equation}
K_2(\phi)=\frac{(1+b_k)^2}{2\left[q^4(1+\zeta_v)^2+(1-b_p)\frac{\lambda}{4}\frac{\phi^4}{M_P^4}-(1+b_p)\frac{m^2\phi^2}{2M_P^4}\right]}.
\label{K2 phi}
\end{equation}

The $\phi$-field equation in the Einstein frame reads
\begin{eqnarray}
&& 
\frac{1}{\sqrt{-\tilde{g}}}\partial_{\mu}\Bigg(\frac{b_k-\zeta}{(1+\zeta)\Omega}\sqrt{-\tilde{g}}\tilde{g}^{\mu\nu}\partial_{\nu}\phi\Bigg)
+\frac{b_p+\zeta}{(1+\zeta)^2\Omega^2}\lambda\phi^3+\frac{b_p-\zeta}{(1+\zeta)^2\Omega^2}m^2\phi
  \nonumber
\\
&&
+\xi R(\tilde{g})\frac{\phi}{\Omega}=0,
\label{phi eq Ein with zeta and bk-zeta}
\end{eqnarray}
where the last term  appears in the equation due to non-mimimal coupling.
The expression for the scalar curvature $R(\tilde{g})$ is obtained from the Einstein equations (\ref{grav eq Ein})
\begin{equation}
R(\tilde{g})=
\frac{1}{M_P^2}\Bigl(2\frac{K_1(\phi)}{\Omega}X_{\phi}-4U_{eff}^{(tree)}(\phi)\Bigr).
\label{R in Ein vua K1}
\end{equation}
 The performed procedure, which also includes the substitution of $R(\tilde{g})$ into eq.(\ref{phi eq Ein with zeta and bk-zeta}), means that we are dealing with an explicitly formulated self-consistent system "Higgs field $+$ gravity".

In the vicinity of the vacuum  where $\zeta\approx\zeta_v= 1$ and $\langle\phi\rangle_{cf} =\sigma$,  eq.(\ref{phi eq Ein with zeta and bk-zeta}) reduces to
\begin{equation}
\frac{1}{\sqrt{-\tilde{g}}}\partial_{\mu}\bigl(\sqrt{-\tilde{g}}\tilde{g}^{\mu\nu}\partial_{\nu}\phi\bigr)+V'_{eff,vac}(\phi)=0
\label{phi eq via Vprime}
\end{equation}
where  the derivative $V'_{eff,vac}(\phi)$
of the Higgs field  potential in the vicinty of the  vacuum has the  form \footnote{In eqs.(\ref{phi eq via Vprime}) and (\ref{Vprime near vac}), the corrections $\propto\frac{\sigma}{M_P}$  are  omitted. In section \ref{The essence of the cosmological concept} we  find the value of $\sigma$, from which it follows
 that $\frac{\sigma^2}{M_P^2}\sim 10^{-27}$.} 
\begin{eqnarray}
V'_{eff,vac}(\phi)&=&\frac{1}{(1+\zeta_v)(b_k-\zeta_v)}\bigl[(b_p+\zeta_v)\lambda\phi^2
+(b_p-\zeta_v)m^2\bigr]\phi
\label{Vprime near vac}
\end{eqnarray}
One should pay special attention that in order to obtain equation (\ref{phi eq via Vprime}) in canonical form, it was necessary to devide eq.(\ref{phi eq Ein with zeta and bk-zeta}), considered near the vacuum, by the constant $\frac{b_k-\zeta_v}{1+\zeta_v}$.
 It is seen  that if $\zeta_v>b_p$, SSB occurs with
 the following expression for the VEV of the classical scalar field $\phi$ in the CF
\begin{equation}
\sigma^2=\frac{\zeta_v-b_p}{b_p+\zeta_v}\cdot\frac{m^2}{\lambda}.
\label{s2}
\end{equation}
Eqs. (\ref{phi eq via Vprime}) and (\ref{Vprime near vac})  can be rewritten in  the  typical  SM form 
 \begin{equation}
\frac{1}{\sqrt{-\tilde{g}}}\partial_{\mu}\bigl(\sqrt{-\tilde{g}}\tilde{g}^{\mu\nu}\partial_{\nu}\phi\bigr)+\lambda_{(cf)}\phi^3-m_{(cf)}^2\phi=0,
\label{h eq}
\end{equation}
where the expressions in the CF for the parameters of the quartic self-coupling and the mass squared near a vacuum appear
\begin{equation}
\lambda_{(cf)}=\frac{b_p+\zeta_v}{(1+\zeta_v)(b_k-\zeta_v)}\lambda,    \qquad   m_{(cf)}^2=\frac{\zeta_v-b_p}{(1+\zeta_v)(b_k-\zeta_v)}m^2.
\label{lambda and m2 SM}
\end{equation}

The cosmological constant $\Lambda$, which is defined as the energy density in the vacuum state $\sigma$ is  $\Lambda=U_{eff}(\sigma)$. The expression  $U_{eff}^{(tree)}(\phi)$ in eq.(\ref{V eff}) was obtained using the value of the integration constant ${\mathcal M}=2(qM_P)^4$ (see eq.(\ref{M 2sqrt})), that is, by setting $\delta=0$. In fact,  playing with possible values of $|\delta|\ll 1$, one can obtain any value of $\Lambda$ in the interval, for example, 
$0\leq\Lambda\lesssim (electroweak \, scale)^4$. 

Instead of the system of the Einstein equation (\ref{grav eq Ein}) with the energy-momentum tensor
(\ref{Tmn 0 phi K1 K2})-(\ref{K2 phi}) and the field $\phi$-equation (\ref{phi eq Ein with zeta and bk-zeta}), it is more convenient to proceed with the
TMT-effective action, the variation of which gives these equations. As usual, if $\tilde{g}^{\alpha\beta}\phi_{,\alpha}\phi_{,\beta}>0$,  the TMT-effective energy-momentum tensor $T_{\mu\nu}^{(eff)}$ can be rewritten in the form of a perfect fluid. Then the pressure density plays the role of the matter Lagrangian  in the  action, and we arrive
at the {\em tree level TMT-effective action}
\begin{equation}
S_{eff}^{(TMT)}=\int \left(-\frac{M_P^2}{2}R(\tilde{g})+L_{eff}^{(tree)}(\phi, X_{\phi})\right)\sqrt{-\tilde{g}}d^4x,
\label{Seff}
\end{equation}
with the following tree level TMT-effective Lagrangian
\begin{equation}
L_{eff}^{(tree)}(\phi, X_{\phi}))=K_1(\phi)\frac{X_{\phi}}{\Omega}-\frac{1}{2}K_2(\phi)\frac{X_{\phi}^2}{M_P^4}-U_{eff}^{(tree)}(\phi).
\label{Leff}
\end{equation}
Note that $L_{eff}^{(tree)}(\phi, X_{\phi}))$ has the form typical for K-essence models \cite{Kess1}-\cite{Kess4}. 
This is why the variation of the action (\ref{Seff}) with respect to $\phi$ gives the equation  (\ref{phi eq Ein with zeta and bk-zeta}) with 
non-canonical kinetic term.

  In ref. \cite{1-st paper}, after redefining the field $\phi$ using the relation 
\begin{equation}
\phi=\frac{M_P}{\sqrt{\xi}}{\sinh}\left(\sqrt{\xi}\frac{\varphi}{M_P}\right),
\label{phi via canonical varphi}
\end{equation}
the TMT-effective potential $U_{eff}(\phi)$ was expressed in terms of $\varphi$ 
\begin{equation}
U_{eff}^{(tree)}(\varphi) =\frac{\lambda M_P^4}{4\xi^2}{\tanh}^4z\cdot F(z), \quad \text{where} \,\, z=\sqrt{\xi}\frac{\varphi}{M_P},
\label{Veff varphi tanh no fine tun 1}
\end{equation}
\begin{equation}
F(z)=\frac{(\zeta_v+b_p)q^4+
\frac{\lambda}{16\xi^2}\sinh^4z-\frac{\lambda m^2}{4\xi M_P^2}\sinh^2z+\frac{m^4}{4\lambda M_P^4}
-2 (\zeta_v-b_p)q^4\frac{\xi m^2}{\lambda M_P^2}\cdot \sinh^{-2}z }
{(1+\zeta_v)^2q^4+(1-b_p)\frac{\lambda}{4\xi^2}{\sinh}^4z -(1+b_p)\frac{m^2}{2 M_P^2\xi}\sinh^2z}.
\label{Veff varphi tanh no fine tun 2}
\end{equation}
 The model parameter $b_p\approx 0.5$. 
It turns out that $U_{eff}^{(tree)}(\varphi)$ can have one or two plateaus, depending of the value of the model parameter $q^4$. 
In the case of choosing parameters as in eq.(\ref{V1=V2=10 16}), $U_{eff}^{(tree)}(\varphi)$ has one plateau of height 
\begin{equation}
U_{eff}^{(tree)}|_{plateau}\approx \frac{\lambda}{8\xi^2} M_P^4 
\label{Veff plateau}
\end{equation}
\begin{figure}
\includegraphics[width=13.0cm,height=9cm]{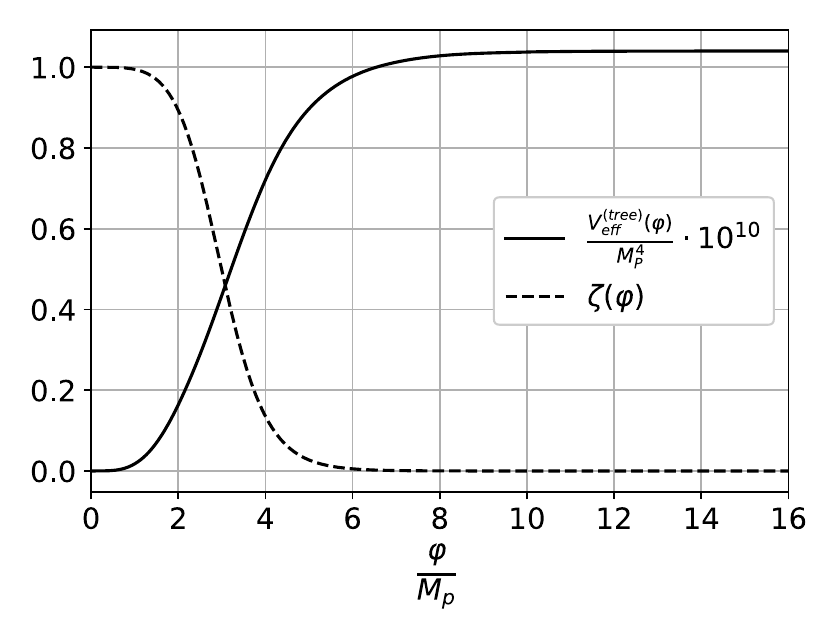}
\caption{The solid curve is a plot of the TMT-effective potential $10^{10}U_{eff}^{(tree)}(\varphi)/M_P^4$ defined by eqs.(\ref{Veff varphi tanh no fine tun 1})
 and (\ref{Veff varphi tanh no fine tun 2}), has one plateau due to the choice of $q^4=3\cdot 10^{-10}$, i.e. $|V_1|=|V_2|\approx (10^{16}GeV)^4$, while other model parameters are  $\lambda = 2.3\cdot 10^{-11}$, \,
 $\xi=\frac{1}{6}$, \, $b_p=\frac{1}{2}(1+\delta_p)$ \, $\delta_p= 10^{-8}$, \, $m=0.7GeV$.
The dashed line is  the graph of the scalar function $\zeta(\varphi)$ defined by the constraint (\ref{1+ zeta fine tun phi M}) represented via $\varphi$, where the terms $\propto \frac{X_{\varphi}}{M_P^4}$ are assumed to be negligible compared to the other terms. The $\zeta(\varphi)$ curve intersects $\zeta=0$ at $\varphi\approx 14M_P$; the intersection point is very sensitive to the choice of  $\delta_p$.}
\label{fig1}
\end{figure}
and its
 shape  is shown in figure \ref{fig1}. The study of   the slow-roll inflation was carried out with the choice of a non-minimal coupling constant to the scalar curvature $\xi=\frac{1}{6}$. The resulting inflation parameters are consistent with Planck's observational data if $\frac{\lambda}{8\xi^2}\sim 10^{-10}$, that is if the parameter $\lambda$ in the primordial action (\ref{S-H 12})  is of order $\lambda\sim 10^{-11}$. Such a tiny value of the primordial self-coupling constant of the Higgs field obviously contradicts the SM. This problem is well known in Higgs inflation models and is usually solved by choosing a giant value of $\xi\sim 10^{4}-10^{8}$.
In the model of ref. \cite{1-st paper}, the key to solving this problem is that in our TMSM the TMT-effective expression for $\lambda$ near vacuum, $\lambda_{cf}$, eq.(\ref{lambda and m2 SM}), contains a factor $( b_k-\zeta_v)^{-1}$, which can compensate for the exceptional smallness of $\lambda$.
 It is important that this is achieved by a completely natural selection of the very small deviation of the parameter $b_k$ from  $\zeta_v=1$. 
 In section \ref{Bosons  in LPP near vacuum} we will show that for comparison with what is known near the vacuum of the GWS theory, the results should be presented in LPPF. Then, to ensure the value of $\lambda_{v}\sim 0.1$ required by the GWS theory, we should choose $b_k$ such that $b_k-1\approx 10^{-5}$. It will also be shown there that $m\approx 0.7 \, GeV$, that is $\frac{m^2}{M_P^2}\sim 8\cdot 10^{-38}$.

\section{Fermions in the primordial TMSM action}
\label{Fermions in the primordial TMSM action}

To include fermions in the TMSM, we start by adding the appropriate fermion sector terms to the primordiall TMT action (\ref{S-gr-H-VB}) describing the gravitational and bosonic sector of the TMSM. This will be done in accordance with the basic idea of TMT, that is, involving two volume elements $\sqrt{-g}d^4x$ and $\Upsilon d^4x$ in the primordial action. 

\subsection{Leptons}
\label{leptons in primordial}

We will study a model including 3 generations of leptons
\begin{equation}
L_l=\frac{1-\gamma_5}{2} \left(\begin{matrix} \nu^{(l)} \\ l \end{matrix}\right);
\qquad l_R=\frac{1+\gamma_5}{2} l ; \qquad  \nu^{(l)}_R=\frac{1+\gamma_5}{2}\nu^{(l)}; \qquad  l=e, \mu, \tau
\label{GWS fermions 3}
\end{equation}
To construct a generally coordinate invariant kinetic terms of the action for fermionic sector we have to use the  covariant operator
\begin{equation}
\nabla_{\mu}=D_{\mu}+\frac{1}{2}\omega_{\mu}^{ik}\sigma_{ik},
\label{nabla  f}
\end{equation}
where $D_{\mu}$  defined by Eq.(\ref{Nabla SM}), $\sigma_{ik}=\frac{1}{2}(\gamma_i\gamma_k -\gamma_k\gamma_i)$, \, $\omega_{\mu}^{ik}$
 is the affine spin-connection
\begin{equation}
\omega_{\mu}^{ik}=\frac{1}{2}\Bigl[ V^{i\lambda}\bigl(\partial_{\mu} V^k_{\lambda}+\Gamma^{\beta}_{\mu\lambda} V^k_{\beta}\bigr)-
(i \leftrightarrow k)\Bigr]
\label{affine spin-conn}
\end{equation}
 and $V_{k}^{\mu}$ is vierbein ($i, k$ are Lorentz indeces).  

Turn now to  the Yukawa coupling terms responsible for the fermion mass generation after SSB. The standard method is that for each generation of charged leptons, the Yukawa coupling constants $y_e$, $y_{\mu}$, $y_{\tau}$ are selected to ensure agreement with the experimental data.
As is well known, this aspect of SM contains a flaw indicating insufficient completeness of the theory:  in order to obtain the required values of the masses of charged leptons,  the corresponding Yukawa coupling constants must vary more than three orders of magnitude 
$y_e:y_{\mu}:y_{\tau}=m_e:m_{\mu}:m_{\tau}$.  In the quark sector, when trying to describe, for example, the masses of up-quarks through their Yukawa coupling to the Higgs field, the range of scatter of the corresponding Yukawa constants is 5 orders of magnitude. The situation with the choice of the value of the Yukawa coupling constant necessary to obtain the top quark mass is also  very strange. In the framework of the GWS theory where particle masses arise as a result of the SSB, the spread of Yukawa coupling constants required for the observed fermion mass hierarchy looks unnatural for a theory that claims to correctly describe nature.
In this regard, the SM extension implemented in TMSM provides an unexpected opportunity to circumvent this hierarchy problem.
 It turns out that to construct a realistic TMSM with three generations of leptons, it is possible to suppose that {\bf Yukawa coupling constant $y^{(ch)}$ is chosen to be universal for all charged leptons}. Then the problem of the mass hierarchy is solved simply by using the freedom to choose the coefficients in the linear combination of the densities of the volume measures $\sqrt{-g}$ and $\Upsilon$ when constructing the volume elements. Similar to selecting signs  in the primordial action of the bosonic sector, eqs.(\ref{S-H 12}) and (\ref{S gauge}), for charged leptons we choose  volume elements $(b_l\sqrt{-g}-\Upsilon)d^4x$ for kinetic terms  and $(b_l\sqrt{-g}+\Upsilon)d^4x$ for the Yukawa coupling terms  in the primordial action. The basis for the success of the idea that makes it possible to obtain 
the masses of charged leptons with one universal  Yukawa coupling constant in the primordial action lies in the possibility
of such a choice of the parameters $b_l$ ($l=e, \mu, \tau$ ) 
that the deviations of their values from unity are small.

To simplify the first attempt to implement the conceived idea, in this paper we completely ignore neutrino mixing and limit ourselves to considering only the electron neutrino $\nu_{e}$, focusing on a fundamentally new effect that TMSM gives us: explaining the exceptional smallness of the electron neutrino mass. For brevity, we will further omit the subscript $e$ in $\nu_{e}$. For the  right isoscalar  neutrino $\nu_R$ we choose the volume elements $(b_e\sqrt{-g}-\Upsilon)d^4x$ and $(b_{\nu}\sqrt{-g}+\Upsilon)d^4x$  in the primordial action for the kinetic term and for the Yukawa coupling term  respectively. The neutrino Yukawa coupling constant $y_{\nu}$ will be chosen defferent from the one for the charged leptons: $y_{\nu}\neq y^{(ch)}$.

Following these ideas, we adopt  the following  general coordinate invariant and $SU(2)\times U(1)$ gauge-invariant primordial TMSM action for the lepton sector
\begin{eqnarray}
S_{(lept)}&=&\int d^4x\sum_{l=e,\mu,\tau} (b_l\sqrt{-g}-\Upsilon)\frac{i}{2} \bigl[\overline{L_l}\gamma^{\mu}\nabla_{\mu}L_l -(\nabla_{\mu}\overline{L_l})\gamma^{\mu}L_l+
\overline{l}_R\gamma^{\mu}\nabla_{\mu}l_R -(\nabla_{\mu}\overline{l}_R)\gamma^{\mu}l_R\bigr]
\nonumber
\\
&-&\int d^4x\sum_{l=e,\mu,\tau} (b_l\sqrt{-g}+\Upsilon)\cdot y^{(ch)}\left(\overline{L}_lH l_R +
\overline{l}_RH^{\dag}L_l\right)
\nonumber
\\
&+&\int d^4x (b_e\sqrt{-g}-\Upsilon)\frac{i}{2}\bigl[\overline{\nu_R}\gamma^{\mu}\nabla_{\mu}\nu_R -(\nabla_{\mu}\overline{\nu_R})\gamma^{\mu}\nu_R\bigr]
\nonumber
\\
&-&\int d^4x(b_{\nu}\sqrt{-g}+\Upsilon)\cdot y_{\nu}\left(\overline{L}_eH_c \nu_R +
\overline{\nu_R}H_c^{\dag}L_e\right),
\label{S l3}
\end{eqnarray}
where $\gamma^{\mu}=V_{k}^{\mu}\gamma^{k}$, \, $\gamma^{k}$ are Dirac matrices; $H_c$ is the charged
 conjugated to the Higgs isodoublet $H$. 

\subsection{Quarks}
\label{quarks in primordial}

We will consider three generations of quarks of the electroweak $SU(2)\times U(1)$ gauge invariant theory, and use the following notations 
for the left $SU(2)$ doublets 
\begin{equation}
L_q=\frac{1-\gamma_5}{2} \left(\begin{matrix} q \\ K_{qd}d+K_{qs}s+K_{qb}b \end{matrix}\right),
 \qquad q=u, c, t  
\label{left doublet q}
\end{equation}
and for the right singlets of the up-quarks $u$, $c$, $t$
\begin{equation}
 r_q=\frac{1+\gamma_5}{2}q,   \qquad q=u, c, t .
\label{right singlets up quarks}
\end{equation}
To preserve the general principles of the construction of our TMSM, we must assume that, like leptons, the quarks Yukawa coupling constants in the primordial  action must also be universal: one for all up-quarks and perhaps another for all down-quarks (with the corresponding modification caused by the Kobayashi-Maskawa mixing matrix $K_{qd_j}$, $q=u, c, t; \, j=1,2,3;  \, d_1=d, \, d_2=s, \, d_3=b.$). Again, as in the case of leptons, the implementation of such an approach to describing quarks taking into account quark mixing would require too many detailed calculations, which could become an obstacle to demonstrating the main issues of the model under study. Therefore, in this paper we ignore the mass generation of down quarks and concentrate only on some of the specific fundamental issues that distinguish TMSM from SM+Gravity. Namely, the description of the up-quarks and their mass generation in TMSM is one of them. That's why with this approach we won't need to use right singlets of the down-quarks.

The general coordinate invariant and $SU(2)\times U(1)$ gauge-invariant primordial TMSM action  with the universal Yukawa coupling constant $y^{(up)}$,
which allows one to study  the up-quarks,  is chosen as follows\footnote{Of course, as usual in the electroweak model, a standard choice of charges for all fermions is assumed, as well as three color degrees of freedom for the quarks.}
\begin{eqnarray}
S_{(quarks)}&=&\int d^4x\sum_{q=u,c,t}\Biggl[ (b_q\sqrt{-g}-\Upsilon)\frac{i}{2} \Biggl(\overline{L_q}\gamma^{\mu}\nabla_{\mu}L_q
 -(\nabla_{\mu}\overline{L_q})\gamma^{\mu}L_q
\label{Sq}
\\
&+&\overline{r_q}\gamma^{\mu}\nabla_{\mu}r_q -(\nabla_{\mu}\overline{r_q})\gamma^{\mu}r_q\Biggr)
- (b_q\sqrt{-g}+\Upsilon)\cdot y^{(up)}\left(\overline{L_q}H_c r_q +
\overline{r_q}H_c^{\dag}L_q\right)\Biggr].
\nonumber
\end{eqnarray}

\section{Higgs+gravity sector as a cosmological background 
\\
of the tree level TMSM}
\label{Higgs-grav background}

\subsection{General view}
\label{Back General view}

TMSM differs fundamentally from  conventional models in the way the Higgs sector of the electroweak SM is used to study cosmology.
Returning to the Higgs+gravity  sector of TMSM investigated in ref. \cite{1-st paper} and briefly described in section \ref{summary of 1-st paper}, we argue that eq.(\ref{phi eq Ein with zeta and bk-zeta}) together with the Einstein equations (\ref{grav eq Ein}) with the TMT effective energy-momentum tensor
(\ref{Tmn 0 phi K1 K2})-(\ref{K2 phi}), or, equivalently, the TMT-effective action (\ref{Seff})-(\ref{Leff}) are applicable over the entire range of values of 
$\phi\geq \sigma$  allowed by classical cosmology.
Thus, {\em TMSM provides us with a unique opportunity to study cosmology in the interval of $\phi$ from  $\phi \approx\sigma$ up to $\phi\gg M_P$, corresponding to the energy scale of inflation. while remaining in the tree approximation of the electroweak SM.}
The possibility of Higgs inflation with a small non-minimal coupling constant, together with the ability to provide the Higgs field self-coupling constant necessary for the electroweak SM, which was demonstrated in  paper \cite{1-st paper}, shows the effectiveness of this approach. A key role in the ability of TMSM to realize such a cosmological scenario is played by the scalar $\zeta$, determined by the constraint (\ref{1+ zeta fine tun phi M}), which will now be conveniently used in the following equivalent form
\begin{equation}
\zeta(\phi,X_{\phi})
=\frac{2q^4\zeta_v(1+\zeta_v)-(2b_p-1)\frac{\lambda\phi^4}{4M_P^4}-
(2b_p+1)\frac{m^2\phi^2}{2M_P^4}+(1+b_k)\Omega \cdot\frac{X_{\phi}}{M_P^4}}
{2q^4(1+\zeta_v)+\frac{\lambda\phi^4}{4M_P^4}-\frac{m^2\phi^2}{2M_P^4}-(1+b_k)\Omega \cdot\frac{X_{\phi}}{M_P^4}}.
\label{zeta for radical}
\end{equation}

In the  model of the Universe constructed and studied in ref. \cite{1-st paper}, we completely ignored the gauge and fermion fields of the electroweak SM, as well as the $SU(2)\times U(1)$ gauge symmetry. Equally important, in studying cosmological evolution, instead of describing the Higgs isodoublet as a local quantum field that can be represented as
\begin{equation}
H(x)= \frac{1}{\sqrt{2}}\left(\begin{matrix} 0 \\ \phi(x) \end{matrix}\right) 
\label{Higgs in unitary}
\end{equation}
 in the unitary gauge, we used the cosmological average (\ref{cosm aver Higgs  doublet}) over the entire space of a spatially flat FLRW Universe, that is
\begin{equation}
\langle \phi(x)\rangle_{cosm.av.}=\phi(t)
\label{cosm av}
\end{equation}

We now {\em define the background scalar field $\phi_{back}(t)$ at a certain stage of cosmological evolution} as the function $\phi(t)$ that is obtained by solving the cosmological equations corresponding to that stage of evolution
\begin{equation}
\langle H(x)\rangle_{cosm.av.}|_{back}=\frac{1}{\sqrt{2}}\left(\begin{matrix} 0 \\ \phi_{back}(t) \end{matrix}\right) 
\label{H cosm av, back}
\end{equation}
Continuing, as in paper \cite{1-st paper}, to study the evolution of the cosmological background considered in the CF, we will be able to further clarify the idea
of describing particle physics in the LPPF on a cosmological background.
  It is evident that the cosmological evolution considered in the CF can be divided into stages, such as
 inflation in the slow-roll regime, the last stage of inflation, the post-inflationary stage, the stage of approaching vacuum.
To implement TMSM, which includes all SM fields, \underline{it seems logical to treat the model constructed}
\underline{ in the 
CF as a background on which all SM fields are considered locally}: in a certain interval of cosmic time $\Delta t$ (such as, for example, the duration of one of the listed stages of cosmological evolution) and in an arbitrarily large region of space. Using general coordinate invariance, the theory of the electroweak SM fields in this four-dimensional spacetime can be described in arbitrary local coordinates $({\mathsf x})$. 
In what follows, we will use the notation ${\mathcal V}_4$  for this four-dimensional space-time\footnote{In studying quantum corrections we will follow the paper \cite{Markkanen}, where the 1-loop effective potential is evaluated on an arbitrary gravitational background. However, in the final application of the results to cosmology we will restrict ourselves to considering the two simplest (and opposite in the history of the Universe) cases: Minkowski space and de Sitter space. With this in mind, when we need to compare the description of the $\phi$ field in coordinates $(t, \vec{x})$ used in the Friedmann metric 
(\ref{FRW metric})  with the description in coordinates $({\mathsf x})$ used in ${\mathcal V}_4$, we will be able to choose the time coordinate of  $({\mathsf x})$ so that it coincides with the cosmic time $t$.}.

The describtion of a background  in the CF implies that we  use  the coordinates $(t, \vec{x})$ and metric (\ref{FRW metric}) and we know:

1) the classical field $\phi_{back}(t)$, which fills all  space, drives  the cosmological evolution and satisfies eq.(\ref{phi eq Ein with zeta and bk-zeta});

2) the scalar $\zeta=\zeta_{back}$ defined  by eq.(\ref{zeta for radical}), where  $\phi$ is replaced by $\phi_{back}(t)$, and the corresponding replacement 
is made for $X_{\phi}$;

3) The Ricci tensor and scalar curvature obtained from the Einstein eqs.(\ref{grav eq Ein}) using the TMT-effective energy-momentum tensor, eqs.(\ref{Tmn 0 phi K1 K2})-(\ref{K2 phi}).
In particular,  the scalar curvature $R(\tilde{g})$ is defined by eq.(\ref{R in Ein vua K1}).

 In this paper, we will limit ourselves to only studying how the electroweak TMSM is realized at two background stages: at the stage of approaching a vacuum and at the stage of inflation in the slow-roll regime. But before we do this, we will need a more detailed study of the relevant background stages.

\subsection {Some features of the evolution of the cosmological background}
\label{some features of back}

As noted above,, the scalar function $\zeta$ plays a key role in the fact that the TMSM studied in the CF is capable of describing the evolution of the Universe at the classical level over the entire range of energy scales. Therefore, at each stage of evolution, we must know the values of $\zeta$, which are found using constraint (\ref{zeta for radical}). When 
$\phi$ is only a few orders of magnitute larger than $\sigma$, the contribution of  terms $\propto\frac{X_{\phi}}{M_P^4}$ to the value of $\zeta$ can obviously be neglected. 
In appendix \ref{append} is shown that during inflation in the slow-roll regime, the contribution of the terms $\propto\frac{X_{\phi}}{M_P^4}$ to the constraint is subdominant.  But immediately after the end of inflation, the contribution of the terms  $\propto\frac{X_{\phi}}{M_P^4}$  to the constraint can be significant. In this paper we focus on studying the new effects that TMSM predicts for the inflationary era and for the stage of approaching the energy scale of the electroweak SM. Therefore, leaving for future study the predictions of TMSM for the intermediate stage of cosmological evolution, we can consider 
the contribution of  the terms $\propto\frac{X_{\phi}}{M_P^4}$ to  the constraint (\ref{zeta for radical})  to be subdominant or negligible.
In this context, the monotonic dependence of $\zeta$ on $\phi$ under the condition $X_{\phi} = 0$, shown in figure \ref{fig1} (using $\varphi$ instead of $\phi$) is correct in the regions $\phi\gg M_P$ and $\phi\ll M_P$. But in the intermediate region the curve is only illustrative. since it is possible that due to the contribution of $X_{\phi}$ to the constraint, immediately after the end of inflation the function
$\zeta(t)$ does not change monotonically. 
Working within the framework of this approach,  we see  that a monotonic decrease in the homogeneous field $\phi(t)$ in a huge range from $\phi\gg M_P$ at the inflationary era to $\phi\approx \sigma$  entails such a change in $\zeta$ that  $\zeta\gtrapprox 0$ for $\phi\gg M_P$ and $\zeta\approx \zeta_v =1$ for $\phi\approx \sigma\lll M_P$.  The following facts are of key importance for the results of this paper: 

1) Both in the era of inflation and at the stage of approaching a vacuum, $\zeta$ increases monotonically. 

2) When $\phi$ is close enough to $\sigma$, it is naturally to use that $\frac{\lambda\sigma^4}{M_P^4}=2.3\cdot 10^{-11}\frac{\sigma^4}{M_P^4}\lll q^4\zeta_v(1+\zeta_v)=6\cdot 10^{-10}$ and $\frac{m^2\sigma^2}{M_P^4}\lll 10^{-10}$, where  we used the  values of $q^4$ and $\lambda$ given in section \ref{summary of 1-st paper}. Indeed, from the results obtained in 
sections \ref{Bosons  in LPP near vacuum} and \ref{The essence of the cosmological concept} it follows
that $\frac{m^2}{M_P^2}\sim 10^{-38}$ and  $\frac{\sigma^2}{M_P^2}\sim 10^{-27}$. Therefore, even if $\phi$ is several tens of times larger than $\sigma$, the terms in the constraint
(\ref{zeta for radical}), proportional to $\lambda$ and $m^2$, can be neglected. Moreover, since $\phi$ monotonically approaches $\sigma$, it can be argued that the contribution of the terms $\propto\frac{\dot{\phi}^2}{M_P^4}$ to the constraint (\ref{zeta for radical}) is negligible.
Once the field $\phi$ reaches the oscillatory regime near $\sigma$, we can expect the oscillation amplitude to be only a few orders of magnitude larger than $\sigma$, and the ratio of the characteristic oscillation frequency to $M_P$ is negligible compared to unity. That is, in the oscillatory regime, the contribution of the terms $\propto\frac{\dot{\phi}^2}{M_P^4}$ to the constraint (\ref{zeta for radical}) is also negligible.  Thus, when $\phi$ approaches $\sigma$ or oscillates around $\sigma$, we can  assert that $\zeta=1$ with exceptional accuracy.

3)
It is not excluded that   $\zeta$ can be taken equal to unity with very high accuracy in a much more wide range  of $\sigma<\phi\lesssim 10^{15}GeV$, if the contribution of $(1+b_k)\Omega \cdot\frac{X_{\phi}}{M_P^4}\lesssim 2\Bigl(1+\mathcal{O}\big(10^{-7}\bigr)\Bigr)\frac{X_{\phi}}{M_P^4}$ to the constraint (\ref{zeta for radical}) in this range occurs to be negligible. The study of cosmological dynamics in this  range of $\phi$ is complicated by the need in numerical solutions which take into account the K-essense type structure. This goes well beyond the purposes of this paper and will be done somewhere.

4)  An increase in $\zeta$ from $\zeta\approx 0$ to $\zeta=1$ inevitably leads to  a change in the sign of the mass term $\propto (b_p-\zeta)m^2\phi$ in the $\phi$-equation (\ref{phi eq Ein with zeta and bk-zeta}). This new TMSM effect enables spontaneous symmetry breaking (SSB) near vacuum, providing an answer to the mystery of the origin of the negative mass term in the Higgs potential. It would be interesting to know at what stage in the evolution of the cosmological background this effect occurs. According to the iilustrative curve $\zeta(\varphi)$ in figure \ref{fig1}, this can be at $\varphi\approx 3M_P$. However this plot was obtained by completely ignoring the contributions of the terms $\propto X_{\phi}$ to the constraint (\ref{zeta for radical}).  The value $\varphi\approx 3M_P$ lies in the interval corresponding to post-inflationary stage, where the contributions of the terms $\propto X_{\phi}$ can be significant. Therefore,  to clarify the question at what stage of evolution the effect of changing the sign of the mass term occurs, a numerical solution of cosmological equations is required, which will be done somewhere later.

5) 
The cosmological background formed during the stage of slow-roll inflation was studied in detail in ref. \cite{1-st paper}. In addition to what was done there, as part of the description in the CF, in the next section we will also need to know both  the expression for the scalar curvature $R ( \tilde{g})$ and the behavior of $\zeta(\phi)$ during slow-roll inflation. For the chosen model parameters, the latter occurs at $\phi \gtrsim 14.2M_P$, which corresponds to 
$\varphi\gtrsim 6M_P$, where the kinetic terms in the TMT effective energy-momentum tensor, eq.(\ref{Tmn 0 phi K1 K2}), are negligible with respect to the TMT effective potential $U_{eff}^{(tree)}(\phi)$, eq.(\ref{V eff}), which, in turn, is almost flat.
  Thus, at the stage of slow-roll inflation, the expressions for the Ricci tensor and the scalar curvature in the Einstein frame are reduced to 
\begin{equation}
R_{\mu\nu}(\tilde{g})|_{\phi> 14.2M_P}\approx -\tilde{g}_{\mu\nu}\frac{1}{M_P^2}U_{eff}^{(tree)}(\phi), \qquad R(\tilde{g})|_{\phi> 14.2M_P}\approx -\frac{4}{M_P^2}U_{eff}^{(tree)}(\phi),
\label{R in Ein via Ueff during infl}
\end{equation}
where $U_{eff}^{(tree)}(\phi)$ given by eq.(\ref{V eff}), up to relative corrections of the order $\lesssim  3\cdot 10^{-2}$, is equal to a constant (see also eqs.(\ref{Veff varphi tanh no fine tun 1}) and (\ref{Veff plateau}))
\begin{equation}
U_{eff}^{(tree)}(\phi)|_{\phi> 14.2M_P}\approx \frac{\lambda M_P^4}{8\xi^2}.
\label{Ueff in Ein via Ueff during infl}
\end{equation}
Therefore we get that    
\begin{equation}
R(\tilde{g})|_{\phi> 14.2M_P}\approx -\frac{\lambda}{2\xi^2}M_P^2= -18\lambda M_P^2,
\label{R in Ein during infl}
\end{equation}
where $\xi=\frac{1}{6}$ was used.
In the appendix \ref{append} we analytically found an upper bound for the function $\zeta(\varphi)$; representing it also through $\phi$, we obtain
 \begin{equation}
0<\zeta\approx \frac{32q^4}{\lambda/8\xi^2}\cdot e^{-\sqrt\frac{8}{3}\frac{\varphi}{M_P}}\approx 209\cdot\Bigl(\frac{M_P}{\phi}\Bigr)^4\lesssim 5\cdot 10^{-3}.
\label{zeta during infl}
\end{equation}
 Thus, during slow-roll inflation, a monotonic increase in $\zeta$ can begin from a value arbitrarily close to zero;  a monotonic decrease in $\phi(t)$ to 
$\phi\approx 14.2M_P$ is accompanied by a monotonic increase in $\zeta$, limited from above by the value $\zeta\approx 5\cdot 10^{-3}$.

\section{Towards a cosmologically modified copy of the SM in the LPPF
\\
 on an arbitrary cosmological background.}
\label{Towards}

Following the concept   of the cosmological realization of the TMSM formulated in the introductory section \ref{sec:intro}, particle physics should be considered in the LPPF as a {\em quantized field theory} studied on a cosmological background, which, in turn, is described in the CF as a  model of {\em classical field theory}.

\subsection{Specificity of the description
in the LPPF on a cosmological background}
\label{descrition in LPP on back}

In accordance with the general principles of TMT, to obtain the physical results of the model, we must start from the primordial action, which is the sum 
of the actions described by eqs.(\ref{S-H 12}), (\ref{S gauge}), (\ref{S l3}), (\ref{Sq})
\begin{equation}
S^{(particles)}_{primordial}=S_H+S_g+S_{(lept)}+S_{(quarks)},
\label{S particles primord}
\end{equation}
 and follow all the procedures. When starting to study physics in the LPPF on a cosmological background, we must take into account the following fundamental differences from the general principles of TMT:

1) Integration in the primordial action $S^{(particles)}_{primordial}$ must be performed  over the spacetime  ${\mathcal V}_4$ using the coordinates $({\mathsf x})$. 

2) In all terms of the Lagrangian density in the action (\ref{S particles primord}), the background value of $\zeta=\zeta_{back}$ should be substituted for the ratio $\Upsilon/\sqrt{-g}$; this means that variation with respect to  functions  $\varphi_a$ is excluded from the TMT procedure.

3) By limiting ourselves to studying the model of electroweak interaction on the cosmological background, we will neglect the possible back reaction of local gravitational effects on background gravity.
Therefore, the variation of the metric is also excluded from the TMT procedure.  Since both the metric and the functions $\varphi_a$ do not vary, the vacuum term (\ref{S_vac}) does not participate in the principle of least action. 

4) According to the  Palatini formulation, the variation with respect to the affine connection  has already been performed, and by the Weyl transformation 
(\ref{gmunuEin}) of the metric to the Einstein frame,  the background Ricci tensor and scalar curvature have been obtained. Therefore,  when starting to study the model of electroweak interaction on a cosmological background, the primordial action (\ref{S particles primord})  should be represented in the Einstein frame. 

5) Based on the definition described by eq.(\ref{H cosm av, back}),  in what follows we will use the notation $\phi_{back}(t)$ or simply $\phi_{back}$ for the background field to avoid confusion with the real part $\phi({\mathsf x})$ of the neutral component of the Higgs isodoublet $H({\mathsf x})$ in the arbitrary gauge considered in ${\mathcal V}_4$.
{\em The Higgs isodoublet $H({\mathsf x})$ and all other matter fields are considered as local fields in ${\mathcal V}_4$, which are dynamical variables independent of the background field $\phi_{back}$.}

6) Bearing in mind that the procedure of functional quantization will be applied to the obtained classical theory, it is necessary to redefine all matter fields so as to absorb factors of type $(b_i\pm \zeta)$ in front of the kinetic terms in the Lagrangian densities and thus bring the kinetic terms to a canonical form.

7)  As already mentioned above, we will  study the description of TMSM in the LPPF only at two background stages: at the stage of approaching the vacuum and at the stage of the slow-roll inflation.

\subsection{The SM bosonic sector
\\
  in the LPPF 
 on a cosmological background}
\label{SM bosonic sector in LPP on a back}

Applying the above remarks to the primordial action $S_H+S_g$, eqs.(\ref{S-H 12}) and (\ref{S gauge}), , we use a description of two stages of the cosmological background characterized by 

\begin{equation}
 \phi_{back} =  \begin{cases}
(\approx) \sigma & \text{at the stage  near the vacuum,}\\
\phi_b(t) & \text{which is a solution of cosmological eqs. at  slow-roll inflation},
\end{cases} 
\label{back phi for two cases}
\end{equation}

and, therefore,

\begin{equation}
 \Omega_{back} =  \begin{cases}
 1 & \text{at the stage  near the vacuum,}\\
\Omega_b=1+\xi\phi_b^2/M_P^2 & \text{at the stage of slow-roll inflation},
\end{cases} 
\label{back Omega for two cases}
\end{equation}

\begin{equation}
 \zeta_{back} =  \begin{cases}
 1 & \text{at the stage  near the vacuum,}\\
0   & \text{at the stage of slow-roll inflation (neglecting correction of the order  $\lesssim 5\cdot 10^{-3}$)},
\end{cases} 
\label{back zeta for two cases}
\end{equation}

\begin{equation}
 R_{back} = \begin{cases}
0 & \text{at the stage  near the vacuum,}\\
R_b=-\frac{\lambda}{2\xi^2}M_P^2 & \text{at the stage of slow-roll inflation},
\end{cases} 
\label{back R for two cases}
\end{equation}
where the value of $R_b$  follows from  eq.(\ref{R in Ein during infl}).

Thus, we arrive at the following action of the TMSM bosonic sector described in the LPPF  on a cosmological background
\begin{equation}
S^{(bos \, LPPF)}_{(on \, back)}=\int L^{(bos \, LPPF)}_{(on \, back)} \sqrt{-\tilde{g}}d^4{\mathsf x},
\label{S bos LPP on back}
\end{equation}
where the integration is performed over the space-time ${\mathcal V}_4$ and the Lagrangian has the form
\begin{eqnarray}
L^{(bos \, LPPF)}_{(on \, back)}&=&\frac{b_k-\zeta_{back}}{(1+\zeta_{back})\Omega_{back}}\tilde{g}^{\alpha\beta}\left( \mathit{D}_{\alpha}H\right)^{\dag} \mathit{D}_{\beta}H
 -\frac{b_p+\zeta_{back}}{(1+\zeta_{back})^2\Omega_{back}^2}\lambda |H|^4
\nonumber
\\
&&-\frac{b_p-\zeta_{back}}{(1+\zeta_{back})^2\Omega_{back}^2}m^2 |H|^2
-\frac{\xi}{\Omega_{back}}R_{back}|H|^2
\nonumber
\\
&&-(b_k-\zeta_{back})\frac{1}{4} \tilde{g}^{\mu\alpha}\tilde{g}^{\nu\beta}\left[\mathbf{F}_{\mu\nu}\mathbf{F}_{\alpha\beta}+B_{\mu\nu}B_{\alpha\beta}\right].
 \label{L bos back general}
\end{eqnarray}
Here 
\begin{equation}
\mathit{D}_{\mu}=
\frac{\partial}{\partial{\mathsf x}^{\mu}}
-ig\mathbf{\hat{T} A}_{\mu}-i\frac{g'}{2}\hat{Y}B_{\mu},  
\label{Dmu in V4}
\end{equation}
\begin{equation}
\mathbf{F}_{\mu\nu}=\frac{\partial\mathbf{A}_\nu}{\partial{\mathsf x}^{\mu}} -\frac{\partial\mathbf{A}_\mu}{\partial{\mathsf x}^{\nu}} +
g\mathbf{A}_\mu\times \mathbf{A}_\nu, \quad
B_{\mu\nu}=\frac{\partial B_\nu}{\partial{\mathsf x}^{\mu}}-\frac{\partial B_\mu}{\partial{\mathsf x}^{\nu}}.
\label{Fmunu and Bmunu in V4}
\end{equation}

\subsection{The SM fermionic sector
\\
  in the LPPF 
 on  a cosmological background}
\label{SM fermionic sector in LPP on a back}

Now we need to move on to the action of the TMSM fermion sector in the LPPF on a cosmological background, for the description of which we use the quantities listed in sec.\ref{SM bosonic sector in LPP on a back}.
Applying the analysis given in items 1) - 5) in section \ref{descrition in LPP on back} to the primordial action $S_{(lept)}+S_{(quarks)}$, 
eqs.(\ref{S l3}) and (\ref{Sq}), it is necessary to take into account that in the procedure for finding the background in the CF we used the Palatini formalism with the subsequent Weyl transformation (\ref{gmunuEin}) of the metric to the Einstein frame. In this case, the affine connection 
$\Gamma^{\lambda}_{\mu\nu}$ is also transformed into the Christoffel connection $\{^{\lambda}_{\mu\nu}\}$ of metric $\tilde{g}_{\mu\nu}$, which must be taken into account in the definition of the spin-connection by replacing $\Gamma^{\lambda}_{\mu\nu}\to \{^{\lambda}_{\mu\nu}\}$. And since the transformation of the metric entails the vierbeins transformation $V_k^{\mu}=\sqrt{(1+\zeta_b)\Omega_b}\tilde{V}_k^{\mu}$, we obtain that when moving to the description of fermions in the LPPF, the following transformation takes place for the spin-connection
\begin{equation}
\omega_{\mu}^{ik}=\frac{1}{2}\Bigl[ V^{i\lambda}\bigl(\partial_{\mu} V^k_{\lambda}+\Gamma^{\beta}_{\mu\lambda} V^k_{\beta}\bigr)-
(i \leftrightarrow k)\Bigr] \to \tilde{\omega}_{\mu}^{ik}=\frac{1}{2}\Biggl[ \tilde{V}^{i\lambda}\Biggl(\frac{\partial\tilde{V}^{k}_{\lambda}}{\partial{\mathsf x}^{\mu}}+\{^{\beta}_{\mu\lambda}\}\tilde{V}^{k}_{\beta}\Biggr)-(i \leftrightarrow k)\Biggr].
\label{affine spin-conn}
\end{equation}
Therefore, the operator $\nabla_{\mu}$, eq.(\ref{nabla  f}), transforms to the operator
\begin{equation}
  \tilde{\nabla}_{\mu}=\frac{\partial}{\partial{\mathsf x}^{\mu}}
-ig\mathbf{\hat{T} A}_{\mu}-i\frac{g'}{2}\hat{Y}B_{\mu}+\frac{1}{2}\tilde{\omega}_{\mu}^{ik}\sigma_{ik},
\label{Nabla SM in Ein}
\end{equation}
which acts in the space-time ${\mathcal V}_4$.
Then we arrive at the following action of the TMSM fermion sector, described in the LPPF on a cosmological background 
\begin{equation}
S^{(ferm \, LPPF)}_{(on \, back)}=\int L^{(ferm \, LPPF)}_{(on \, back)} \sqrt{-\tilde{g}}d^4{\mathsf x}, 
\label{Sferm LPP on back}
\end{equation}
\begin{equation}
  L^{(ferm \, LPPF)}_{(on \, back)}=
L^{(lept \, LPPF)}_{(on \, back)}+L^{(quarks \, LPPF)}_{(on \, back)}
\label{Lferm LPP on back}
\end{equation}
with the following Lagrangians for leptons and up-quarks\footnote{The notation $\tilde{\gamma}^{\mu}=\tilde{V}_{k}^{\mu}\gamma^{k}$ is used.}
\begin{eqnarray}
&&L^{(lept \, LPPF)}_{(on \, back)}=
\nonumber
\\
&&\sum_{l=e,\mu,\tau} \frac{b_l-\zeta_{back}}{\bigl((1+\zeta_{back})\Omega_{back}\bigr)^{3/2}}\cdot \frac{i}{2} \Bigl[\overline{L_l}\tilde{\gamma}^{\mu}\tilde{\nabla}_{\mu}L_l -(\tilde{\nabla}_{\mu}\overline{L_l})\tilde{\gamma}^{\mu}L_l+
\overline{l_R}\tilde{\gamma}^{\mu}\tilde{\nabla}_{\mu}l_R -(\tilde{\nabla}_{\mu}\overline{l_R})\tilde{\gamma}^{\mu}l_R\Bigr]
\nonumber
\\
&-&\sum_{l=e,\mu,\tau}  \frac{b_l+\zeta_{back}}{(1+\zeta_{back})^2\Omega_{back}^2}\cdot y^{(ch)}\left(\overline{L_l}H l_R +
\overline{l_R}H^{\dag}L_l\right)
\nonumber 
\\
&+&\frac{b_e-\zeta_{back}}{\bigl((1+\zeta_{back})\Omega_{back}\bigr)^{3/2}}\cdot\frac{i}{2}\bigl[\overline{\nu_R}\tilde{\gamma}^{\mu}\tilde{\nabla}_{\mu}\nu_R -(\tilde{\nabla}_{\mu}\overline{\nu_R})\tilde{\gamma}^{\mu}\nu_R\bigr]
\nonumber
\\
&-&\frac{b_{\nu}+\zeta_{back}}{(1+\zeta_{back})^2\Omega_{back}^2}\cdot y_{\nu}\left(\overline{L_e}H_c \nu_R +
\overline{\nu_R}H_c^{\dag}L_e\right),
\label{L lept LPP on back}
\end{eqnarray}
\begin{eqnarray}
&&L^{(quarks \, LPPF)}_{(on \, back)}=
\sum_{q=u,c,t} \Biggl[\frac{b_q-\zeta_{back}}{\bigl((1+\zeta_{back})\Omega_{back}\bigr)^{3/2}}\cdot \frac{i}{2} \Biggl(\overline{L_q}\tilde{\gamma}^{\mu}\tilde{\nabla}_{\mu}L_q -(\tilde{\nabla}_{\mu}\overline{L_q})\tilde{\gamma}^{\mu}L_q+
\nonumber
\\
&+&\overline{r_q}\tilde{\gamma}^{\mu}\tilde{\nabla}_{\mu}r_q -(\tilde{\nabla}_{\mu}\overline{r_q})\tilde{\gamma}^{\mu}r_q\Biggr)
- \frac{b_q+\zeta_{back}}{(1+\zeta_{back})^2\Omega_{back}^2}\cdot y^{(up)}\left(\overline{L_q}H_c r_q +
\overline{r_q}H_c^{\dag}L_q\right)\Biggr].
\label{L quarks LPP on back}
\end{eqnarray}

\subsection{Concluding remarks}
\label{Concluding remarks on quant}

The sum of the actions  $S^{(bos \, LPPF)}_{(on \, back)}+S^{(ferm \, LPPF)}_{(on \, back)}$ presented in sections \ref{SM bosonic sector in LPP on a back} and \ref{SM fermionic sector in LPP on a back} describe the SM fields in the LPPF on  an {\em arbitrary cosmological background}.
Following the concept of cosmological realization of TMSM, the Lagrangian $L^{(bos \, LPPF)}_{(on \, back)}+ L^{(ferm \, LPPF)}_{(on \, back)}$
 should be reduced to the canonical form of the electroweak model, which will make it possible to apply the functional quantization method. This is achieved by appropriate, depending on the background stage, redefinitions of fields and model parameters. In the general case, i.e., with an arbitrary cosmological background, this task is difficult to accomplish, since it requires rather cumbersome numerical solutions. In the next two sections we will see that for two important cosmological backgrounds these difficulties can be avoided analytically.  Namely, we will study the TMSM in the LPPF at two opposite stages of the evolution of the cosmological background: a) near the vacuum; b) at the stage of slow-roll inflation. For this, we will use the results collected in eqs.(\ref{back phi for two cases})-(\ref{back R for two cases}).

\section{The GWS theory as the  implementation of the TMSM
\\
 in the LPPF 
on the cosmological background near vacuum}
\label{TMSM in LPP near vacuum}

In section \ref{summary of 1-st paper}, where we reviewed the main results of the paper \cite{1-st paper}, a vacuum with zero cosmological constant was described in the CF. Near this vacuum, using results collected in section \ref{SM bosonic sector in LPP on a back}, one can take with extraordinary accuracy $\zeta_{back}=1$, \, $\Omega_{back}=1$ and the action
 $S^{(bos)}_{(LPP \, on \, back)}+S^{(ferm)}_{(LPP \, on \, back)}$ can be represented in Minkowski space by replacing $\tilde{g}_{\alpha\beta}$ with $\eta_{\alpha\beta}$. One can also replace   $\tilde{\gamma}^{\mu}$ with $\gamma^{\mu}$ and $\tilde{\nabla}_{\mu}$ with $D_{\mu}$.

\subsection{The TMSM bosonic sector
  in the LPPF 
\\
on the cosmological background near vacuum}
\label{Bosons  in LPP near vacuum}

 Near vacuum the Lagrangian
$L^{(bos \, LPPF)}_{(on \, back)}$   reduces to the following 
\begin{eqnarray}
L_{(on \, back \,near\,vac)}^{(bos \, LPPF)}&=&
 \frac{1}{2}(b_k-1)\eta^{\alpha\beta}\left(\mathit{D}_{\alpha}H\right)^{\dag} \mathit{D}_{\beta}H-\frac{1+b_p}{4}\lambda |H|^4+\frac{1-b_p}{4}m^2 |H|^2\Bigr]
\nonumber
\\
&-&\frac{1}{4}(b_k-1)\eta^{\mu\alpha}\eta^{\nu\beta}\left(\mathbf{F}_{\mu\nu}
\mathbf{F}_{\alpha\beta}+B_{\mu\nu}B_{\alpha\beta}\right).
\label{L bos back near vac}
\end{eqnarray}
To bring the Lagrangian to the canonical form of the electroweak GWS model, we make the following redefinitions:

 the Higgs isodoublet
\begin{equation}
\sqrt{\frac{b_k-1}{2}}\cdot H=\mathcal{H}
\label{H in LPP near vac}
\end{equation}

and the corresponding parameters
\begin{equation}
\lambda_{SM}=\frac{1+b_p}{(b_k-1)^2}\lambda, \qquad m_{SM}^2=\frac{1-b_p}{2(b_k-1)}m^2;
\label{lambda and m2 in LPP near vac}
\end{equation}

the gauge fields and gauge coupling constants 
\begin{equation}
\sqrt{b_k-1}\cdot \mathbf{A}_\mu=\mathcal{A}_\mu, \quad \sqrt{b_k-1}\cdot B_\mu=\mathcal{B}_\mu, \quad 
\frac{g}{\sqrt{b_k-1}}=\mathtt{g}, \quad \frac{g'}{\sqrt{b_k-1}}=\mathtt{g}'
\label{gauge fields in LPP near vac}
\end{equation}

and, correspondently,
\begin{equation}
\sqrt{b_k-1}\cdot\mathbf{F}_{\mu\nu}=\mathcal{F}_{\mu\nu}=\frac{\partial\mathcal{A}_\nu}{\partial{\mathsf x}^{\mu}} -\frac{\partial\mathcal{A}_\mu}{\partial{\mathsf x}^{\nu}} +
\mathtt{g}\mathcal{A}_\mu\times \mathcal{A}_\nu, \quad
\sqrt{b_k-1}\cdot B_{\mu\nu}=\mathtt{B}_{\mu\nu}=\frac{\partial \mathtt{B}_\nu}{\partial{\mathsf x}^{\mu}}-\frac{\partial \mathtt{B}_\mu}{\partial{\mathsf x}^{\nu}}.
\label{gauge redefinitions for Fmmunu and Bmunu}
\end{equation}
\begin{equation}
  \mathit{D}_{\mu} \to \mathcal{D}_{\mu}=\frac{\partial}{\partial{\mathsf x}^{\mu}}
-i\mathtt{g}\mathbf{\hat{T} \mathcal{A}}_{\mu}-i\frac{\mathtt{g}'}{2}\hat{Y}\mathcal{B}_{\mu}
\label{Dmu gauge redefinitions}
\end{equation}
As a result, the Lagrangian $L_{(near\,vac)}^{(bos \, LPPF)}$ takes the form of the Lagrangian of the bosonic sector of  the GWS theory
\begin{eqnarray}
L^{(bos \, LPPF)}|_{(on \, back \,near\,vac)}&=&
 \eta^{\alpha\beta}\left(\mathcal{D}_{\alpha}\mathcal{H}\right)^{\dag}\mathcal{D}_{\beta}\mathcal{H}-V(\mathcal{H})
\nonumber
\\
&-&\frac{1}{4}\eta^{\mu\alpha}\eta^{\nu\beta}\left(\mathcal{F}_{\mu\nu}
\mathcal{F}_{\alpha\beta}+\mathcal{B}_{\mu\nu}\mathcal{B}_{\alpha\beta}\right),
\label{L bos back near vac}
\end{eqnarray}
\begin{equation}
V(\mathcal{H})=\lambda_{SM} |\mathcal{H}|^4-m_{SM}^2 |\mathcal{H}|^2.
\label{V(tilde H)}
\end{equation}
With our choice of $b_p\approx 0.5$, which  was dictated by the requirements of the inflationary regime, the mass term in the potential $V(\mathcal{H})$ of the Higgs field $\mathcal{H}$  has the "wrong" sign. Thus, we have the standard picture of SSB $SU(2)\times U(1)\to U(1)$ as the  $\mathcal{H}$ develops VEV
\begin{equation}
\langle 0|\mathcal{H}|0\rangle= \left(\begin{matrix} 0 \\ \frac{1}{\sqrt{2}}v\end{matrix}\right)
\label{Higgs  doublet unitary}
\end{equation}
with
\begin{equation}
v^2 =\frac{m^2_{SM}}{\lambda_{SM}}=\frac{1-b_p}{1+b_p}\cdot\frac{(b_k-1)m^2}{2\lambda}=\frac{(b_k-1)m^2}{6\lambda}.
\label{v^2 LPP}
\end{equation}
  Having made the gauge transformation, we can move on to the unitary gauge, where the Higgs isodouplet is represented, as usual, in the form
\begin{equation}
\mathcal{H}= \left(\begin{matrix} 0 \\ \frac{1}{\sqrt{2}}(v+h({\mathsf x})) \end{matrix}\right), \qquad \langle 0|h|0\rangle=0.
\label{H SM unitary LPP}
\end{equation}
Then the potential (\ref{V(tilde H)}) reduces to  the standard potential of the Higgs boson $h$
\begin{equation}
V(h)=\frac{1}{4}\lambda_{SM}\bigl(2vh+h^2\bigr)^2-\frac{1}{4}\lambda_{SM}v^4.
\label{V of h LPP}
\end{equation}
Therefore, the expression for the mass squared of the Higgs boson is as follows
\begin{equation}
m^2_h=2\lambda_{SM}v^2=\frac{m^2}{2(b_k-1)}.
\label{m^2 of h LPP}
\end{equation}

Recall that to ensure agreement with the observed CMB data, choosing $\xi=\frac{1}{6}$, as done in \cite{1-st paper}, requires that the primrdial Higgs self-interaction parameter be $\lambda=2.3\cdot 10^{-11}$. On the other hand, it was shown in \cite{1-st paper} that this value of $\lambda$ can be consistent with the electroweak SM considered near the vacuum in the CF if the parameter $b_k$ is very close to unity. But we have not yet been able to find the value of $b_k-1$. Right now, working in LPPF, combining equations (\ref{v^2 LPP}) and (\ref{m^2 of h LPP}), we obtain the relation
\begin{equation}
(b_k-1)^2=3\lambda\frac{v^2}{m^2_h}.
\label{bk-1 2}
\end{equation} 
Substituting the values of $v\approx 246 \, GeV$ and $m_h\approx 125 \, GeV$ known from particle physics, and the value of 
$\lambda\approx 2.3\cdot 10^{-11}$  we get
\begin{equation}
b_k-1\approx 1.6\cdot 10^{-5}.
\label{bk-1}
\end{equation} 
From eqs.(\ref{m^2 of h LPP}) and (\ref{bk-1}) we obtain the estimate 
\begin{equation}
m\approx 0.7 \, GeV. 
\label{m 0.7 GeV}
\end{equation} 
Note that, given this estimate, we could at some stages of cosmological evolution neglect the terms $\propto \frac{m^2}{M_P^2}\sim 8\cdot 10^{-38}$  in the constraint (\ref{zeta for radical}).

As usual in the electroweak SM, we define 
\begin{eqnarray}
&&W^{\pm}_\mu=\frac{1}{\sqrt{2}}(\mathcal{A}^1_\mu\mp i\mathcal{A}^2_\mu), 
\nonumber
\\
&& Z_{\mu}=\sin\theta_W  \, \mathcal{B}_\mu -\cos\theta_W \, \mathcal{A}^3_\mu, \quad A_{\mu}=\cos\theta_W  \, \mathcal{B}_\mu +\sin\theta_W \, \mathcal{A}^3_\mu,
\label{defining W Z A}
\end{eqnarray} 
where the Weinberg angle $\theta_W$ is given by
\begin{equation}
\tan\theta_W=\frac{\mathtt{g}'}{\mathtt{g}}=\frac{g'}{g}.
\label{tan theta W}
\end{equation} 
For  generated by the Higgs phenomenon masses of $W$ and  $Z$-bosons, the  Lagrangian (\ref{L bos back near vac}) gives the standard expressions
\begin{equation}
M_{W}=\frac{\mathtt{g}}{2}\cdot v, \qquad  
M_{Z}=\frac{\sqrt{\mathtt{g}^2+\mathtt{g}^{\prime 2}}}{2}\cdot v, \quad \text{as well as}\quad M_{A}=0 \quad \text{for photon.}
\label{MW and MZ eff}
\end{equation} 
 Agreement with the experimental data obtained, as usual in the GWS model, using the effective four-fermion Lagrangian\footnote{
The canonical SM structure of the fermion Lagrangian in the LPPF (see  eqs.(\ref{L lept in LPP appr vac}) and (\ref{L q in LPP appr vac})) provides a standard expression for the effective four-fermion Lagrangian.}
 is achieved with the  standard VEV $v\approx 246\, GeV$. 
As usual in the electroweak SM, there are relations between $\mathtt{g}$, $\mathtt{g}'$, the Weinberg angle $\theta_W$ and the electric charge $e=\sqrt{4\pi\alpha}$
\begin{equation}
e=\mathtt{g}\sin\theta_W, \quad  e=\mathtt{g}'\cos\theta_W,
\label{e g gprime angle}
\end{equation} 
from which, using the redifinitions of the gauge coupling constants in eqs.(\ref{gauge fields in LPP near vac}) and the  value of $(b_k-1)$ obtained 
above, eq.(\ref{bk-1}), we find  the values of the primordial model parameters $g$ and $g'$ in the primordial action (\ref{S gauge})
\begin{equation}
g\approx 2.6\cdot 10^{-3}, \qquad g'\approx 1.4\cdot 10^{-3}.
\label{g and gprime}
\end{equation}

\subsection{The TMSM fermionic sector
  in the LPPF 
\\
on the cosmological background near vacuum}
\label{fermions near vacuum}

In addition to what was described at the very beginning of section \ref{TMSM in LPP near vacuum}, near vacuum one can substitute $(b_l\pm\zeta_b)=
(b_l\pm 1)$, $(b_{\nu}+\zeta_b)=(b_{\nu}+1)$, $(b_q\pm\zeta_b)=(b_q\pm 1)$. Thus, near vacuum the action for fermions in the LPPF, eqs.(\ref{Sferm LPP on back})-(\ref{L quarks LPP on back}), takes the form
\begin{equation}
S^{(ferm \, LPPF)}_{(on \, back \,near\,vac)}=\int \Bigl(L^{(lept \, LPPF)}_{(on \, back \,near\,vac)}+L^{(quarks \, LPPF)}_{(on \, back \,near\,vac)}\Bigr)d^4{\mathsf x}, 
\label{Sferm LPP on back near vac}
\end{equation}
with the following Lagrangians for leptons and up-quarks
\begin{eqnarray}
&&L^{(lept \, LPPF)}_{(on \, back \,near\,vac)}=
\nonumber
\\
&&\sum_{l=e,\mu,\tau} \frac{b_l-1}{2^{3/2}}\cdot \frac{i}{2} \Bigl[\overline{L_l}\gamma^{\mu}D_{\mu}L_l -D_{\mu}\overline{L_l})\gamma^{\mu}L_l+
\overline{l_R}\gamma^{\mu}D_{\mu}l_R -(D_{\mu}\overline{l_R})\gamma^{\mu}l_R\Bigr]
\nonumber
\\
&-&\sum_{l=e,\mu,\tau}  \frac{b_l+1}{4}\cdot y^{(ch)}\left(\overline{L_l}H l_R +
\overline{l_R}H^{\dag}L_l\right)
\label{L lept back near vac}
\\
&+&\frac{b_e-1}{2^{3/2}}\cdot\frac{i}{2}\bigl[\overline{\nu_R}\gamma^{\mu}\partial_{\mu}\nu_R -(\partial_{\mu}\overline{\nu_R})\gamma^{\mu}\nu_R\bigr]
-\frac{b_{\nu}+1}{4}\cdot y_{\nu}\left(\overline{L_e}H_c \nu_R +
\overline{\nu_R}H_c^{\dag}L_e\right),
\nonumber
\end{eqnarray}
\begin{eqnarray}
&&L^{(quarks \, LPPF)}_{(on \, back \,near\,vac)}=
\sum_{q=u,c,t} \Biggl[\frac{b_q-1}{2^{3/2}}\cdot \frac{i}{2} \Biggl(\overline{L_q}\gamma^{\mu}D_{\mu}L_q -(D_{\mu}\overline{L_q})\gamma^{\mu}L_q+
\nonumber
\\
&+&\overline{r_q}\gamma^{\mu}\,\frac{\partial r_q}{\partial{\mathsf x}^{\mu}} -\frac{\partial \overline{r_q}}{\partial{\mathsf x}^{\mu}}\,\gamma^{\mu}r_q\Biggr)
- \frac{b_q+1}{4}\cdot y^{(up)}\left(\overline{L_q}H_c r_q +
\overline{r_q}H_c^{\dag}L_q\right)\Biggr].
\label{L quarks back near vac}
\end{eqnarray}

\subsubsection{Leptons}
\label{leptons appr vacuum}

Using the results of the bosonic sector study in section \ref{Bosons  in LPP near vacuum}, we introduce redefinitions of the Higgs doublet, eq.(\ref{H in LPP near vac}), gauge bosons and gauge coupling constants, eq.(\ref{gauge fields in LPP near vac}), together with a redefinition of the covariant derivative, 
eq.(\ref{Dmu gauge redefinitions}). These manipulations must be supplemented with the following redefinitions of the lepton fields and the Yukawa coupling constants:
\begin{equation}
\frac{\sqrt{b_l-1}}{2^{3/4}}L_l=L'_l \, , \quad 
\frac{\sqrt{b_l-1}}{2^{3/4}}l_R=l'_R \, , \quad l=e, \mu, \tau ; \qquad
\frac{\sqrt{b_e-1}}{2^{3/4}}\cdot\nu_R=\nu'_R 
\label{l and nuR in LPP}
\end{equation}
\begin{equation}
Y^{(l)}_{SM}=\frac{b_l+1}{(b_l-1)\sqrt{b_k-1}}y^{(ch)},  \quad l=e, \mu, \tau ; \quad Y^{(\nu)}_{SM}=\frac{b_{\nu}+1}{(b_e-1)\sqrt{b_k-1}}y_{\nu}.
\label{Yl and Ynu in LPP}
\end{equation}
Then the Lagrangian  $L^{(lept \, LPPF)}_{(on \, back \,near\,vac)}$ , eq.(\ref{L lept back near vac}),
reduces to the  Lagrangian for leptons in the GWS model:
\begin{eqnarray}
&&L^{(lept \, LPPF)}|_{(on \, back \,near\,vac)}
\label{L lept in LPP appr vac}
\\
&=&\sum_{l=e,\mu,\tau}\Biggl(\frac{i}{2} \Bigl(\overline{L_l'}\gamma^{\mu}\mathcal{D}_{\mu}L_l' -(\mathcal{D}_{\mu}\overline{L_l'})\gamma^{\mu}L_l'+
\overline{l_R'}\gamma^{\mu}\mathcal{D}_{\mu}l_R' -(\mathcal{D}_{\mu}\overline{l_R'})\gamma^{\mu}l_R'\Bigr)
-Y^{(l)}_{SM}\left(\overline{L_l'}\mathcal{H} l_R' +
\overline{l_R'}\mathcal{H}^{\dag}L_l'\right)\Biggr)
\nonumber
\\
&&\hspace{1cm}+\frac{i}{2}\Bigl[\overline{\nu_R'}\gamma^{\mu}\,\frac{\partial\nu_R'}{\partial{\mathsf x}^{\mu}} -
\frac{\partial\overline{\nu_R'}}{\partial{\mathsf x}^{\mu}}\,\gamma^{\mu}\nu_R'\Bigr]
-Y^{(\nu)}_{SM}\left(\overline{L_e'}\mathcal{H}_c \nu_R' +
\overline{\nu_R'}\mathcal{H}_c^{\dag}L_e'\right).
 \nonumber
\end{eqnarray}

\subsubsection{Quarks}
\label{sub quarks appr vac}

Repeating  the manipulations performed for leptons in section \ref{leptons appr vacuum}
one can see that the redefinitions to the LPPF in the bosonic sector (\ref{H in LPP near vac}), (\ref{gauge fields in LPP near vac}), (\ref{Dmu gauge redefinitions}) should be supplemented by the following redefinitions
\begin{equation}
\frac{\sqrt{b_q-1}}{2^{3/4}}L_q=L'_q \, , \quad 
\frac{\sqrt{b_q-1}}{2^{3/4}}r_q=r'_q  ,
\label{L in LPP}
\end{equation}
\begin{equation}
Y^{(q)}_{SM}=\frac{b_q+1}{(b_q-1)\sqrt{b_k-1}}y^{(up)}  \, , \quad q=u,c,t .
\label{y q in LPP}
\end{equation}
Then the Lagrangian $L^{(quarks \, LPPF)}_{(on \, back \,near\,vac)}$,  eq.(\ref{L quarks back near vac}),
reduces to the Lagrangian for up-quarks in the GWS model
\begin{eqnarray}
&&L^{(quarks \, LPPF)}|_{(on \, back \,near\,vac)}
\label{L q in LPP appr vac}
\\
&=&\sum_{q=u,c,t}\Biggl(\frac{i}{2} \Bigl[\overline{L_q'}\gamma^{\mu}\mathcal{D}_{\mu}L_q'
 -(\mathcal{D}_{\mu}\overline{L_q'})\gamma^{\mu}L_q'+
\overline{r_q'}\gamma^{\mu}\mathcal{D}_{\mu}r_q' -(\mathcal{D}_{\mu}\overline{r_q'})\gamma^{\mu}r_q'\Bigr]
-Y^{(q)}_{SM}\Bigl[\overline{L_q'}\mathcal{H}_c r_q' +
\overline{r_q'}\mathcal{H}_c^{\dag}L_q'\Bigr]\Biggr)
\nonumber
\end{eqnarray}

\subsubsection{Fermion masses without the hierarchy problem}
\label{fermion masses}

Production of fermion masses by the SSB of the gauge symmetry  $SU(2)\times U(1)\rightarrow U(1)$ at the last stage of cosmological evolution has the same mechanism as in the tree-level GWS theory.  But instead of mass values obtained  with specially tuned Yukawa constants  as model parameters in the original action, in our TMSM we start from the primordial action with the universal Yukawa coupling constant $y^{(ch)}$ for charged leptons and  the universal Yukawa coupling constant $y^{(up)}$ for for up-quarks. 
As usual in the SM, from eqs.(\ref{L lept in LPP appr vac}), (\ref{L q in LPP appr vac}) we obtain expressions for the fermion masses, which, using notations 
(\ref{Yl and Ynu in LPP}) and (\ref{y q in LPP}), are represented as follows   
\begin{equation}
m_l=Y^{(l)}_{SM}\frac{v}{\sqrt{2}}=\frac{b_l+1}{(b_l-1)\sqrt{b_k-1}}y^{(ch)}\frac{v}{\sqrt{2}};  \quad l=e, \mu, \tau, \quad b_l=b_e, b_{\mu}, b_{\tau},
\label{ml and fl appr vac}
\end{equation}
\begin{equation}
m_{\nu}=Y^{(\nu)}_{SM}\frac{v}{\sqrt{2}}=\frac{b_{\nu}+1}{(b_e-1)\sqrt{b_k-1}}y_{\nu}\frac{v}{\sqrt{2}},
\label{mnu ynu appr vac}
\end{equation}
\begin{equation}
m_q=Y^{(q)}_{SM}\frac{v}{\sqrt{2}}=\frac{b_q+1}{(b_q-1)\sqrt{b_k-1}}y^{(up)}\frac{v}{\sqrt{2}}; 
  \quad q=u, c, t; \quad  b_q=b_u, b_c, b_t. 
\label{mq appr vac}
\end{equation}

Tuning the fermion masses is carried out by choosing the appropriate values of the model parameters $b_l$, $b_{\nu}$, $b_q$, which appear due  to two volume measures in the primordial action - a fundamental feature of TMT. 

\begin{itemize}

\item

\underline{Masses of charged leptons and electron neutrino}

For charged leptons we choose the universal Yukawa coupling constant 
\begin{equation}
y^{(ch)}\approx 10^{-6}.
\label{y 10-6}
\end{equation}
Substituting the values of the masses of charged leptons \cite{Part Data} and $v\approx 246 \, GeV$ into eqs.(\ref{ml and fl appr vac}), we obtain the following values of the corresponding parameters
\begin{equation}
b_e\approx 1.003; \quad b_{\mu}\approx 1+ 10^{-5}; \quad b_{\tau}\approx 1+8\cdot 10^{-7}.
\label{b_l values}
\end{equation}
For the electron neutrino (Dirac, active), taking $m_{\nu}\sim 1 eV$ and choosing $y_{\nu}\approx 10^{-11}$, we obtain
\begin{equation}
b_{\nu}\approx 1.01.
\label{b_nu}
\end{equation}

\item

\underline{Masses of up-quarks}

For up-quarks we choose the  universal Yukawa coupling constant\footnote{Note that the simpler choice of $y^{(up)}=y^{(ch)}\approx 10^{-6}$ does not lead to qualitative changes in the model results.} 
\begin{equation}
y^{(up)}\approx 10^{-5}.
\label{y 10-5}
\end{equation}
Substituting the values of the masses of up-quarks \cite{Part Data} and $v\approx 246 \, GeV$ into eqs.(\ref{mq appr vac}), we obtain the following values of the corresponding parameters
\begin{equation}
b_u\approx 1.05; \quad b_c\approx 1+9\cdot 10^{-5}; \quad b_t\approx 1+6\cdot 10^{-7}.
\label{b_q values}
\end{equation}

\end{itemize}

Thus, the observed fermion mass hierarchy is realized if the values of all $b_i$-parameters are very close to unity, making the proposed approach to solving the fermion mass hierarchy problem quite natural. 
In this regard, it is also worth paying attention to the fact that in the above calculations, the proximity of the value of another model parameter to 1, namely $b_k\approx 1+1.6\cdot 10^{-5}$ \, (see  eq.(\ref{bk-1})), significantly affects the results obtained. To better understand the complexity of the structure (cosmology+particle physics) of the TMSM, it is important to remember that this value of $b_k$ was obtained in section \ref{Bosons  in LPP near vacuum} by matching the constraints imposed by the CMB data on the parameters of the inflation model, on the one hand, with experimental data from particle physics at accelerator energies, on the other hand.

\section{Cosmologically modified copy of the tree-level SM  in the LPPF 
\\
on the cosmological background at the stage of  the slow-roll inflation (up-copy)}
\label{TMSM in LPP on back slow-rol}

The stage of inflation in the slow-roll regime is realized as $\phi\gtrsim 14.2M_P$.
We will study the SM  in the LPPF on the cosmological background at this stage, neglecting relative corrections of the order of $\lesssim 3\cdot 10^{-2}$. There are two reasons for these errors in our use of the Lagrangian (\ref{L bos back general}) and (\ref{Lferm LPP on back}) - (\ref{L quarks LPP on back})).
 The first reason for this error is the neglect of $\zeta_b$ in the combinations $(1+\zeta_b)$, $(b_k-\zeta_b)$ and $(b_p\pm\zeta_b)$, based on the result $\zeta_b\lesssim 5\cdot 10^{-3}$ obtained in the appendix \ref{append} and presented in section \ref{some features of back} in  item 5)\footnote{
The presence of a multiplier $(b_k-\zeta_{back})$ before the last term of Lagrangian (\ref{L bos back general}) does not violate gauge invariance. But the dependence of $\zeta_{back}$ on $\phi_{back}$ leads to the appearance of additional terms in the equations for the Green's functions of gauge fields. However, at the stage of slow-roll inflation $\phi_{back}=\phi_b$ and these terms are proportional to $\dot{\phi_b}$. Therefore, due to the slow change in $\phi_b$, the order of magnitude of their relative contribution turns out to be significantly less than $10^{-2}$. Therefore, neglecting $\zeta_b$ compared to $b_k$ in this case is within the framework of the approximation in which we work.}, see eq.(\ref{zeta during infl}). 
For the case of fermions, the corresponding approximations made using eqs.(\ref{Lferm LPP on back}) - (\ref{L quarks LPP on back}) will be discussed in section \ref{fermions infl slow}. The second source of error is the use of the approximation 
$\Omega_b^{-1}\approx \frac{M_P^2}{\xi\phi_b^2}$, and the relative value of this error is limited from above by $\lesssim 3\cdot 10^{-2}$ at $\phi\gtrsim 14.2M_P$. 
  In addition, we have to use that $b_p\approx 1/2+10^{-8}\approx 1/2$ and $b_k\approx 1+1.6\cdot 10^{-5}\approx 1$. The choice of  the model parameters $b_p$ and $b_k$ was made  in ref. \cite{1-st paper} and in section \ref{Bosons  in LPP near vacuum}  of this paper and  was based on the need to provide the desired properties of inflation and for matching the CMB data with the physics at accelerator energy scales. Now we will see that this choice is also very important  in the tree-level particle physics  at the stage of inflation in the slow-roll regime.

\subsection{Bosonic sector}
\label{Bosonic sector slow-roll}

 Operating with the described precision, we reduce the Lagrangian (\ref{L bos back general}) to the following form 
\begin{eqnarray}
 L_{(on \, back \, \phi_b>14M_P)}^{(bos \, LPPF)}&=&\frac{M_P^2}{\xi\phi_b^2}\tilde{g}^{\alpha\beta}\left(\mathit{D}_{\alpha}H\right)^{\dag}\mathit{D}_{\beta}H
 -\frac{M_P^4}{2\xi^2\phi_b^4}\lambda |H|^4-\frac{M_P^2}{2\xi\phi_b^2}m^2 |H|^2
\nonumber
\\
 &&-\frac{M_P^2}{\phi_b^2}R_b|H|^2-\frac{1}{4} \tilde{g}^{\mu\alpha}\tilde{g}^{\nu\beta}\left[\mathbf{F}_{\mu\nu}\mathbf{F}_{\alpha\beta}+B_{\mu\nu}B_{\alpha\beta}\right],
 \label{L bos back slow roll}
\end{eqnarray}
where the background scalar curvature $R_b= -\frac{\lambda}{2\xi^2} M_P^2$, see eqs.(\ref{R in Ein during infl}) and (\ref{back R for two cases}).

According to what  was explained in  section \ref{Concluding remarks on quant}, we would like first  redefine the Higgs field to bring the kinetic term to the canonical form. For this we use the  redefinition\footnote
{Note that here, in contrast to section \ref{Bosons  in LPP near vacuum}, we leave without redefinition the gauge fields $\mathbf{A}_{\mu}$, $B_{\mu}$ and the gauge coupling parameters $g$, $g'$. This is possible because by reducing eq.(\ref{L bos back general}) to form (\ref{L bos back slow roll}), for the factor before the last term in eq.(\ref{L bos back general}) we can approximately take $(b_k-\zeta_b)\approx 1$ since $|(b_k-\zeta_b)-1|\lesssim 5\cdot 10^{-3}$.}
\begin{equation}
\tilde{H}({\mathsf x})=\frac{M_P}{\sqrt{\xi}\phi_b(t)}H({\mathsf x}),
\label{redif H in slow}
\end{equation}  
First of all note that
\begin{equation}
\frac{M_P^2}{\xi\phi_b^2}\tilde{g}^{ij}\left(\frac{\partial H}{\partial {\mathsf x}^i}\right)^{\dag}\frac{\partial H}{\partial {\mathsf x}^j}
=\tilde{g}^{ij}\left(\frac{\partial \tilde{H}}{\partial {\mathsf x}^i}\right)^{\dag}\frac{\partial \tilde{H}}{\partial {\mathsf x}^j}.
\label{spacial deriv of redif H in slow}
\end{equation}  
Assuming a choice of coordinates where $\tilde{g}^{0i}=0$ ($i=1,2,3$), we are left to analyze how the redefinition (\ref{redif H in slow}) affects the time derivative, which reads
\begin{equation}
\frac{M_P}{\sqrt{\xi}\phi_b}\dot{H}=\dot{\tilde{H}}+\tilde{H}\frac{\dot{\phi}_b}{\phi_b}.
\label{dotH dot tilde H dot phi_b}
\end{equation}
Recall that $\dot{\phi}_b<0$.
It is convenient to use  the redifined operater 
\begin{equation}
\mathit{D}_{\mu}\to \tilde{\mathit{D}}_{\mu} = \begin{cases}
\tilde{\mathit{D}}_k=\mathit{D}_k=
\frac{\partial}{\partial{\mathsf x}^k}
-ig\mathbf{\hat{T} A}_k-i\frac{g'}{2}\hat{Y}B_k, \quad k=1,2,3
\\
\tilde{\mathit{D}}_0=\mathit{D}_0+\frac{\dot{\phi_b}}{\phi_b}=\frac{\partial}{\partial t}+\frac{\dot{\phi_b}}{\phi_b}-ig\mathbf{\hat{T} A}_0-i\frac{g'}{2}\hat{Y}B_0, \quad \mu  =0.
\end{cases} 
\label{D redifened}
\end{equation}
With the redefinitions (\ref{redif H in slow}) and (\ref{D redifened})
  the Lagrangian (\ref{L bos back slow roll}) take the following form
\begin{eqnarray}
L^{(bos \, LPPF)}_{(slow \, roll \, infl)} &=&\tilde{g}^{\alpha\beta}\left(\tilde{\mathit{D}}_{\alpha}\tilde{H}\right)^{\dag}\tilde{\mathit{D}}_{\beta}\tilde{H} -V(\tilde{H})
\nonumber
\\
&-&\frac{1}{4} \tilde{g}^{\mu\alpha}\tilde{g}^{\nu\beta}\bigl[\mathbf{F}_{\mu\nu}\mathbf{F}_{\alpha\beta}+B_{\mu\nu}B_{\alpha\beta}\bigr].
 \label{L bos after redef}
\end{eqnarray}
\begin{equation}
V(\tilde{H})= \frac{1}{2}\lambda |\tilde{H}|^4+\frac{1}{2}m^2 |\tilde{H}|^2+\xi R_b|\tilde{H}|^2. 
\label{V tilde H}
\end{equation}
It is obvious that the appearance of $\dot{\phi_b}/\phi_b$ in $\tilde{\mathit{D}}_0$ shows the possibility of a direct effect of cosmological evolution on the physics described in the LPPF. However, first of all, it is obvious that this does not violate the gauge invariance. Moreover, since the background is in a slow-roll inflation stage, it is natural to assume that the background field $\phi_b$ is a slow dynamical variable compared to $\tilde{H}$, which is subject to quantization at Planck scales and can therefore be considered as a fast variable.
 In such a picture the following strong inequality must hold\footnote{We used  the coordinate system $({\mathsf x})$ in which the time component of the vector $d{\mathsf x}^{\mu}$ coincides with the time component $dt$ of the vector $dx^{\mu}$ in the coordinates used for the Friedmann metric. See also footnote 8.} 
\begin{equation}
\frac{\dot{\phi_b}^2}{\phi_b^2}\ll \frac{|\dot{\tilde{H}}|^2}{|\tilde{H}|^2},
\label{fast and slow var}
\end{equation}

Using the estimate $\frac{m^2}{M_P^2}\approx 8\cdot 10^{-38}$, obtained in section \ref{Bosons  in LPP near vacuum}, we see that the second term in $V(\tilde{H})$ is negligible compared to the last term. 
The potential $V(\tilde{H})$ has a minimum at
\begin{equation}
|\tilde{H}_0|^2=  \frac{1}{2}\tilde{v}^2=- \frac{\xi}{\lambda}R_b=\frac{1}{2\xi}M_P^2 
\label{tilde H2min}
\end{equation}
with
\begin{equation}
 \tilde{v}=\frac{1}{\sqrt{\xi}}M_P=\sqrt{6}M_P
\label{tilde v}
\end{equation}
if $\xi=\frac{1}{6}$.
This means that, {\em up to relative corrections of order $\lesssim 3\cdot 10^{-2}$, during the entire slow-roll inflation stage, the potential $V(\tilde{H})$} of the Higgs field $\tilde{H}({\mathsf x})$ in the LPPF  {\em has a minimum independent of the slowly rolling  background field $\phi_b$}. However, this minimum cannot be regarded as a stationary vacuum state, since the presence of $\dot{\phi_b}/\phi_b$ in $\tilde{\mathit{D}}_0$
makes it, strictly speaking, impossible for the kinetic term 
$\tilde{g}^{\alpha\beta}\left(\tilde{\mathit{D}}_{\alpha}\tilde{H}\right)^{\dag}\tilde{\mathit{D}}_{\beta}\tilde{H}$ to remain exactly zero during slow-roll inflation. The slow change of $\phi_b(t)$ apparently allows one to describe this interesting physical effect as {\em adiabatic}.
However, in this paper we will limit ourselves to merely stating this phenomenon, leaving its study for the future. Therefore, taking into account inequality
(\ref{fast and slow var}), we will neglect the contribution of $\dot{\phi_b}/\phi_b$ to $\tilde{\mathit{D}}_0$ and continue our study in this approximation.

Obviously, the Lagrangian 
(\ref{L bos after redef}) is $SU(2)\times U(1)$ gauge invariant \footnote{Note that  the matrix used to describe the $SU(2)$ gauge transformations near the minimum $\tilde{v}$ of the potential $V(\tilde{H})$ must  differ from the matrix used in the GWS model near the VEV, $v\approx 246\, GeV$.  We are now studying a cosmological modification of the electroweak theory, in which the minimum of the potential is achieved at $\tilde{v}$  of the order of  the Planck scale.
 Therefore, by analogy with the GWS theory, near the minimum of the potential $V(\tilde{H})$ in the matrix of the $SU(2)$  gauge transformations $\tilde{U}$ it is necessary to use $\tilde{v}=\sqrt{6}M_P$ instead of $v$, that is 
\begin{equation}
\tilde{U}=\exp\Big(\frac{i}{\tilde{v}} \tau_a\tilde{\theta}^a({\mathsf x})\Big),
\label{matrix with tilde v}
\end{equation}
where the isovector of "angular" variables $\tilde{\theta}^a$  ($a=1,2,3$) is used.
In accordance with this, the field $\tilde{H}$ near the minimum $\tilde{v}$ can be represented in the form
\begin{equation}
\tilde{H}({\mathsf x})\simeq\exp\Big(-\frac{i}{\tilde{v}} \tau_a\tilde{\chi}^a({\mathsf x})\Big)\frac{1}{\sqrt{2}}\left(\begin{matrix} 0 \\ \tilde{v}+\tilde{h}({\mathsf x}) \end{matrix}\right) \simeq \frac{1}{\sqrt{2}}\left(\begin{matrix} -i(\tilde{\chi}^1-i\tilde{\chi}^2) \\ \tilde{v}+\tilde{h}+i\tilde{\chi}^3 \end{matrix}\right),
\label{tilde H presentation near min}
\end{equation}
 and the last equality is valid for $|\tilde{\chi}^a|/\tilde{v}\ll 1$ and $|\tilde{h}|/\tilde{v}\ll 1$. This parametrization will be convenient  to describe the gauge fixing Lagrangian in section \ref{Quantization procedures}.}. 
Therefore, quite analogous to how it is done in the GWS theory,  one can  go to the unitary gauge, where the Higgs isodoublet $\tilde{H}$ reduces to
\begin{equation}
\tilde{H}= \frac{1}{\sqrt{2}}\left(\begin{matrix} 0 \\ \tilde{v}+\tilde{h}({\mathsf x}) \end{matrix}\right).
\label{tilde Higgs doublet unitary}
\end{equation}
In accordance with this, we can define a new vacuum state $\widetilde{|0\rangle}$ for which  SSB has the form
\begin{equation}
 \widetilde{\langle 0|}\tilde{H}\widetilde{|0\rangle}= \frac{1}{\sqrt{2}}\left(\begin{matrix} 0 \\ \tilde{v} \end{matrix}\right),
\label{VEV tilde Higgs}
\end{equation}
where $\widetilde{\langle 0|}\tilde{h}\widetilde{|0\rangle}=0$. To distinguish these SSB and  vacuum from the SSB and vacuum in the GWS theory, we will use the terms "up-SSB", "up-vacuum" and "up-VEV" for $\tilde{v}$. 
Similarly, to denote the difference from the Higgs boson  $h$ in the GWS theory, for the $\tilde{h}$ boson we will use the term “up-Higgs boson”.
 The potential
of the up-Higgs boson $\tilde{h}$ near the up-vacuum reduces to the form
\begin{equation}
V(\tilde{h})= \frac{\lambda}{8}\bigl(2 \tilde{v}\tilde{h}+\tilde{h}^2\bigr)^2-\frac{\lambda}{8\xi^2} M_P^4,
\label{V tilde h}
\end{equation}
from where it follows the expression for the mass squared of the up-Higgs boson $\tilde{h}$ 
\begin{equation}
m_{\tilde{h}}^2=\lambda \tilde{v}^2=6\lambda M_P^2.
\label{m^2 h LPP}
\end{equation}
Therefore, we get that  $m_{\tilde{h}}\approx 2.9\cdot 10^{13} \, GeV$.

The properties of gauge bosons studied in the LPPF at the stage of slow-roll inflation  differ significantly from their properties in the GWS theory. Therefore, to avoid confusion, we will use the notations $\tilde{W}$, $\tilde{Z}$ and $\tilde{A}$ for them.
 Similar to the GWS theory we define
\begin{eqnarray}
&&\tilde{W}^{\pm}_\mu=\frac{1}{\sqrt{2}}(A^1_\mu\mp iA^2_\mu), 
\nonumber
\\
&& \tilde{Z}_{\mu}=\sin\theta_W  \, B_\mu -\cos\theta_W \, B^3_\mu, \quad \tilde{A}_{\mu}=\cos\theta_W  \, A_\mu +\sin\theta_W \, A^3_\mu,
\label{defining W Z A infl stage}
\end{eqnarray} 
where the Weinberg angle $\theta_W$ is given by eq.(\ref{tan theta W}), that is the same as in the GWS theory.
Using the values of $g$ and $g'$  found in section \ref{Bosons  in LPP near vacuum}, eq.(\ref{g and gprime}),  we obtain from the  Lagrangian (\ref{L bos after redef}) that  masses of $\tilde{W}$ and  $\tilde{Z}$-bosons generated by the  up-SSB are equal to
\begin{equation}
M_{\tilde{W}}=\frac{g}{2}\cdot \tilde{v}\approx 3.2\cdot 10^{-3}M_P \approx 7.8\cdot 10^{15}\, GeV, 
\label{MW infl}
\end{equation} 
\begin{equation}  
M_{\tilde{Z}}=\frac{1}{2}\sqrt{g^2+g^{\prime 2}}\cdot \tilde{v}\approx 3.6\cdot 10^{-3}M_P \approx 8.8\cdot 10^{15}\, GeV 
\label{MZ infl}
\end{equation} 
and $M_{\tilde{A}}=0$ for photon.

The relationship between $g$, $g'$, Weinberg angle $\theta_W$ and electric charge is similar to that in the GWS model.
But a direct consequence of the values of $g$ and $g'$ given by eq.(\ref{g and gprime}) is a “surprising” result: in the up-copy, the electric charge of the $\tilde{W}$ boson is equal to
\begin{equation}
\tilde{e}=g\sin \theta_W=\mathtt{g}\sqrt{b_k-1}\,\sin\theta_W=4\cdot 10^{-3}e \quad \text{or} \quad \tilde{\alpha}=\frac{\tilde{e}^2}{4\pi}=1.6\cdot 10^{-5}\alpha.
\label{e infl}
\end{equation} 

In what follows, to shorten the name "Cosmologically modified copy of the electroweak SM  in the LPPF on the cosmological background at the stage of slow-roll inflation", we, following the terminology introduced after eq.(\ref{VEV tilde Higgs}), will often call this copy \underline{"up-copy"}.

\subsection{Fermionic sector of up-copy}
\label{fermions infl slow}

Let's start with the action $S^{(ferm \, LPPF)}_{(on \, back)}$, eqs.(\ref{Sferm LPP on back})-(\ref{L quarks LPP on back}). First of all, we draw attention
to the fact  that for any real differentiable function  $f({\mathsf x})$ and for any  fermion field $\Psi$ the following equality holds
\begin{equation}
 f({\mathsf x})\Bigl[\overline{\Psi}\tilde{\gamma}^{\mu}\tilde{\nabla}_{\mu}\Psi -(\tilde{\nabla}_{\mu}\overline{\Psi})\tilde{\gamma}^{\mu}\Psi\Bigr]=
\bigl(f^{1/2}\overline{\Psi}\bigr)\tilde{\gamma}^{\mu}\tilde{\nabla}_{\mu}\Bigl(f^{1/2}\Psi\Bigr) -\tilde{\nabla}_{\mu}\Bigl(f^{1/2}\overline{\Psi}\Bigr)\tilde{\gamma}^{\mu}\bigl(f^{1/2}\Psi\bigr).
\label{absorbing f(x) in kinetic terms}
\end{equation}
Consequently, all kinetic terms  in the Lagrangians $L^{(lept \, LPPF)}_{(on \, back)}$ and $L^{(quarks \, LPPF)}_{(on \, back)}$ are reduced to  canonical form  using redifinitions
\begin{equation}
{\tilde L_l}=\frac{\sqrt{b_l-\zeta_b}}{\bigl((1+\zeta_b)\Omega_b\bigr)^{3/4}}L_l, \quad {\tilde l_R}=\frac{\sqrt{b_l-\zeta_b}}{\bigl((1+\zeta_b)\Omega_b\bigr)^{3/4}}l_R, \quad  {\tilde \nu_R}=\frac{\sqrt{b_e-\zeta_b}}{\bigl((1+\zeta_b)\Omega_b\bigr)^{3/4}}\nu_R,
\label{leptions redefin at infl}
\end{equation}
\begin{equation}
{\tilde L_q}=\frac{\sqrt{b_q-\zeta_b}}{\bigl((1+\zeta_b)\Omega_b\bigr)^{3/4}}L_q, \qquad {\tilde r_q}=\frac{\sqrt{b_q-\zeta_b}}{\bigl((1+\zeta_b)\Omega_b\bigr)^{3/4}}r_q
\label{quarks redefin at infl}
\end{equation}
Considering the Yukawa coupling terms of  the Lagrangians $L^{(lept \, LPPF)}_{(on \, back)}$ and $L^{(quarks \, LPPF)}_{(on \, back)}$, we have to replace the Higgs isodoublet $H$ using the redefinition (\ref{redif H in slow}). The latter was carried out  in an approximation in which we neglected relative corrections of the order $\lesssim  3\cdot 10^{-2}$. Therefore, when studying the fermion sector, we can limit ourselves to the same accuracy.  Taking into account eq.(\ref{zeta during infl}) and the results described by eqs.(\ref{b_l values}), (\ref{b_nu}), (\ref{b_q values}) obtained in sec.(\ref{fermion masses}),  one can make replacements according to the following approximate equalities\footnote{Considering that for the $\tau$-lepton the parameter $b_{\tau}$ exceeds 1 by 0.07, we note that in this case the relative correction is approximately equal to $0.07$.}
 $(b_l\pm\zeta_b)\approx b_l\approx 1$, \, $(b_{\nu}+\zeta_b)\approx b_{\nu}\approx 1$ \ and $(b_q \pm\zeta_b)\approx b_q\approx 1$.

As a result of the described redefinitions,  the action $S^{(ferm \, LPPF)}_{(on \, back)}$, eqs.(\ref{Sferm LPP on back})-(\ref{L quarks LPP on back}),
in the case of up-copy (i.e. when the background is in the stage of slow-roll inflation) takes the form
\begin{equation}
S^{(ferm \, LPPF)}_{(slow\,roll\,infl)}=\int\Bigl(L^{(lept \, LPPF)}_{(slow\,roll\,infl)} +L^{(quarks \, LPPF)}_{(slow\,roll\,infl)} \Bigr)\sqrt{-\tilde{g}}d^4{\mathsf x}
\label{action ferm on back slow-roll infl}
\end{equation}
\begin{eqnarray}
L^{(lept \, LPPF)}_{(slow\,roll\,infl)}&=&\sum_{l=e,\mu,\tau}\Biggl(\frac{i}{2} \Bigl(\overline{\tilde{L}_l}\tilde{\gamma}^{\mu}\tilde{\nabla}_{\mu}\tilde{L}_l -(\tilde{\nabla}_{\mu}\overline{\tilde{L}_l})\tilde{\gamma}^{\mu}\tilde{L}_l+
\overline{\tilde{l}_R}\tilde{\gamma}^{\mu}\tilde{\nabla}_{\mu}\tilde{l}_R -(\tilde{\nabla}_{\mu}\overline{\tilde{l}_R})\tilde{\gamma}^{\mu}\tilde{l}_R\Bigr)
\nonumber
\\
&&\hspace{2.5cm}- y^{(ch)}\left(\overline{\tilde{L}_l}\tilde{H}\tilde{l}_R +
\overline{\tilde{l}_R}\tilde{H}^{\dag}\tilde{L}_l\right)\Biggr)
\label{L lept in LPP slow-roll infl}
\\
&&\hspace{1cm}+\frac{i}{2}\bigl[\overline{\tilde{\nu}_R}\tilde{\gamma}^{\mu}\frac{\partial\tilde{\nu}_R}{\partial{\mathsf x}^{\mu}}
 - \frac{\partial\overline{\tilde{\nu}_R}}{\partial{\mathsf x}^{\mu}}\tilde{\gamma}^{\mu}\tilde{\nu}_R\bigr]
- y_{\nu}\left(\overline{\tilde{L}_e}\tilde{H}_c\tilde{\nu}_R +
\overline{\tilde{\nu}_R}\tilde{H}_c^{\dag}\tilde{L}_e\right).
 \nonumber
\end{eqnarray}
\begin{eqnarray}
L^{(quarks \, LPPF)}_{(slow\,roll\,infl)}&=&\sum_{q=u,c,t}\Biggl(\frac{i}{2} \bigl[\overline{\tilde{L}_q}\tilde{\gamma}^{\mu}\tilde{\nabla}_{\mu}\tilde{L}_q -(\tilde{\nabla}_{\mu}\overline{\tilde{L}_q})\tilde{\gamma}^{\mu}\tilde{L}_q+
\overline{\tilde{r}_q}\tilde{\gamma}^{\mu}\tilde{\nabla}_{\mu}\tilde{r}_q -(\tilde{\nabla}_{\mu}\overline{\tilde{r}_q})\tilde{\gamma}^{\mu}\tilde{r}_q\bigr]
\nonumber
\\
&&\hspace{2cm}-y^{(up)}\bigl[\overline{\tilde{L}_q}\tilde{H}_c \tilde{r}_q +
\overline{\tilde{r}_q}\tilde{H}_c^{\dag}\tilde{L}_q\bigr]\Biggr).
 \label{L q in LPP slow-roll infl}
\end{eqnarray}

Using our choices for $y^{(ch)}$, $y^{(up)}$ and $y_{\nu}$ made in section \ref{fermion masses}, we find  that, up to relative corrections of the order of $\lesssim 3\cdot 10^{-2}$, the fermion masses generated by the up-SSB  of the $S\tilde{U}(2)\times U(1)$ gauge symmetry are
\begin{equation}
m_{\tilde{e}}\approx m_{\tilde{\mu}} \approx m_{\tilde{\tau}}\approx y^{(ch)}\frac{\tilde{v}}{\sqrt{2}}\approx \sqrt{3}\cdot 10^{-6}M_P,
 \label{equal masses of leptons and nuetr}
\end{equation}
\begin{equation}
m_{\tilde{u}}\approx m_{\tilde{c}} \approx m_{\tilde{t}}\approx  y^{(up)}\frac{\tilde{v}}{\sqrt{2}}\approx \sqrt{3}\cdot 10^{-5}M_P,
 \label{equal masses of quarks}
\end{equation}
\begin{equation}
 m_{\tilde{\nu}}\approx y_{\nu}\frac{\tilde{v}}{\sqrt{2}}\approx \sqrt{3}\cdot 10^{-11}M_P.
\label{up neutr mass}
\end{equation}

\subsection{Summary of the up-copy results (at  the tree-level)}
\label{Coupling constants and particle masses} 

Sum of the actions with the Lagrangians (\ref{L bos after redef}) for the bosonic sector and (\ref{L lept in LPP slow-roll infl}, (\ref{L q in LPP slow-roll infl}) for the fermionic sector describes the tree-level up-copy (a cosmologically modified copy of the tree-level electroweak SM, considered in the LPP frame, which is realized on the cosmological background at the inflation stage in the slow-roll regime). {\em The structure of this action coincides with the structure of the action for SM in an external gravitational field. However, the vacuum, parameters and masses of particles differ significantly.} 
In the next sections we will use many of the results obtained above. For ease of reference, we summarize them here.  

As can be seen from  eq.(\ref{V tilde H}), the coupling constant of the quartic selfinteraction of the Higgs field $\tilde{H}$ is equal to $\lambda/2\approx 1.1 \cdot10^{-11}$. The Higgs field mass term  of the "wrong" sign, responsible for the up-SSB, is created by the background scalar curvature $R_b$ and is equal to $\xi R_b|\tilde{H}|^2\approx -3\lambda M_P^2|\tilde{H}|^2$. The up-VEV is equal to $\tilde{v}=\sqrt{6}M_P$.

An important issue for the up-copy  is the values of the gauge coupling parameters. From the definitions (\ref{Dmu in V4}), (\ref{Nabla SM in Ein}) and (\ref{D redifened}) of the operator $\tilde{\nabla}_{\mu}$ it follows that the gauge coupling constants  coincide with the  corresponding model parameters  $g$ and $g'$ in the primordial action (see eqs.(\ref{Nabla SM}) and (\ref{nabla f})); their values $g\approx 2.6\cdot 10^{-3}$, \,  $g'\approx 1.4\cdot 10^{-3}$ were found at the end of section \ref{Bosons  in LPP near vacuum}. The electric charge $\tilde{e}=g\sin \theta_W\approx 4\cdot 10^{-3}e$ was represented by eq.(\ref{e infl}).

An effect similar to that described for $g$ and $g'$ was also found for the Yukawa coupling constants $y^{(ch)}$, $y_{\nu}$ and $y^{(up)}$ of leptons and up-quarks in the Lagrangians (\ref{L lept in LPP slow-roll infl}) and (\ref{L q in LPP slow-roll infl}), where these constants coincide with the corresponding 
{\em universal} Yukawa coupling constants in the primordial actions (\ref{S l3}) and (\ref{Sq}). 
As we will see in section \ref{Effective potential}, the smallness of the values of $h^{(ch)}\approx 10^{-6}$, \, $y_{\nu}\approx 10^{-11}$ and $h^{(up)}\approx 10^{-5}$, as well as the smallness of $g$ and $g'$ will be of decisive importance for the  up-vacuum stability.

The masses of the $\tilde{W}$ and $\tilde{Z}$ bosons are of the order of $10^{15}-10^{16}\,GeV$, see eqs.(\ref{MW infl}) and (\ref{MZ infl}). The up-Higgs boson mass was obtained in sec.\ref{Bosonic sector slow-roll}, eq.(\ref{m^2 h LPP}). Interestingly, the found  mass of the up-Higgs boson $m_{\tilde{h}}\approx 2.9\cdot 10^{13} \, GeV$ agrees with the inflaton mass predicted by the baryogenesis models \cite{GUT baryogenesis}. However, it is obvious that with such values of coupling constants and masses, the processes of elementary particle physics  predicted by the TMSM in the up-copy may differ significantly from those when using the conventional SM+Gravity  up to energies of the scale of $10^{15}-10^{16}\,GeV$. But this extremely important and interesting aspect of TMSM is beyond the scope of this paper.

\section{Some quantitative results concerning
\\
  the concept of cosmological realization of TMSM}
\label{The essence of the cosmological concept}

The results obtained in sections \ref{Bosons  in LPP near vacuum} and \ref{Bosonic sector slow-roll} make it possible to supplement the formulation of 
the concept of cosmological realization of the TMSM with very important and interesting quantitative content.

\underline{At the stage of slow-roll inflation},  the potential $V(\tilde{H})$, eq.(\ref{V tilde H}), of the Higgs field $\tilde{H}({\mathsf x})$ in the LPPF  has a minimum at $\tilde{H}_0$, defined by eq.(\ref{tilde H2min}). The surprising effect mentioned after eq.(\ref{tilde v}) is that {\em during the  entire process of slow-roll inflation} 
$\tilde{H}_0$ is  independent\footnote{Let us recall that all quantitative statements are valid in approximation, when we neglect relative corrections of the order of $\lesssim 3\cdot 10^{-2}$.} of the slowly rolling  background field $\phi_b(t)$. 
This  effect  has clear explanation. The up-SSB occurs
due to the term $\xi R_b|\tilde{H}|^2$ which appears in the potential  $V(\tilde{H})$, eq.(\ref{V tilde H}), as a mass term with "wrong" sign. In general, the background scalar curvature $R_b$ is a function of the background field $\phi_b$ and therefore the position of the minimum of the potential depends on
$\phi_b$.  But the slow-roll inflation, described in the CF,  takes place when 
$\phi\geq 14.2 M_P$, where the TMT-effective potential $U_{eff}^{(tree)}(\phi)$ is flat  with high accuracy (see eqs.(\ref{V eff}),  (\ref{Veff varphi tanh no fine tun 1}) and figure \ref{fig1}). Besides, in the regime of slow-roll inflation, the kinetic part of the TMT-effective energy-momentim tensor is negligible compared to $U_{eff}^{(tree)}$. Therefore, it follows from the Einstein equations that the background scalar curvature $R_b$  appears to be constant with high accuracy (see eqs.(\ref{R in Ein via Ueff during infl})-(\ref{R in Ein during infl}) and (\ref{back R for two cases})).

Let us now move on to the analysis of the correspondence between some physical quantities of the TMSM considered in the CF and in the LPPF at the final stage of cosmological evolution, that is, \underline{near the vacuum}.
Following the prescriptions of the concept of the cosmological realization of the TMSM, we obtained two different descriptions of the vacuum state of the Higgs field: one description in CF, which was studied in ref. \cite{1-st paper} and briefly discussed in section \ref{summary of 1-st paper}; the other description in LPPF, which was studied in section \ref{Bosons  in LPP near vacuum}. From eqs. (\ref{cosm aver Higgs doublet}), (\ref{def of sigma}) and (\ref{s2}) it follows that the vacuum state in CF can be described as follows:
\begin{equation}
\langle H\rangle_{cosm. \, aver.}|_{(vac \, in \, CF)}= \left(\begin{matrix} 0 \\ \frac{1}{\sqrt{2}}\sigma\end{matrix}\right), \quad \text{where} \quad
\sigma=\Biggl(\frac{\zeta_v-b_p}{\zeta_v+b_p}\cdot\frac{m^2}{\lambda}\Biggr)^{1/2}.
\label{cosm aver Higgs  doublet in vac}
\end{equation}
When describing particle physics in the LPPF on this background vacuum state, we must take into account that the Higgs field vacuum state in the LPPF is described by eq.(\ref{Higgs  doublet unitary}). Therefore, in eqs.(\ref{cosm aver Higgs  doublet in vac}) and (\ref{Higgs  doublet unitary}) we have different descriptions of the same vacuum. Noting that by redefinition (\ref{H in LPP near vac}), $H$ and $\mathcal{H}$ differ by a constant factor ${\sqrt\frac{b_k-1}{2}}$, we obtain that the relation between the corresponding VEV's is
\begin{equation}
\sigma =\frac{v\sqrt{2}}{\sqrt{b_k-1}}\approx 353.6 \, v.
\label{sigma via v}
\end{equation} 
Here it is probably worth recalling the "technical" reasons leading to the difference in the descriptions of the vacuum state in the CF and LPPF.
In the CF, to reduce the $\phi$-equation (\ref{phi eq Ein with zeta and bk-zeta}) to the equation near  vacuum in the form (\ref{phi eq via Vprime}) (with the derivative of the potential (\ref{Vprime near vac})) it was necessary to devide all terms of the equation by  $(b_k-\zeta_v)$. 
In the LPPF,  to reduce the Lagrangian (\ref{L bos back near vac}) near vacuum  to the canonical form of the Lagrangian of the GWS theory, it was necessary  to use the redifinition (\ref{H in LPP near vac}) of the Higgs doublet $H$. As a result,  the model parameters $\lambda$ and $m^2$ are devided by different powers of  $(b_k-\zeta_v)$ (see eq.(\ref{lambda and m2 in LPP near vac})), which leads to a VEV different from that obtained in the CF.

\section{Quantization based on the concept of cosmological realization of TMSM}
\label{Quantization}

{\em In TMSM we operate in accordance with the concept of cosmological realization of TMSM.} The essence of the concept is that the electroweak SM manifests itself in the form of \underline{cosmologically modified copies of the GWS theory  described in the LPPF}.
{\em Each of these copies is realized on  a cosmological background  at a certain stage of its classical evolution} and   \underline{the cosmological background is described in the CF}. 
In general, the structure of cosmologically modified copies may differ significantly from the structure of the GWS theory. However,  in the previous sections we discovered that not only at the stage of cosmological evolution at the energy scales of accelerator physics  the tree-level GWS theory is reproduced, but also at the stage of slow-roll inflation the corresponding copy of the SM differs from the tree-level GWS theory only in the values of coupling constants, masses and VEV.
In the context of cosmology, these two copies of the electroweak SM are separated from each other by a time interval of the order of the lifetime of our Universe. Therefore, {\em  two significantly different vacuum states in these copies of the electroweak SM, as well as the physical processes occurring in them, are not able to directly influence each other. Consequently, the quantization of these copies must also be performed as independent procedures.}

\subsection{Some features of quantization procedures}
\label{Quantization procedures}

\paragraph{At the  stage near vacuum}
The results of section \ref{TMSM in LPP near vacuum} show that {\em the tree-level TMSM considered in the LPPF {\em on the cosmological background near vacuum}
coincides with the tree-level GWS theory}  with the Lagrangian
\begin{eqnarray}
L_{GWS}
=\Bigl(L^{(bos \, LPPF)}+L^{(lept \, LPP)}+L^{(quarks \, LPPF)}\Bigr)\Bigm|_{(on \, back \,near\,vac)}
\label{L tot LPP near vac}
\end{eqnarray}
where the three Lagrangians on the right side are given by eqs.(\ref{L bos back near vac}),  (\ref{L lept in LPP appr vac}) and (\ref{L q in LPP appr vac}).
Therefore, by adding the gauge-fixing  and ghosts terms to the Lagrangian (\ref{L tot LPP near vac}), we obtain an electroweak SM Lagrangian ready for the standard Faddeev-Popov quantization  procedure (see, for example,  \cite{Peskin}).  Thus, near the SM vacuum, the quantization of the TMSM in the LPPF does not differ from the quantization of the GWS theory either in procedure or in results. 

\paragraph{At the  stage of the slow-roll inflation (up-copy)}
	Let us now move on to discuss the cosmologically modified copy of the GWS theory in the LPPF  on the cosmological background at the  stage of the slow-roll inflation.   In section \ref{TMSM in LPP on back slow-rol} we found the contributions of the bosonic and fermionic sectors to the tree-level action of the up-copy, and here they can be conveniently presented as follows
\begin{equation}
S_{(up \, copy)}==\int  L_{(up \, copy)}\,\sqrt{-\tilde{g}}d^4{\mathsf x}
\label{S up-copy}
\end{equation}
\begin{equation}
L_{(up \, copy)}=L^{(bos \, LPPF)}_{(slow \, roll \, infl)}+L^{(lept \, LPPF)}_{(slow\,roll\,infl)}+L^{(quarks \, LPPF)}_{(slow\,roll\,infl)},
\label{L up-copy}
\end{equation}
where the Lagrangians on the right side are given by  eqs.(\ref{L bos after redef}), (\ref{L lept in LPP slow-roll infl}) and (\ref{L q in LPP slow-roll infl}).
This action has the same structure as the action of the electroweak SM in  curved spacetime, various aspects of quantization of which are well studied in the literature \cite{QFT in Curved-1}-\cite{QFT in Curved-15}, \cite{Markkanen}. One of the  differences of the up-copy from the convensional electroweak SM in  curved spacetime consists in  the values of coupling constants, masses and VEV. Recall that for the case of up-copy at the tree-level,  the corresponding set of values is presented in
 section \ref{Coupling constants and particle masses}. Another peculiarity of the up-model is related to the external (background) curvature invariants  that need to be added to the action (\ref{S up-copy}) when constructing\footnote{For brevity, we will skip the step of rewriting the tree level action (\ref{S up-copy}) as a bare action where the $0$-index must be added to all fields and parameters}
 the bare action \cite{Birrell Davies}, \cite{QFT in Curved-8}. 
To ensure multiplicative renormalizability of the theory, it is necessary to add the action of an external gravitational field in the form \cite{QFT in Curved-8}
 \begin{equation}
S_{(back \,curv)}=\int \sqrt{-\tilde{g}}d^4{\mathsf x}\Bigl[-V_0-a_{01}R+a_{02}R^2+a_{03}R_{\mu\nu}R{\mu\nu}
+a_{04}R_{\mu\nu\alpha\beta}R^{\mu\nu\alpha\beta}+a_{05}\tilde{\square}R  \Bigr],
\label{S curv inv}
\end{equation} 
where $V_0$, $a_{01}$, ...$a_{05}$ are the bare parameters. When studying up-copy, the scalar curvature of the external gravitational field is the background curvature $R_b(\tilde{g})$. In addition, since we are operating in an approximation in which relative corrections of the order of $\lesssim 3\cdot 10^{-2}$ have been neglected, from eqs.(\ref{R in Ein via Ueff during infl})-(\ref{R in Ein during infl}), (\ref{back R for two cases}) it follows that the background space-time can be considered as a de Sitter space with a constant Hubble parameter $H^2=-\frac{1}{12} R_b=const .$.
Therefore,  the  invariants
 \begin{equation}
R_{\mu\nu}R{\mu\nu}=\frac{1}{4}R_b^2, \qquad R_{\mu\nu\alpha\beta}R^{\mu\nu\alpha\beta}=\frac{1}{6}R_b^2
\label{Constancy curv inv}
\end{equation}
are also constats.
Continuing to neglect relative corrections of order $\lesssim 3\cdot 10^{-2}$, we also obtain $\tilde{\square}R_b=0$.
 Thus, adding $S_{(back \,curv) }$ is equivalent to adding an inessential constant to the constructed bare Lagrangian of the up-copy.

To apply functional quantization to the model, 
the gauge-fixing  and ghosts terms must be added to the Lagrangian $L_{(up \, copy)}$  as usual.
To avoid confusion with the non-minimal coupling constant $\xi$ when using $R_{\xi}$ gauges, we parameterize the gauge by $\omega$. 
Taking into account the general coordinate invariance, the gauge-fixing term added to the Lagrangian $L_{(up \, copy)}$ should be chosen as
\begin{equation}
L_{G}=-\frac{1}{2}(G)^2, \qquad G^j=\frac{1}{\sqrt{\omega}}\Biggl(\frac{1}{\sqrt{-\tilde{g}}}
\frac{\partial\bigl(\sqrt{-\tilde{g}}A^{j\mu}\bigr)}{\partial{\mathsf x}_{\mu}}-\omega g F^j_a \tilde{\chi}^a\Biggr), \quad  j=1,2,3,4;  \quad a=1,2,3,
\label{G2 and Gj}
\end{equation}
where we used the parametrization defined in footnote 15  and  introduced  the  definitions \cite{Peskin}, \cite{Markkanen} 
\begin{equation}
gF^j_a=\frac{\tilde{v}}{2} \left( \begin{array}{ccc}
g & 0 & 0 \\ 
0 & g & 0 \\
0 & 0 & g \\
0 & 0 & -g'
\end{array} \right),
\qquad (A^{1\mu}, A^{2\mu}, A^{3\mu})=\mathbf{A}^{\mu},  \qquad  A^{4\mu}=B^{\mu}.
\label{Fja}
\end{equation}
Then, similarly to how it happens in the GWS theory \cite{Peskin}, the dangerous part in the kinetic term of the Higgs field, mixing $A^{j\mu}$ and 
$\tilde{\chi}^a$, is canceled out by the corresponding term in $L_{G}$. 
The ghost contribution  to the Lagrangian differs from its expression  in the GWS theory \cite{Peskin} by an appropriate modification to satisfy the requirements of general coordinate invariance. But we will not need to use it explicitly later in this paper.

\subsection{One-loop effective potential in the up-copy of the GWS theory}
\label{Effective potential}

There are at least two reasons why accounting the quantum effects of the Standard Model is fundamentally important in Higgs inflation models. First, in order to avoid conflicts with the CMB data, one should make sure that quantum corrections do not have a significant effect on the flat shape of the potential. Secondly, the vacuum stability  must be ensured in some way. A huge number of articles are devoted to solving these issues within the framework of conventional theory in 1-loop and higher approximations, improved by the RG method. This is, in particular, motivated by the fact that the stability of the vacuum is extremely sensitive to the experimental values of the masses of the top quark and the Higgs boson. In contrast to this, in this section we will see that in the TMSM under consideration, the quantum corrections 
in the one-loop approximation leaves the vacuum stable and the slow-roll inflation regime is preserved.

The most systematic and exhaustive calculation of the effective potential in the 1-loop approximation for the SM in an arbitrary external gravitational field is apparently presented in paper \cite{Markkanen}. In particular, for the case of the SM in de Sitter space, this work includes the results of numerical calculations of the RG improved 1-loop potential, accounting $\beta$-functions. 
At first glance, the discussion in section \ref{Quantization procedures} suggests that the up-copy has a canonical structure in the sense that its functional quantization and the computation of the RG-improved 1-loop effective potential are not significantly different  from what is done with SM in curved spacetime in general and in ref.\cite{Markkanen} in particular. But the computational procedure runs into a serious theoretical problem:  unlike  the GWS theory, in the up-copy we do not have experimentally verified values of the physical observables that are needed as boundary conditions in solving the differential equations for the running couplings. However, a more detailed analysis  of the situation shows that there are two  peculiarities of the up-copy that allow us to circumvent this problem. First,  the tree level up-copy   is formulated at energy
scale $\sim M_{\tilde{W}}\sim 10^{15}-10^{16}GeV$. Therefore, the renormalization parameter for determining physical observables must be chosen to be approximately equal to $\mu_0\approx 10^{15}-10^{16}GeV$. Then extrapolation  to the Planck scale requires increasing the renormalization scale by only 2-3 orders of magnitude, from $\mu_0$ to $\mu\approx M_P\sim 10^{18}GeV$. Consequently, the logarithm determining the behavior of the running coupling constants can receive an addition of no more than $\ln\frac{\mu}{\mu_0}\lesssim 6.9$.  As an example illustrating the situation, let us  consider the running $SU(2)$ gauge coupling $g(\mu)$, for which the one-loop $\beta$-function is $\beta_g=-\frac{1}{(4\pi)^2}\frac{19}{6}g^3$, that is
\begin{equation}
g(\mu)=g(\mu_0)\Bigl(1+\frac{19}{6(4\pi)^2}g^2(\mu_0)\ln\frac{\mu}{\mu_0}\Bigr)^{-1/2}.
\label{g2 of ln}
\end{equation}
The second peculiarity of the up-copy is that the coupling constants 
 in the tree-level Lagrangian (\ref{L up-copy}), (\ref{L bos after redef}), (\ref{L lept in LPP slow-roll infl}) and (\ref{L q in LPP slow-roll infl})
 are unusually small in comparison with those in the tree-level GWS model.
 In particular, in the up-copy,  $g\approx 2.6\cdot 10^{-3}$, as it is given by eq.(\ref{g and gprime}).
Therefore, even if we assume that the physical
\footnote{It should be noted that in fact, unlike the GWS model, when studying the up-copy we do not have sufficiently convincing grounds to call
the values of running couplings near the up-vacuum "physical".}
coupling $g(\mu_0)$ is substantially larger than $g$, say, $g(\mu_0)\sim 10^{-2}$, then it follows from (\ref{g2 of ln}) that the relative difference between  $g(\mu)$ on the Planck energy scale $\mu\approx M_P$ and $g(\mu_0)$ is proportional to  $g^2(\mu_0)$
\begin{equation}
\frac{|g(M_P)-g(\mu_0)|}{g(\mu_0)}\approx \frac{19}{12(4\pi)^2}g^2(\mu_0)\ln\frac{\mu}{\mu_0}\sim 7\cdot 10^{-6}.
\label{relative diff in g(mu)}
\end{equation}
A similar analysis can be performed for the relative changes of other running couplings of the up-copy\footnote{By choosing $\mu_0\approx 10^{15}-10^{16}GeV$, we followed  the situation familiar from the GWS theory, where the Higgs boson mass is about 1.5 times  larger than  masses of the gauge bosons $W$ and $Z$ and about 2 times less then VEV, and so $M_{W}$ is a natural choice to describe the typical electroweak energy scale. In contrast, in TMSM the up-VEV exceeds $M_P$ by $\sqrt{6}$ times and masses of the gauge bosons 
$\tilde{W}$ and  $\tilde{Z}$ are about two orders of magnitude  larger than  the mass of the Higgs boson $\tilde{h}$.  So in TMSM there is some ambiguity in the choice of the typical TMSM energy scale. However, if we choose $\mu_0\sim m_{\tilde{h}}\sim 10^{13}GeV$ instead of $\mu_0\approx 10^{15}-10^{16}GeV$, the above results change slightly: we have $\ln(M_P/\mu_0)=11.5$ (instead of 6.9) and on the right-hand side of 
eq.(\ref{relative diff in g(mu)}) we get $2.9\cdot 10^{-6}$.}. As a result, we obtain estimates of the same order of magnitude or even many orders of magnitude smaller than the estimates on the right-hand side of 
eq.(\ref{relative diff in g(mu)}). Important conclusions follow from all this: 1) using the RG  improved  technique in the up-copy, we remain within the applicability of perturbation theory; 2) working in the up-copy with an effective potential in the 1-loop approximation, we can completely neglect the dependence of the running couplings on the renormalization parameter. It is worth adding to this discussion a reminder that the description of the up-copy at the classical level was obtained in an approximation where we neglect relative corrections of the order of $\lesssim 3\cdot 10^{-2}$.

The results of the work \cite{Markkanen} can be directly applied to find the one-loop effective Higgs potential in the up-copy of the electroweak SM.
To explicitly indicate that we are dealing with an effective  Higgs potential in the up-copy, in the usual expansion of the real part of the neutral component of the Higgs doublet $\tilde{H}$ about the classical field, for the latter we will use the notation $\tilde{\phi}$
\begin{equation}
\tilde{H}({\mathsf x})= \frac{1}{\sqrt{2}}\left(\begin{matrix} -i\bigl(\tilde{\chi}^1-i\tilde{\chi}^2\bigr) \\ \tilde{\phi}+\tilde{h}+i\tilde{\chi}^3 \end{matrix}\right).
\label{tilde H presentation about class field}
\end{equation}
 In the 1-loop effective potential, the mean field $\tilde{\phi}$ enters through the following combinations (we use here Table 1 in ref. \cite{Markkanen}):
\begin{equation}
\mathcal{M}_{\tilde{h}}^2=\frac{1}{2}m^2+\frac{3}{2}\lambda\tilde{\phi}^2-\Bigl(\xi+\frac{1}{6}\Bigr)R_b \, ,
\label{Mh2}
\end{equation}
\begin{equation}
\mathcal{M}_{\tilde{W}}^2|_{(1,2)}=\frac{g^2}{4}\tilde{\phi}^2-\frac{1}{12}R_b \quad \text{and} \quad 
\mathcal{M}_{\tilde{W}}^2|_{(3)}=\frac{g^2}{4}\tilde{\phi}^2+\frac{1}{6}R_b \, ,
\label{MW}
\end{equation}
\begin{equation}
\mathcal{M}_{\tilde{Z}}^2|_{(1,2)}=\frac{g^2+g'^2}{4}\tilde{\phi}^2-\frac{1}{12}R_b \quad \text{and} \quad 
\mathcal{M}_{\tilde{Z}}^2|_{(3)}=\frac{g^2+g'^2}{4}\tilde{\phi}^2+\frac{1}{6}R_b \, ,
\label{MZ}
\end{equation}
\begin{eqnarray}
&&\mathcal{M}_{\tilde{f}}^2=\frac{1}{2} y_{\tilde{f}}^2\tilde{\phi}^2-\frac{1}{12}R_b \, , \quad \text{where} 
\label{Mf}
\\
 y_{\tilde{f}}=y^{(ch)} \, \, \text{for} \, \, \tilde{f}=\tilde{e}, \tilde{\mu}, \tilde{\tau} ; \qquad &&  y_{\tilde{f}}=y_{\nu}  \,\, \text{for} \,\, \tilde{f}= \tilde{\nu} ; \qquad y_{\tilde{f}}=y^{(up)} \, \, \text{for} \, \,\tilde{f}=\tilde{u}, \tilde{c}, \tilde{t}
\nonumber
\end{eqnarray}
\begin{equation}
\mathcal{M}_{\tilde{\chi}\tilde{W}}^2=\frac{1}{2}m^2+\frac{1}{2}\lambda\tilde{\phi}^2+\omega_{\tilde{W}}\frac{g^2}{4}\tilde{\phi}^2-\Bigl(\xi+\frac{1}{6}\Bigr)R_b \, ,
\label{chi w}
\end{equation}
\begin{equation}
\mathcal{M}_{\tilde{\chi}\tilde{Z}}^2=\frac{1}{2}m^2+\frac{1}{2}\lambda\tilde{\phi}^2+\omega_{\tilde{Z}}\frac{g^2+g'^2}{4}\tilde{\phi}^2-\Bigl(\xi+\frac{1}{6}\Bigr)R_b \, ,
\label{chi z}
\end{equation}
\begin{equation}
\mathcal{M}_{c_{\tilde{W}}}^2=\omega_{\tilde{W}}\frac{g^2}{4}\tilde{\phi}^2+\frac{1}{6}R_b \, ,
\label{c W}
\end{equation}
\begin{equation}
\mathcal{M}_{c_{\tilde{Z}}}^2=\omega_{\tilde{Z}}\frac{g^2+g'^2}{4}\tilde{\phi}^2+\frac{1}{6}R_b \, ,
\label{c Z}
\end{equation}
where separate gauge fixing parameters $\omega_{\tilde{W}}$ and $\omega_{\tilde{Z}}$ are chosen 
for $\tilde{W}^{\pm}$ and $\tilde{Z}$  contributions; $c_{\tilde{W}}$ and $c_{\tilde{Z}}$ are the corresponding ghosts. 
According to the conclusions formulated in the previous paragraph, in all the listed $\mathcal{M}^2$-s, 
the dependence of the renormalized quantities on the renormalization scale can be neglected. 
In addition, due to the applicability of perturbation theory, it can be argued that 
the relative differences between the physical (renormalized) quantities 
 and the corresponding classical quantities are small.

Since the  up-VEV is $\tilde{v}=\sqrt{6}M_P$, we will be interested in $\tilde{\phi}^2\geqslant 6M_P^2$. Taking into account the value of $R_b=-18\lambda M_P^2$ and using the values of the parameters summarized in section \ref{Coupling constants and particle masses}, the following estimates can be obtained
\begin{equation}
\frac{3}{2}\lambda\tilde{\phi}^2 \geqslant 9\lambda M_P^2 \gtrapprox \Big|\Bigl(\xi+\frac{1}{6}\Bigr)R_b\Big|\approx 6\lambda M_P^2 \, ,
\label{lambda phi2 gsim lambda MP2}
\end{equation}
 \begin{equation}
\frac{g^2}{4}\tilde{\phi}^2\sim \frac{g^2+g'^2}{4}\tilde{\phi}^2 \gtrapprox \mathcal{O}(1)10^{-6}M_P^2 \ggg \frac{1}{6}|R_b|=3\lambda M_P^2\approx  7\cdot 10^{-11} M_P^2 \, ,
\label{g2 phi2 >> lambda MP2}
\end{equation}
\begin{equation}
\frac{1}{2}(y^{(ch)})^2\tilde{\phi}^2 \gtrapprox 3\cdot 10^{-12}M_P^2 \quad \text{and} \quad \frac{1}{12}|R_b|\approx 3.5\cdot 10^{-11} M_P^2\, ,
\label{ych2 > lambda MP2}
\end{equation}
\begin{equation}
\frac{1}{2}(y^{(up)})^2\tilde{\phi}^2 \gtrapprox 3\cdot 10^{-10}M_P^2 > \frac{1}{12}|R_b|\approx 3.5\cdot 10^{-11}  M_P^2
\label{yq gsim lambda MP2}
\end{equation}
and similar estimates are valid in eqs.(\ref{chi w})-(\ref{c Z}).  
 These estimates will be useful below to simplify some terms in the one-loop effective potential. In particular, according to (\ref{g2 phi2 >> lambda MP2}), the contribution of $R_b$ can be neglected compared to the contributions of the gauge fields.

In the tree-level potential (\ref{V tilde H}) represented in terms of $\tilde{\phi}$ 
\begin{equation}
V(\tilde{\phi})= \frac{1}{8}\lambda \tilde{\phi}^4+\Bigl(\frac{1}{4}m^2+\frac{1}{2}\xi R_b\Bigr)\tilde{\phi}^2 
\label{V tilde phi}
\end{equation}
the contribution of the term $\propto m^2\sim 10^{-37}M_P^2$ is negligible compared to $\propto |\xi R_b|\sim\lambda M_P^2\sim 10^{-11}M_P^2$.
Using (\ref{V tilde phi}) as the boundary function, after some algebra we obtain the effective potential in the one-loop approximation, which can be conveniently represented in the following form
\begin{equation}
V_{eff}(\tilde{\phi})=  \frac{\lambda_{eff}}{8}\tilde{\phi}^4+\frac{1}{2}\xi_{eff} R_b\tilde{\phi}^2+\mathcal{O}(1)\lambda^2M_P^4, \quad \text{where} \quad 
\xi_{eff}=\frac{1}{6}(1 +\delta_{\xi}),
\label{Veff tilde phi}
\end{equation}
and $\lambda_{eff}$ and $\delta_{\xi}$ are given by the following expressions
\begin{eqnarray}
\lambda_{eff} =\lambda+\frac{1}{8\pi^2}&\Biggl[&\frac{9}{4}\lambda^2\Biggl(\Bigl(1-\frac{6M_P^4}{5\tilde{\phi}^4}\Bigr)\ln\frac{3\lambda(\tilde{\phi}^2+4M_P^2)}{2\mu^2}-\frac{3}{2}\Biggr)+\Delta^{(4)}_{\omega}
    \label{lambda eff}
\\
&+&\frac{3}{8}g^4\Bigl(\ln\frac{g^2\tilde{\phi}^2}{4\mu^2}-\frac{5}{6}\Bigr)+\frac{3}{16}(g^2+g'^2)^2\Bigl(\ln\frac{(g^2+g'^2)\tilde{\phi}^2}{4\mu^2}-\frac{5}{6}\Bigr)
\nonumber
\\
&-&3\cdot (y^{(ch)})^4\Biggl(\Bigl(1-\frac{57}{10}\frac{\lambda^2}{(y^{(ch)})^4}\frac{M_P^4}{\tilde{\phi}^4} \Bigr)\ln\frac{\frac{1}{2}(y^{(ch)})^2\tilde{\phi}^2 +\frac{3}{2}\lambda M_P^2}{\mu^2}-\frac{3}{2}\Biggr)
\nonumber
\\
&-&9\cdot (y^{(up)})^4\Biggl(\Bigl(1-\frac{57}{10}\frac{\lambda^2}{(y^{(up)})^4}\frac{M_P^4}{\tilde{\phi}^4} \Bigr)
\ln\frac{\frac{1}{2}(y^{(up)})^2\tilde{\phi}^2 +\frac{3}{2}\lambda M_P^2}{\mu^2}-\frac{3}{2}\Biggr)\Biggr],
\nonumber
\end{eqnarray}

\begin{eqnarray}
\delta_{\xi} =-\frac{1}{(4\pi)^2}&\Biggl[&3\lambda\Bigl(\ln\frac{3\lambda(\tilde{\phi}^2+4M_P^2)}{2\mu^2}-\frac{3}{2}\Bigr)-\Delta^{(2)}_{\omega}
\label{m2 eff} 
\\
&-&\frac{3}{2} g^2\Bigl(\ln\frac{g^2\tilde{\phi}^2}{4\mu^2}-\frac{7}{6}\Bigr)
-\frac{9}{2}(g^2+g'^2)\Bigl(\ln\frac{(g^2+g'^2)\tilde{\phi}^2}{4\mu^2}-\frac{7}{6}\Bigr)
 \nonumber  
\\
&+&6\cdot (y^{(ch)})^2\Bigl(\ln\frac{\frac{1}{2}(y^{(ch)})^2\tilde{\phi}^2 +\frac{3}{2}\lambda M_P^2}{\mu^2}-\frac{3}{2}\Bigr)
\nonumber
\\
&+&12 \cdot (y^{(up)})^2\Bigl(\ln\frac{\frac{1}{2}(y^{(up)})^2\tilde{\phi}^2 +\frac{3}{2}\lambda M_P^2}{\mu^2}-\frac{3}{2}\Bigr)\Biggr].
 \nonumber 
\end{eqnarray}
We choosed $\omega_{\tilde{W}}=\omega_{\tilde{Z}}=\omega$ and for the Landau gauge ($\omega=0$)
\begin{equation}
\Delta^{(4)}_{\omega=0}=\frac{3}{4}\lambda^2\Bigl(\ln\frac{\lambda(\tilde{\phi}^2+12M_P^2)}{2\mu^2}-\frac{3}{2}\Bigr)  \quad \text{and}\quad
\Delta^{(2)}_{\omega=0}=18\lambda\Bigl(\ln\frac{\lambda(\tilde{\phi}^2+12M_P^2)}{2\mu^2}-\frac{3}{2}\Bigr). 
\label{DELTA omega 0}
\end{equation}
 For the Feynman gauge ($\omega=1$), $\Delta^{(4)}_{\omega=1}=0$ and
\begin{eqnarray}
\Delta^{(2)}_{\omega=1}=9 g^2\Bigl(\ln\frac{g^2\tilde{\phi}^2}{4\mu^2}-\frac{3}{2}\Bigr)
+\frac{9}{2}(g^2+g'^2)\Bigl(\ln\frac{(g^2+g'^2)\tilde{\phi}^2}{4\mu^2}-\frac{3}{2}\Bigr).
\label{DELTA omega 1} 
\end{eqnarray}
The terms $\propto\frac{M_P^4}{\tilde{\phi}^4}$ added to $1$ in brackets before logarithms in the 1-st,  3-rd and 4-th lines in eq.(\ref{lambda eff}) 
arise due to the influence of the background de Sitter space on the 1-loop effective potential in the form  of the terms
 $\propto R_b^2\ln (\mathcal{M}_{\tilde{h}}^2/\mu^2)$ and $\propto R_b^2\ln (\mathcal{M}_{\tilde{f}}^2/\mu^2)$.
Using estimates (\ref{g2 phi2 >> lambda MP2}) valid for $\tilde{\phi}^2\geqslant 6M_P^2$, in the arguments of the logarithms of the gauge field contributions in eqs.(\ref{lambda eff}) and (\ref{m2 eff}), we omitted $6\lambda M_P^2\sim 10^{-10}M_P^2$ compared to $g^2\tilde{\phi}^2\sim (g^2+g'^2)\tilde{\phi}^2\sim 10^{-5}\tilde{\phi}^2\gtrapprox 6\cdot 10^{-5}M_P^2$.  For the same reason, in the parentheses of the factor before the logarithm in the first line of eq.(\ref{lambda eff}),  $\frac{6M_P^4}{5\tilde{\phi}^4}\leqslant \frac{1}{30}$ can be omitted compared to 1. 

Now we will estimate the coefficients in front of the logarithms of the one-loop contributions of the SM fields to $\lambda_{eff}$:

\begin{itemize}

\item

For the Higgs contribution  we get that the factor before the logarithm is 
\begin{equation}
C_{(Higgs)}=\frac{9}{32\pi^2}\lambda^2\sim 10^{-22}.
\label{Higgs to lambda eff} 
\end{equation}

\item
For the charged leptons contribution to $\lambda_{eff}$, in the 3-rd line of (\ref{lambda eff}), inserting $y^{(ch)}\approx 10^{-6}$ (see section \ref{Coupling constants and particle masses}  and eq.(\ref{y 10-6}))  we get that the factor before the logarithm is 
\begin{equation}
C_l=-\frac{3}{8\pi^2}\cdot (y^{(ch)})^4\Bigl(1-\frac{57}{10}\frac{\lambda^2}{(y^{(ch)})^4}\frac{M_P^4}{\tilde{\phi}^4} \Bigr)
\approx -4\cdot 10^{-26}\Bigl(1-3\cdot 10^3\frac{M_P^4}{\tilde{\phi}^4} \Bigr).
\label{ch to lambda eff} 
\end{equation}
The function $C_l$ changes from $C_l\approx 3\cdot 10^{-24}$ at $\tilde{\phi}=\tilde{v}=\sqrt{6}M_P$ to $C_l\approx -4\cdot 10^{-26}$ as
$\tilde{\phi}\gg\sqrt{6}M_P$. At $\tilde{\phi}\approx 11.2M_P$ the function $C_l$ crosses $0$ changing sign from positive to negative.

\item

For the up-quarks contribution in the 4-th line, inserting $y^{(up)}\approx 10^{-5}$ (see  section \ref{Coupling constants and particle masses} and  eq.(\ref{y 10-5})) yields $0.99\lesssim\bigl(1-\frac{57}{10}\frac{\lambda^2}{(y^{(up)})^4}\frac{M_P^4}{\tilde{\phi}^4}\bigr)\leqslant 1$ and
\begin{equation}
C_q=-\frac{9}{8\pi^2}\cdot (y^{(up)})^4\Bigl(1-\frac{57}{10}\frac{\lambda^2}{(y^{(up)})^4}\frac{M_P^4}{\tilde{\phi}^4} \Bigr)
\approx -10^{-21}.
\label{q to lambda eff} 
\end{equation}

\item

  The coefficients in front of the logarithms of the one-loop contributions  of the gauge fields to $\lambda_{eff}$ is
\begin{equation}
C_g=\frac{3}{64\pi^2}g^4\approx \frac{3}{128\pi^2}(g^2+g'^2)^2\approx 2\cdot 10^{-13}.
\label{g to lambda eff} 
\end{equation}

\end{itemize}
Thus, the contributions of gauge fields to the one-loop effective quartic self-coupling $\lambda_{eff}$ significantly dominate the contributions of all other SM fields: $C_g\approx 10^8\cdot\max\{C_{(Higgs)},\, C_l, \, C_q\}$.
Similarly, the coefficients in front of the logarithms of the gauge boson contributions to $\delta\xi$ are approximately $2\cdot 10^{-8}$, which is about $10^5$ times larger than the corresponding contributions of all other SM fields.

All this has \underline{a very important consequences:}

\begin{itemize}

\item

It is well known that in the context of the ordinary (non-Higgs) inflationary model it is rather dangerous if the inflaton field couples to the SM gauge fields, since its tree coupling constant $\lambda\sim 10^{-11}$ can acquire large radiative corrections $\sim g^4\sim g'^4$, where $g$ and $g'$ are the GWS gauge coupling constants. In the Higgs inflation model studied in ref. \cite{1-st paper} and analyzed further in the present paper, the tree coupling constant 
$\lambda$ is of the same order ($\lambda\approx 2.3\cdot 10^{-11}$), and the Higgs field playing the role of the inflaton couples to the SM gauge fields. However, this happens in the up-copy of the electroweak SM, where $g^4\sim g'^4\sim 10^{-11}$, giving  $C_g\sim 10^{-13}$. That is why this interaction has little effect  on $\lambda_{eff}$. The same applies to $\xi_{eff}$.

\item

The demonstrated  significant dominance of gauge field contributions to one-loop corrections allows us to assert that at the stage of slow-roll inflation \underline{the up-minimum} of the one-loop effective potential
$V_{eff}(\tilde{\phi})$ in the up-copy of electroweak SM described in the LPPF \underline{is absolute}; therefore, \underline{the up-vacuum of the up-copy of the electroweak SM} \underline{is stable}.

\end{itemize}

The constant term in eq.(\ref{Veff tilde phi}) comes from contributions to the 1-loop corrections of the form $\propto R_b^2\ln\frac{R_b}{\mu^2}$. Using the results of a detailed study summarized in Table 2 of ref. \cite{Markkanen} and choosing a natural value of the renormalization parameter $\mu$ of the order of $M_{\tilde{W}}$, i.e., 
$\mu^2\approx 10^{-5}M_P^2$, we obtained a constant $\mathcal{O}(1)\lambda^2M_P^4$ in $V_{eff}(\tilde{\phi})$. 

With the choice of  the renormalization parameter $\mu\approx M_{\tilde{W}}\approx 3\cdot 10^{-3}M_P$, for further applications it is convenient to use the 
effective potential  in the 1-loop approximation $V_{eff}(\tilde{\phi})$, eq.(\ref{Veff tilde phi}), with the  following expressions for $\lambda_{eff}$ and 
$\xi_{eff}$
\begin{equation}
\lambda_{eff}=\lambda\Bigl(1+1.7\cdot10^{-2}\ln\frac{\tilde{\phi}^2}{\tilde{v}^2}-1.4\cdot 10^{-2}\Bigr)
\label{lambda eff num}
\end{equation}
\begin{equation}
\xi_{eff}=\xi\Bigl(1+7.8\cdot10^{-8}\ln\frac{\tilde{\phi}^2}{\tilde{v}^2}-3\cdot 10^{-9}\Bigr)
\label{xi eff num}
\end{equation}
Numerical calculations show that radiative corrections in the 1-loop approximation shift the position of the absolute minimum of the effective potential
$V_{eff}(\tilde{\phi})$  from $\tilde{v}=\frac{M_p}{\sqrt{\xi}}=\sqrt{6}M_p\approx 2.45M_P$ at the tree level to $\tilde{v}_{*}\approx 0.99\tilde{v} \approx 2.42M_P$. But given that our entire tree-level up-copy study was implemented with a precision that neglects relative corrections of the order of $\lesssim 3\cdot 10^{-2}$, we can ignore this  shift and use $\tilde{v}$ instead of $\tilde{v}_{*}$ in what follows. By operating with such precision, we will also ignore   the numbers $1.4\cdot 10^{-2}$ and $3\cdot 10^{-9}$ in $\lambda_{eff}$ and $\xi_{eff}$, respectively. For the same reason, for the up-vacuum energy density, we can take its tree-level value following from eq.(\ref{V tilde h})
\begin{equation}
V_{eff}(\tilde{v})\approx V(\tilde{v})= -\frac{\lambda}{8\xi^2} M_P^4\approx -10^{-10} M_P^4,
\label{V eff vac}
\end{equation}
Here we also neglected  the  constant $\mathcal{O}(1)\lambda^2M_P^4$ in the formula (\ref{Veff tilde phi}), which gives a  relative correction of the order of $\lambda \sim 10^{-11}$. The curvature of $V_{eff}(\tilde{\phi})$ in the up-vacuum is equal to $V''_{eff}(\tilde{v})= 6\lambda M_P^2\approx 1.4\cdot 10^{-10}M_P^2$ (cf. formula (\ref{m^2 h LPP})).

\subsection{On the impact of 1-loop radiative corrections
\\
 on Higgs inflation in the slow-roll regime.}
\label{impact}

Now we want to estimate the extent to which quantum corrections can influence slow-roll inflation.  Within the concept of the cosmological realization of the TMSM, inflation should be described in the CF, but  the one-loop effective potential $V_{eff}(\tilde{\phi})$, eqs.(\ref{Veff tilde phi}), (\ref{lambda eff num}), (\ref{xi eff num}) was obtained in the LPPF. In addition, cosmological evolution consists not only of the inflation stage: as we know from the results obtained in paper \cite{1-st paper} within the tree-level TMSM, in addition to the known subsequent stages of evolution, the possibility of previous stages was discovered. Of fundamental importance for the change of physical parameters during evolution is the behavior of the scalar $\zeta$, which is a necessary component in the description of the cosmological background and its evolution carried out in the CF. But working in the LPPF, we deprive ourselves of the opportunity to find quantum corrections to the constraint (\ref{zeta for radical}), and therefore do not know the influence of quantum corrections on $\zeta$. It follows from this analysis that from a more general point of view, that is, not limiting ourselves only to the inflation stage, after finding the quantum corrections in the LPPF, we must perform the inverse transition from the LPPF to the CF. This will allow us to reconstruct the primordial action for the 
Higgs$+$gravity system taking into account the quantum corrections. And by applying the TMT procedure to this "quantum effective primordial TMT  action", we can find all the cosmological equations and constraint that take into account the quantum corrections.
The reconstruction of the quantum effective primordial TMT action of the Higgs+gravity system is implemented in detail in section \ref{Reconstruction of the TMT effective potential}, the result of which is given by eq.(\ref{S-Higgs+gr eff}). The remarkable feature of the quantum effective primordial TMT action in the 1-loop approximation is that it has the same form as the classical primordial TMT action (\ref{S-Higgs+gr class}), where the only difference consists in the replacement of $\lambda$ and $\xi$ with
$\lambda_{eff}$ and $\xi_{eff}$, that is
\begin{eqnarray}
S_{H+gr}^{(eff)}&=&\int d^4x\Biggl[(\sqrt{-g}+\Upsilon)\left( -\frac{M_P^2}{2}\right)\left(1+\xi_{eff}\frac{2|H|^2}{M_P^2}\right)R(\Gamma,g)
+\Biggl(\sqrt{-g}+\frac{\Upsilon^2}{\sqrt{-g}}\Biggr)q^4M_P^4
\nonumber
\\
&+&(b_k\sqrt{-g}-\Upsilon) g^{\alpha\beta}\left(\partial_{\alpha}H\right)^{\dag} \partial_{\beta}H
 -(b_p\sqrt{-g}+\Upsilon)\lambda_{eff} |H|^4\Biggr].
\label{S-Higgs+gr eff in main text}
\end{eqnarray}
Here the Higgs mass term is omitted for reasons explained in section \ref{Reconstruction of the TMT effective potential}. The explicit $\phi$-dependence of $\lambda_{eff}$ and $\xi_{eff}$, eqs.(\ref{lambda eff num}) and (\ref{xi eff num}), according to the method of their derivation, is only applicable to the study of inflation provided that the model considered at the tree level ensures the flatness of the potential and hence describes slow-roll inflation. Therefore, we act according to the following algorithm: we will perform all steps of the TMT procedure  with the quantum effective primordial TMT action (\ref{S-Higgs+gr eff in main text}) without using the explicit form of $\lambda_{eff}$ and $\xi_{eff}$. Only at the end, when applying the TMT effective  equations to the inflationary stage, we will use the expressions that follow from (\ref{lambda eff num}) and (\ref{xi eff num}) after substituting the relation
$\tilde{\phi}({\mathsf x})=\tilde{v}\frac{\phi(x)}{\phi_b(t)}$ based on the redifinition (\ref{redif H in slow}) (see eqs.(\ref{tilde phi via phi}) and (\ref{from V4 back to M4}) and the footnote 8)
\begin{equation}
\lambda_{eff}=\lambda\Biggl(1+1.7\cdot10^{-2}\ln\frac{\phi^2}{\phi_b^2}\Biggr)
\label{lambda eff num in terms of phi}
\end{equation}
\begin{equation}
\xi_{eff}=\xi\Bigl(1+7.8\cdot10^{-8}\ln\frac{\phi^2}{\phi_b^2}\Bigr)
\label{xi eff num in terms of phi}
\end{equation}
It is worth paying special attention to the fact that two fields $\phi$ are involved in the description of quantum corrections: the field variable $\phi(x)$ and the background function $\phi_b(t)$. Namely, the ratio of these functions is the argument of the logarithms in $\lambda_{eff}$ and $\xi_{eff}$, which makes the structure of quantum corrections in TMSM fundamentally different from quantum corrections in generally accepted theories.
 The appearance of $\phi_b(t)$ in the place where a certain constant parameter with the dimension of mass usually stands is a direct consequence of the fact that quantum corrections for parameters characterizing the field $\tilde{\phi}$ are calculated in the LPPF, and the background on which they are calculated is described in the CF using $\phi_b(t)$. This is {\em another manifestation of the adiabatic effect} caused by the slowly varying classical background field 
$\phi_b(t)$ and  discussed after eq.(\ref{tilde v}).

 Continuing to operate with action 
(\ref{S-Higgs+gr eff in main text} ) up to relative corrections of the order $\lesssim 3\cdot 10^{-2}$ (just as we did with the classical TMT action), we notice that the found quantum corrections to $\lambda$ become significant only when $\phi\gtrsim 2.4\phi_b$. 
However, quantum corrections to $\xi$ become significant when $\phi>\phi_b\cdot \exp(10^5)$, which is unrealistically large. Therefore, we can safely take
$\xi_{eff}=\xi=\frac{1}{6}$.

As a result of performing all steps of the TMT procedure with the quantum effective primordial TMT action (\ref{S-Higgs+gr eff in main text}) and using the integration constant $\mathcal{M}=2q^4M_P^4$ (see Appendix \ref{About the integration constant}), we arrive at equations, a potential, a Lagrangian, and an action that should be called "doubly-effective". Indeed, first, we start with the quantum primordial TMT action (\ref{S-Higgs+gr eff in main text}), which we already call effective because it takes into account quantum corrections in the 1-loop approximation. Second, following the terminology we used when working with the classical primordial TMT action, the equations, potential, Lagrangian, and action obtained as a result of performing the TMT procedure should also be called TMT effective. To indicate "doubly-effective" in formulas we will use the subscript d.eff.

It can be shown that the doubly-effective picture near the vacuum state $\sigma$ does not differ significantly from what we learned studying the classical Higgs+gravity system in CF in paper \cite{1-st paper}, and what was summarized in section \ref{summary of 1-st paper} of the preset paper. Some further relevant discussions can be found in
 the first paragaph of section \ref{Quantization procedures}  and in Appendix \ref{About the integration constant}. 

Let us now turn to  our main goal in this section - the study of the influence of quantum corrections on Higgs inflation in the slow-roll regime. The form of the doubly effective potential plays a decisive role here. The simplest and most convenient way to obtain the results we are interested in is to find the 
doubly-effective energy-momentum tensor and then use the pressure density as the doubly-effective Lagrangian $L_{d.eff}$. Adding Einstein's gravitational term, we obtain the doubly-effective action $S_{d.eff}$. 
 Similar to the tree-level model of ref. \cite{1-st paper}, we represent the doubly effective action in a form suitable for studying Higgs inflation (that is expressed via $\varphi$ by the formula (\ref{phi via canonical varphi})). Note that the doubly-effective action and Lagrangian, $S_{d.eff}$ and $L_{d.eff}$, should be compared with the classical effective action and Lagrangian, eqs.(\ref{Seff}) and (\ref{Leff}).
Neglecting the contributions proportional to $\frac{m^2}{M_P^2}\sim 10^{-37}$, we arrive at the following doubly-effective action
\begin{equation}
S_{d.eff}=\int \left(-\frac{M_P^2}{2}R(\tilde{g})+L_{d.eff}(\varphi,X_{\varphi})\right)\sqrt{-\tilde{g}}d^4x
\label{Sd.eff varphi}
\end{equation}
and the doubly-effective Lagrangian 
\begin{equation}
L_{d.eff}(\varphi,X_{\varphi})=K_{1,eff}(\varphi)X_{\varphi}-\frac{1}{2}K_{2,eff}(\varphi)\frac{X_{\varphi}^2}{M_P^4}
-U_{d.eff}(\varphi),
\label{Leff varphi}
\end{equation}
where the doubly-effective potential reads
\begin{equation}
U_{d.eff}(\varphi) =\frac{\lambda_{eff} M_P^4}{4\xi^2}{\tanh}^4z\cdot F_{eff}(\varphi),
\label{Ud.eff varphi tanh}
\end{equation}
\begin{equation}
F_{eff}(\varphi)=\frac{(\zeta_v+b_p)q^4+
\frac{\lambda_{eff}}{16\xi^2}\sinh^4z}
{(1+\zeta_v)^2q^4+(1-b_p)\frac{\lambda_{eff}}{4\xi^2}{\sinh}^4z },
\label{Feff varphi tanh}
\end{equation}
\begin{equation}
z=\sqrt{\xi}\frac{\varphi}{M_P}, \qquad X_{\varphi}=\frac{1}{2}\tilde{g}^{\alpha\beta}\varphi_{\alpha}\varphi_{\beta}.
\label{z X}
\end{equation}
The functions $K_{1,eff}$ and $K_{2,eff}$, eqs.(\ref{K1 phi}) and (\ref{K2 phi}), represented through $\varphi$ are defined as follows
\begin{equation}
K_{1,eff}(\varphi)=\frac{2q^4(1+\zeta_v)(b_k-\zeta_v)+(2b_p+b_k-1)\frac{\lambda_{eff}}{4\xi^2}\sinh^4z}
{2\Bigl[q^4(1+\zeta_v)^2+(1-b_p)\frac{\lambda_{eff}}{4\xi^2}\sinh^4z\Bigr]},
\label{K1eff varphi}
\end{equation}
\begin{equation}
K_{2,eff}(\varphi)=\frac{(1+b_k)^2\cosh^4z}{2\Bigl[q^4(1+\zeta_v)^2+(1-b_p)\frac{\lambda_{eff}}{4\xi^2}\sinh^4z\Bigr]}.
\label{K2eff varphi}
\end{equation}
The last expression for $K_{2,eff}(\varphi)$ takes into account that $X_{\phi}=\cosh^2z\cdot X_{\varphi}$.

\begin{figure}

\includegraphics[width=13.0cm,height=8cm]{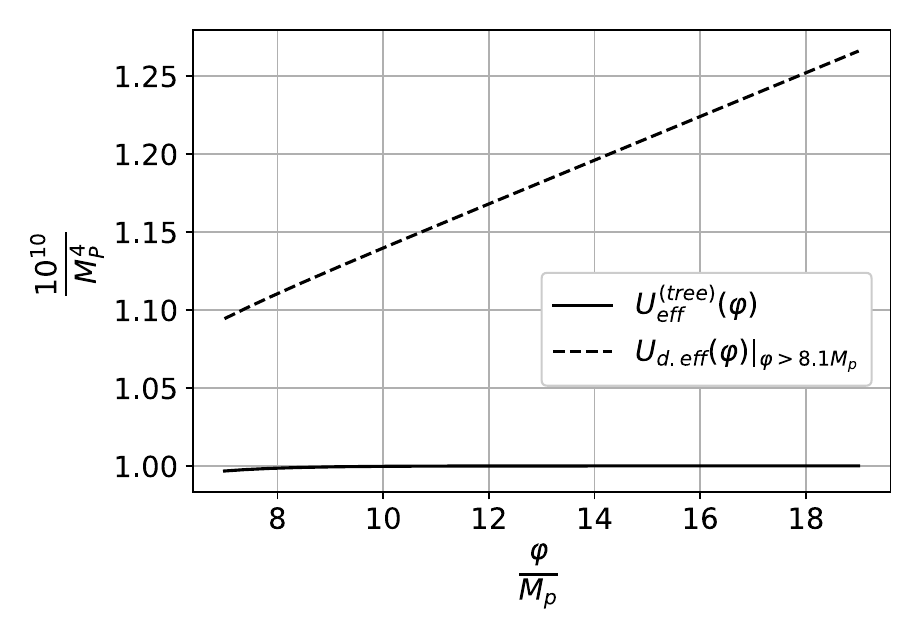}
\caption{Fragments of the graphs  of  the classical TMT-effective  potential $U_{eff}^{(tree)}(\varphi)$ and the doubly-effective potential
 $U_{d.eff}(\varphi)$ for $\varphi\gtrsim 8.1M_P$.}
\label{fig2}
\end{figure}

From (\ref{lambda eff num in terms of phi}) it follows that $\lambda_{eff}$ in terms of $\varphi$ takes the form
\begin{equation}
\lambda_{eff}=\lambda\Bigl[1+1.4\cdot10^{-2}\Bigl(\frac{\varphi}{M_P}-\frac{\varphi_b}{M_P}\Bigl)\Bigr].
\label{lambda eff num in terms of varphi}
\end{equation}
The above-mentioned condition $\phi>2.4\phi_b$, under which quantum corrections to $\lambda$ become significant, when expressed in terms of the field $\varphi$, has the form $\frac{\varphi}{M_P}-\frac{\varphi_b}{M_P}\gtrsim 2.1$. Therefore, if $0\leq\frac{\varphi}{M_P}-\frac{\varphi_b}{M_P}\lesssim 2.1$, then the doubly-effective action (\ref{Sd.eff varphi}) leads to results indistinguishable, within relative corrections $\lesssim 3\cdot 10^{-2}$, from those obtained in sections 3.2 and 5 of the paper \cite{1-st paper} for tree-level picture of the initial stage of cosmological evolution. For this reason, we can focus on studying inflation, where
 the inequality $\varphi> \varphi_b+2.1M_P>8.1M_P$ holds, and we will neglect $\varphi_b$  compared to $\varphi$.
In such a case,  for the above listed functions we get with very high accuracy: $F_{eff}(\varphi)=\frac{1}{2}$, \, $K_{1,eff}(\varphi)=1$, 
\begin{equation}
K_{2,eff}(\varphi)=\frac{1.93\cdot 10^{10}}{1+1.4\cdot10^{-2}\cdot\frac{\varphi}{M_P}},
\label{K2eff varphi>6MP}
\end{equation}
   and the  potential  (\ref{Ud.eff varphi tanh}) reduces to the following
\begin{equation}
U_{d.eff}(\varphi)|_{\varphi\gtrsim 8.1M_P}= \frac{\lambda M_P^4}{8\xi^2}
\Bigl(1+1.4\cdot10^{-2}\cdot\frac{\varphi}{M_P}\Bigr) \Bigl(1-8e^{-\sqrt{\frac{2}{3}}\frac{\varphi}{M_P}} \Bigr).
\label{varphi >6MP}
\end{equation}
Two graphs in figure \ref{fig2} allow to compare the shape of $U_{d.eff}(\varphi)|_{\varphi\gtrsim 8.1M_P}$ with the shape of the classical TMT-effective potential
$U_{eff}^{(tree)}(\varphi)$.

As was shown in section 6.3 of paper \cite{1-st paper}, for $\varphi>6M_P$, the contribution of $K_2(\varphi)$ to the first and second flatness conditions of the classical TMT-effective potential turns out to be negligibly small.  It follows from (\ref{K2eff varphi>6MP}) that quantum corrections to $K_2$ lead to its decrease: $K_{2,eff}(\varphi)<\tilde{K}_2(\varphi)$, where $\tilde{K}_2(\varphi)$ is defined by formula (6.4) in paper \cite{1-st paper}. Therefore, the results of the study of the impact of the K-essence type  structure on the slow-roll inflation in paper \cite{1-st paper} are also applicable to the action (\ref{Sd.eff varphi}) with the Lagrangian (\ref{Leff varphi}).
Then taking into account that $K_{1,eff}(\varphi)=1$ it is easy to see by direct calculation that the flatness conditions imposed on $U_{d.eff}(\varphi)$ are satisfied with a large margin
\begin{equation}
\epsilon|_{\varphi\gtrsim 8.1M_P}=\frac{M_P^2}{2}\Biggl(\frac{U^\prime_{d.eff}}{U_{d.eff}}\Biggr)^2\approx\frac{10^{-4}}{\Bigl(1+1.4\cdot10^{-2}\cdot\frac{\varphi}{M_P}\Bigr)^2}\ll 1, 
\label{epsilon}
\end{equation} 
\begin{equation}
|\eta|_{\varphi\gtrsim 8.1M_P}=M_P^2\Bigl|\frac{U''_{d.eff}}{U_{d.eff}} \Bigr|\approx \frac{16}{3}\cdot e^{-\sqrt{\frac{2}{3}}\frac{\varphi}{M_P}} \ll 1.
\label{eta}
\end{equation} 
Therefore, we can claim that, as in the classical model of ref. \cite{1-st paper}, taking into account quantum corrections in the 1-loop approximation does not change the fact that inflation quickly switches to a slow-roll regime.

\section{Discussion and Conclusion}
\label{Discussion}

The main results of the paper are obtained in accordance with the concept of the cosmological realization of the TMSM. As was already explained in the formulation of the main idea in the introductory section \ref{sec:intro}, the new effects of classical field theory discovered in paper \cite{1-st paper} in the description of cosmological evolution arise naturally due to the scalar $\zeta=\frac{\Upsilon}{\sqrt{-g}}$. This fact is the main reason leading to the necessity of such fundamental theoretical changes in the application of the TMSM compared to conventional models. Now, analyzing the results of  the present work, we want to draw attention to the fact that not only the ratio of the two volume measures, but also the TMT-structure of the  primordial action allows us to obtain new important results in a completely natural way. Indeed, keeping in mind the question of naturalness, we note that due to the TMT structure, TMSM contains a set of primordial parameters $b_i$, ($i=k, p, e, \mu, \tau, \nu, u, c, t$), which enter the volume elements through combinations $(b_i\sqrt{-g}\pm\Upsilon)d^4x$. Most of the important results of both this paper and the paper \cite{1-st paper} are achieved due to small deviations of the primordial parameters $b_i$ from 1 (with the exception\footnote{In paper \cite{1-st paper} it was shown that the choice of  $b_p=0.5(1+10^{-8})$ affects the constraint in such a way that the resulting function $\zeta(\varphi)$ does not contradict the fulfillment of the conditions for the onset of inflation.} of $b_p$, which has a very small deviation from 0.5).
Considering that the parameters $b_i$ characterize the relative contributions of the densities of the volume measures $\sqrt{-g}$ and $\Upsilon$ in each term of the primordial action, it can be assumed that small deviations of the values of the parameters $b_i$ from 1 are a consequence of the violation of symmetry of the type $\sqrt{-g}\Longleftrightarrow\pm\Upsilon$ inherent in some more fundamental theory. The presence of such a possibility allows us to conclude that the considered TMSM fully satisfies the requirements of naturalness. For a complete understanding of the grounds for the assertion of naturalness, it seems useful to recall the main results, paying particular attention to how the values of the parameters $b_i$ were obtained, as well as 
to the close interdependence between the results related to the slow-roll inflation and the results related to particle physics at accelerator energies. 

\hspace{1cm} 1)
As for the value of the parameter  $b_k$, which is present in the kinetic term of the primordial Lagrangian density of the Higgs  field as a factor in the combination 
$(b_k\sqrt{-g}-\Upsilon) = \sqrt{-g}(b_k-\zeta)$, eqs.(\ref{S-H 12}) and (\ref{S gauge}), it is fundamentally important that its value
$b_k\approx 1+1.6\cdot 10^{-5}$ strictly follows from the need to ensure agreement between: 

\hspace{0.2cm} a) the value of the nonminimal coupling constant 
$\xi=\frac{1}{6}$ chosen in \cite{1-st paper}, in combination with the value of the primordial Higgs selfcoupling parameter $\lambda =2.3\cdot 10^{-11}$ obtained by comparing with  the CMB data, on the one hand; 

\hspace{0.2cm}  b) the TMT-effective values of the parameters $m|_{(near \, vac)}$ and $\lambda|_{(near \, vac)}$ near the vacuum that lead to the  VEV of the Higgs field and the Higgs boson mass ($v\approx 246 GeV$ and $m_h\approx 125 GeV$) required in the GWS theory, on the other hand.

\hspace{1cm} 2)
The parameter $b_k$ is also present in the kinetic terms of the primordial Lagrangian density of gauge fields as a factor in the same combination
$(b_k\sqrt{-g}-\Upsilon)$, eq. (\ref{S gauge}). As a result, due to the small factor $b_k-1\sim 10^{-5}$, it follows from the requirements of the GWS theory for the values of the gauge coupling constants that the TMT-effective gauge coupling constants at the stage of slow-roll inflation become equal to their primordial values which are unusually small: $g\sim g'\sim 10^{-3}$ (see section \ref{Bosons  in LPP near vacuum} and eq.(\ref{g and gprime})). Therefore, the coefficients in front of the logarithms of the one-loop contributions of the gauge fields to the effective Higgs self-coupling parameter in the one-loop approximation turn out to be of the order of $10^{-13}$, which is approximately two orders of magnitude smaller than the tree coupling constant $\lambda\sim 10^{-11}$. This is why the well-known problem that arises in models containing the interaction of inflaton and gauge fields is absent in TMSM.

\hspace{1cm} 3)
The choice of the parameters $b_l$, $l=e, \,\mu, \,\tau,$ and $b_q$, $q=u, \, c, \, t$, very close to 1, in combination with the closeness of $b_k$ to 1, allows to achieve the correct values of the charged leptons and up-quarks masses, while the primordial Yukawa constants are the same for all charged leptons ($10^{-6}$) and similarly for all up-quarks ($10^{-5}$). Similar to the situation with the gauge coupling parameters, the TMT-effective Yukawa coupling parameters at the stage of slow-roll inflation  turn out to be practically equal to their primordial values. Thus, this  approach not only solves the problem of the fermion mass hierarchy, but also provides a subdominant fermion contribution to the 1-loop effective potential, which guarantees the stability of the vacuum during inflation. Moreover, taking into account the result of paper \cite{1-st paper} that normal classical cosmological evolution starts with inflation, one can formulate the result differently: the Higgs vacuum is stable throughout the early stage of normal classical cosmological evolution.

The results of this paper together with the results of paper \cite{1-st paper} demonstrate that the TMSM provides the possibility of establishing very non-trivial connections between physics at energies typical for the GWS theory and inflationary energies, giving answers to a number of questions in both the GWS theory and cosmology. Among them: the possibility of Higgs inflation with small $\xi$; the initial conditions for the onset of inflation; the mechanism for the emergence of a mass term with the wrong sign in the Higgs potential; the solution to the problem of the hierarchy of fermion masses.
Here it seems appropriate to discuss in more detail the question of the naturalness of the TMSM, comparing it with conventional models. As a justification for the possibility of using an unnaturally large constant of non-minimal coupling $\xi$, the authors of paper \cite{MetricHI-8} refer to the well-known, but accepted as inevitable, problem of the hierarchy of fermion masses. Indeed, a large quark mass ratio $m_t/m_u \simeq  10^5$ seems to be as unnatural as a large value of $\xi$. However, in the TMSM, both of these problems of unnaturalness are avoided. It is important to add to this discussion that the masses of fermions, like other particles, in both the SM and the TMSM arise as a result of SSB. Therefore, when discussing the naturalness of fermion mass ratios, it is more correct to consider the Yukawa coupling constants rather than the masses themselves. And in the TMSM studied in this paper, the Yukawa coupling constants are chosen to be universal for all generations of charged leptons and all generations of up-quarks.

The TMT structure of the model allows taking into account in the primordial action a vacuum-like term of the form $-\frac{\Upsilon^2}{\sqrt{-g}}V_2$ in addition to the usual $-\sqrt{-g}V_1$. Without the parameter $V_2$ it would be impossible to provide fine tuning to a small (or zero) value of  CC by choosing the integration constant ${\mathcal M}\approx 2\sqrt{V_1V_2}$. 
 The appearance as a result of the TMT procedure of a TMT-effective energy-momentum tensor that provides an adequate description of the inflation would also be impossible without  the parameter $V_2$.

It is worth noting that in this article, when studying the classical up-copy and 1-loop quantum corrections to it, we neglected relative contributions that are $\lesssim 3\cdot 10^{-2}$. If the comparison with observational data requires more accurate results for the  stage of slow-roll inflation, then we will have to abandon the approximations $(1+\xi\frac{\phi^2}{M_P^2})^{-1}\approx \frac{M_P^2}{\xi\phi^2}$ and $\zeta\approx 0$ that we used. There are no fundamental obstacles to doing this, but the calculations will become significantly more complicated.

The possible effect of quantum corrections on the very initial stages of classical evolution with pathological dynamics discovered in paper \cite{1-st paper} will be studied elsewhere.

In this paper, as well as in ref. \cite{1-st paper}, we have not touched upon the problem of preheating. 
In this regard, it should be noted that the approaches to preheating after the end of inflation developed in models \cite{Preheat 1} - \cite{Preheat 3} are not applicable to the model studied in this paper and in ref. \cite{1-st paper}. The reason is that for the effects studied in refs. \cite{Preheat 1} - \cite{Preheat 3}, a very large value of $\xi$ is of fundamental importance, while we believe that the small value of $\xi$ is the main advantage of our model.
It should also be noted that in the model we are studying, additional computational difficulties may arise when describing processes soon after the end of inflation, since the scalar $\zeta$ acquires a significant dependence on the kinetic energy. Finally, based on the summary collected in section \ref{Coupling constants and particle masses}, it should be emphasized that the particle physics processes predicted by  TMSM for the up-copy may differ significantly from those known using  SM up to energies of the order of $10^{15}-10^{16}\,GeV$. But we leave the study of all these extremely important questions for the future.

\acknowledgments
The author expresses gratitude to Gregory Gabadadze for the idea that the scalar function $\zeta$ should be used, along with curvature, to describe the background on which quantum corrections of matter fields are studied. At the beginning of the work, the author had very useful and stimulating discussions with Alexander Kamenshchik, for which the author is very grateful to him.

\appendix

\section{Kinetic energy and $\zeta$ during inflation in the slow-roll regime}
\label{append}

 Taking into account the estimate for the value of $m^2$  obtained in section \ref{Bosons  in LPP near vacuum}, eq.(\ref{m 0.7 GeV}), the terms $\propto \frac{m^2}{M_P^2}\sim 10^{-37}$ can be omitted in the constraint  (\ref{zeta for radical})  when considering inflation.. Therefore we can start from the constraint rewritten as follows
\begin{equation}
\zeta
=\frac{2q^4\zeta_v(1+\zeta_v)-(2b_p-1)\frac{\lambda\phi^4}{4M_P^4}+(1+b_k)\Omega \cdot\frac{X_{\phi}}{M_P^4}}
{2q^4(1+\zeta_v)+\frac{\lambda\phi^4}{4M_P^4}-(1+b_k)\Omega \cdot\frac{X_{\phi}}{M_P^4}}.
\label{zeta without m2}
\end{equation}
It is convenient to study the role of the contribution of $X_{\phi}$ to the constraint separately during inflation and after it ends.

When studying inflation it is more appropriate instead of the Higgs field $\phi$ to proceed with the field $\varphi$ defined by the relation
 (\ref{phi via canonical varphi}).
The constraint (\ref{zeta without m2}) represented through $\varphi$ becomes
\begin{eqnarray}
\zeta(\varphi, X_{\varphi})&&
\label{zeta via varphi}
\\
&=&\frac{2q^4\zeta_v(1+\zeta_v)-(2b_p-1)\frac{\lambda}{4\xi^2}\sinh^4(\sqrt{\xi}\frac{\varphi}{M_P})+
(1+b_k)\cosh^4(\sqrt{\xi}\frac{\varphi}{M_P}) \cdot\frac{X_{\varphi}}{M_P^4}}
{2q^4(1+\zeta_v)+\frac{\lambda}{4\xi^2}\sinh^4(\sqrt{\xi}\frac{\varphi}{M_P})-(1+b_k)\cosh^4(\sqrt{\xi}\frac{\varphi}{M_P}) \cdot\frac{X_{\varphi}}{M_P^4}},
\nonumber
\end{eqnarray}
where
\begin{equation}
 X_{\varphi}=\frac{1}{2}\tilde{g}^{\alpha\beta}\varphi_{\alpha}\varphi_{\beta},
\label{z X}
\end{equation}
When $\sqrt{\xi}\varphi\gg M_P$ the  TMT effective potential (\ref{Veff varphi tanh no fine tun 1}), (\ref{Veff varphi tanh no fine tun 2}) has a plateau-like shape 
\begin{equation}
U_{eff}^{(tree)}(\varphi) \approx \frac{\lambda M_P^4}{16(1-b_p)\xi^2}{\tanh}^4(\sqrt{\xi}\frac{\varphi}{M_P})\approx
\frac{\lambda M_P^4}{8\xi^2}\Bigl(1-8e^{-\sqrt{\frac{8}{3}}\frac{\varphi}{M_P}}\Bigr).
\label{1-st plateau}
\end{equation}
In this appendix we will analyze the function $\zeta(\varphi, X_{\varphi})$ on the flat segment of the potential $U_{eff}(\varphi)$. The latter is realized for 
$\varphi\gtrsim 6M_P$, where the constraint (\ref{zeta via varphi}) reduces to the following  
\begin{equation}
\zeta(\varphi,X_{\varphi})=\frac{32q^4e^{-4\sqrt{\xi}\frac{\varphi}{M_P}}-\frac{\lambda}{8\xi^2}\delta_p
+\frac{X_{\varphi}}{M_P^4}}
{32q^4e^{-4\sqrt{\xi}\frac{\varphi}{M_P}}+\frac{\lambda}{8\xi^2}-\frac{X_{\varphi}}{M_P^4}},
\label{constr for initial}
\end{equation}
Let's remember that  $\lambda\approx 2.3\cdot 10^{-11}$, \, $\xi=\frac{1}{6}$, \, $q^4=3\cdot 10^{-10}$, \, $b_p=\frac{1}{2}(1+\delta_p)$, \, $\delta_p\approx 10^{-8}$,
that is $\frac{\lambda}{8\xi^2}\approx 10^{-10}$ and $32q^4e^{-4\sqrt{\xi}\frac{\varphi}{M_P}} \lesssim 5.3\cdot 10^{-13}$ for $\varphi\gtrsim 6M_P$. In the context of the FLRW Universe, we are limited to $X_{\varphi}$ as kinetic energy, that is, $X_{\varphi}=\frac{1}{2}\dot{\varphi}^2$. 
In section 6.1 of ref.\cite{1-st paper} it  was shown that
the classical cosmological evolution starts with a normal dynamics only if the initial value $\dot{\varphi}_{in}$ satisfies the condition 
$\frac{1}{2}\dot{\varphi}_{in}^2< 0.17\cdot U_{eff}^{(tree)}(\varphi_{in})$. Moreover,  under this condition, cosmological evolution begins with inflation.
A detailed study carried out in sec. 6.3 of ref.\cite{1-st paper}  showed that in the spatially flat FLRW universe, the Friedmann and  field $\varphi$ equations  in  the slow-roll approximation with very high accuracy coincide with the usual ones
\begin{equation}
H^2=\frac{1}{3M_P^2}U_{eff}^{(tree)}(\varphi), \qquad  3H\frac{d\varphi}{dt}=-U_{eff}^{(tree)'}(\varphi),
\label{Frid slow}
\end{equation}
 From these equations it follows that in the slow-roll approximation the following equality is satisfied
\begin{equation}
X_{\varphi}=\frac{1}{2}\Bigl(\frac{d\varphi}{dt}\Bigr)^2=\frac{M_P^2}{6U_{eff}}(U_{eff}^{(tree)'})^2.
\label{phi dot 2}
\end{equation}
Thus, we find  that for $\varphi > 6M_P$  the contribution of $X_{\varphi}$ to the constraint (\ref{constr for initial}) is bounded from above
according to the following estimate 
\begin{equation}
\frac{X_{\varphi}}{M_P^4}\approx 128\lambda \cdot e^{-2\sqrt{\frac{8}{3}}\frac{\varphi}{M_P}} <9\cdot 10^{-18}.
\label{estimate for Xvarphi}
\end{equation}
Then, using the estimates mentioned after eq.(\ref{constr for initial}), we see that 
the first term in the numerator of eq.(\ref{constr for initial}) is leading and  is about $10^{5}$ times larger than the second and third terms.
We conclude that in the interval of $\varphi > 6M_P$, where $U_{eff}^{(tree)}(\varphi)$ is flat  and the slow-roll approximation is valid,  the scalar $\zeta$ can vary within the range
\begin{equation}
0<\zeta\approx\frac{32q^4}{\lambda/8\xi^2}\cdot e^{-\sqrt{\frac{8}{3}}\frac{\varphi}{M_P}} \lesssim 5\cdot 10^{-3}.
\label{zeta on flat segment}
\end{equation}

\section{Reconstruction of the quantum effective  primordial  TMT action
\\
 in the 1-loop approximation}
\label{Reconstruction of the TMT effective potential}

 The  potential $V_{eff}(\tilde{\phi})$, eq.(\ref{Veff tilde phi}),  was obtained in the LPPF when the Higgs field $\tilde{H}$, together with other  electroweak SM fields, is studied in the up-copy, i.e. on the background formed during slow-roll inflation.
To reconstruct the quantum effective primordial  TMT action, {\em we must perform in reverse order} all the steps that were performed to go from the description in the CF to the description in the LPPF.

\underline{The first step} of this procedure is that in the Higgs part of the Lagrangian (\ref{L bos after redef}) the Higgs field $\tilde{H}$, relation (\ref{tilde H presentation about class field}), is represented in the unitary gauge
\begin{equation}
\tilde{H}({\mathsf x})= \left(\begin{matrix} 0 \\ \frac{1}{\sqrt{2}}\tilde{\phi}({\mathsf x})\end{matrix}\right),
\label{tilde H in unitary}
\end{equation}
and  the potential $V(\tilde{H})$ is replaced by the 1-loop effective potential $V_{eff}(\tilde{\phi})$, eq.(\ref{Veff tilde phi}).

\underline{In the second  step}, proceeding in the unitary gauge, we must return from the description in terms of $\tilde{\phi}$ to the original $\phi$  using the redefinition  (\ref{redif H in slow}), which it is convenient to represent as\footnote{Recall (see eqs.(\ref{Higgs in unitary})-(\ref{H cosm av, back})) that $\phi_b(t)$ is the background field that is obtained in the CF (in the FLRW Universe) by cosmological averaging the Higgs field $\phi(x)$ with the subsequent requirement that it, together with the scalar curvature $R_b=-12\dot{a}/a\approx -18\lambda M_P^2$, satisfy the cosmological equations of the  classical TMSM in the regime of slow-roll inflation.} 
\begin{equation}
\tilde{H}({\mathsf x})=\left(\begin{matrix} 0 \\ \frac{1}{\sqrt{2}}\tilde{\phi}({\mathsf x})\end{matrix}\right)= \frac{M_P}{\sqrt{\xi}\phi_b(t)}H({\mathsf x})=
\tilde{v}\left(\begin{matrix} 0 \\ \frac{1}{\sqrt{2}}\frac{\phi({\mathsf x})}{\phi_b(t)}\end{matrix}\right),
\label{redif H in back to primord}
\end{equation}  
from where we get the relation
\begin{equation}
\tilde{\phi}({\mathsf x})=\tilde{v}\frac{\phi({\mathsf x})}{\phi_b(t)}.
\label{tilde phi via phi}
\end{equation}
 It is appropriate to recall here that the effective action $\Gamma [\tilde{\phi}]$ is an essentially non-local object, which is usually calculated by expanding in a series in derivatives of the mean field. To obtain an effective action, in the expansion of which one can restrict oneself to the kinetic term to the first power and which in the limit $\hbar\to 0$ coincides with the classical action, it is customary to assume that the mean field varies slowly in space-time.  In the LPPF of the model under consideration, the corresponding local approximation has the form of an expansion\footnote{Note that the value of the up-copy anomaleous dimension $\gamma_{(up)}$ is dominated by the contributions of the gauge fields and is of the order of $|\gamma_{(up)}|\sim\frac{g^2}{16\pi^2}\sim10^{-7}$. Therefore, continuing to neglect corrections of order $\lesssim 3\cdot 10^{-2}$, we can set $Z(\tilde{\phi})=1$.}
\begin{equation}
\Gamma [\tilde{\phi}]=\int \sqrt{-\tilde{g}}d^4{\mathsf x}\Bigl[-V_{eff}(\tilde{\phi})+\frac{1}{2}Z(\tilde{\phi})\tilde{g}^{\alpha\beta}\frac{\partial \tilde{\phi}}{\partial {\mathsf x}^{\alpha}}\frac{\partial \tilde{\phi}}{\partial {\mathsf x}^{\beta}}+...\Bigr].
\label{expansion of Gamma}
\end{equation}
As $\tilde{\phi}>\tilde{v}$ increases, the first term of the effective potential $V_{eff}$, eq.(\ref{Veff tilde phi}), becomes dominant and
\begin{equation}
V_{eff}(\tilde{\phi}) \approx\frac{\lambda_{eff}}{8}\tilde{\phi}^4> \frac{\lambda_{eff}}{8\xi^2}M_P^4\gtrsim \frac{\lambda}{8\xi^2}M_P^4\approx U_{eff}^{(tree)}(\phi)|_{\phi> 14.2M_P}\approx 10^{-10}M_P^4 \, ;
\label{leading term in Veff}
\end{equation}
here $U_{eff}^{(tree)}(\phi)|_{\phi> 14.2M_P}$ is equal to the plateau height of the classical (tree-level) TMT effective potential (see eqs.
(\ref{Veff varphi tanh no fine tun 1}), (\ref{Veff plateau})) and (\ref{Ueff in Ein via Ueff during infl})), where slow-roll inflation is realized. 
The regime of the slow-roll inflation, considered at the tree level, implies that $\frac{1}{2}(\dot{\phi}_b(t))^2<9\cdot 10^{-18}M_P^4\lll U_{eff}^{(tree)}(\phi)|_{\phi> 14.2M_P}$, where the estimate (\ref{estimate for Xvarphi}) was used.
 Comparing this with the fact that in the expansion
(\ref{expansion of Gamma}) there is no reason to assume that the kinetic term is much smaller than $V_{eff}(\tilde{\phi})$, we conclude that 
$\tilde{g}^{\alpha\beta}\frac{\partial \tilde{\phi}}{\partial {\mathsf x}^{\alpha}}\frac{\partial \tilde{\phi}}{\partial {\mathsf x}^{\beta}}\gg \frac{1}{2}(\dot{\phi}_b(t))^2$. Therefore, substituting $\tilde{H}({\mathsf x})$, which is given by (\ref{redif H in back to primord}) (and (\ref{tilde phi via phi})), into the kinetic term of the Higgs part of the Lagrangian (\ref{L bos after redef}), yields
\begin{equation}
\tilde{g}^{\alpha\beta}\frac{\partial \tilde{\phi}}{\partial {\mathsf x}^{\alpha}}\frac{\partial \tilde{\phi}}{\partial {\mathsf x}^{\beta}}=\tilde{v}^2\tilde{g}^{\alpha\beta}\frac{\partial}{\partial {\mathsf x}^{\alpha}}\Biggl(\frac{\phi({\mathsf x})}{\phi_b(t)}\Biggr)
\frac{\partial}{\partial {\mathsf x}^{\beta}}\Biggl(\frac{\phi({\mathsf x})}{\phi_b(t)}\Biggr)\approx 
\frac{M^2}{\xi\phi^2_b}\tilde{g}^{\alpha\beta}\frac{\partial\phi({\mathsf x})}{\partial {\mathsf x}^{\alpha}}
\frac{\partial\phi({\mathsf x})}{\partial {\mathsf x}^{\beta}},
 \label{comparing kinetic terms}
\end{equation}
where the last approximate equality was obtained using the adiabaticity condition
 (see discussions after eqs.(\ref{tilde v}) and (\ref{xi eff num in terms of phi})).
Thus,  plugging $\tilde{H}({\mathsf x})$
into the Higgs part of the Lagrangian (\ref{L bos after redef}), in which $V(\tilde{H})$, eq.(\ref{V tilde H}), is replaced by $V_{eff}(\tilde{H})$, we obtain the following effective Lagrangian (instead of the tree level Lagrangian (\ref{L bos back slow roll}))
\begin{eqnarray}
 L_{(on \, back \, \phi_b>14M_P)}^{(Higgs, \, eff)}&=&\frac{M_P^2}{\xi\phi_b^2}\tilde{g}^{\alpha\beta}
\frac{\partial H}{\partial {\mathsf x}^{\alpha}}^{\dag}\frac{\partial H}{\partial {\mathsf x}^{\beta}}
 -\frac{M_P^4}{2\xi^2\phi_b^4}\lambda_{eff} |H|^4
\nonumber
\\
&-&\xi_{eff}\frac{M_P^2}{\xi\phi_b^2}R_b|H|^2 +\mathcal{O}(1)\lambda^2M_P^4,
 \label{L bos back slow roll eff}
\end{eqnarray}
 Given that we will ultimately be interested in slow-roll inflation, and the estimate 
$\frac{m^2}{M_P^2}\sim 8\cdot 10^{-38}$ obtained in section \ref{Bosons  in LPP near vacuum}, we  ignored the term $\propto m^2|H|^2$.

\underline{In the third step}, we have to reconstruct (taking into account radiative corrections) the main features of the structure of the Lagrangian
$L^{(bos \, LPPF)}_{(on \, back)}$, eq.(\ref{L bos back general}), valid for an arbitrary cosmological background. To do this, we need to return from the approximations described at the beginning of section \ref{Bosonic sector slow-roll} and reconstruct the original structure using the following replacements (the inverse of the approximations made there):
$ \frac{M_P^2}{\xi\phi_b^2}\to \frac{1}{\Omega_{(back)}}$ ; \, $1\to (1+\zeta_{(back)})$; \, in the kinetic term $1\to (b_k-\zeta_{(back)})$; \, in the selfcoupling term $\frac{1}{2}\to (b_p+\zeta_{(back)})$. 
As a result, we obtain the 1-loop effective Lagrangian density for the Higgs field in the LPPF on an arbitrary cosmological background given by the field $\phi_{(back)}$, the scalar $\zeta_{back}$ and the scalar curvature $R_{(back)}$
\begin{eqnarray}
\sqrt{-\tilde{g}}L^{(Higgs \, LPPF)}_{(on \, back)}|_{(eff)}&=&
\sqrt{-\tilde{g}}\Bigl[\frac{b_k-\zeta_{(back)}}{(1+\zeta_{(back)})}\Omega_{(back)}\tilde{g}^{\alpha\beta}
\frac{\partial H}{\partial {\mathsf x}^{\alpha}}^{\dag}\frac{\partial H}{\partial {\mathsf x}^{\beta}}
\label{L Higgs back general eff}
\\
&-&\frac{b_p+\zeta_{(back)}}{(1+\zeta_{(back)})^2\Omega_{(back)}^2}\lambda_{eff} |H|^4
 -\frac{\xi_{eff}}{\Omega_{(back)}}R_{(back)}|H|^2+\mathcal{O}(1)\lambda^2M_P^4\Bigr].
\nonumber
\end{eqnarray}
We have retained the notations $\lambda_{eff}$ and $\xi_{eff}$ here, keeping in mind that the final result of our manipulations will be applied to the slow-roll inflation, for which $\lambda_{eff}$ and $\xi_{eff}$ are determined by the formulas (\ref{lambda eff num}) and ((\ref{xi eff num}).

\underline{In the final, fourth step} of reconstructing the effective primordial TMT action we must {\em pass from the description in LPPF to the description in CF.}
Given how the reverse transition was performed in sections \ref{Back General view} and \ref{descrition in LPP on back}, we do the following:

(a) we return from the description in coordinates $({\mathsf x})$ in the four-dimensional space-time  ${\mathcal V}_4$  to the description in the FLRW Universe in the coordinates of the Friedman metric (\ref{FRW metric}), that is
\begin{equation}
H({\mathsf x})=\left(\begin{matrix} 0 \\ \frac{1}{\sqrt{2}}\phi({\mathsf x})\end{matrix}\right) \quad \text{is replaced with} \quad
H(x)=\left(\begin{matrix} 0 \\ \frac{1}{\sqrt{2}}\phi(x)\end{matrix}\right),
\label{from V4 back to M4}
\end{equation} 
and respectively
\begin{equation}
\frac{\partial H(\mathsf x)}{\partial {\mathsf x}^{\alpha}} \quad \text{is replaced with} \quad
\partial_{\alpha} H(x)
\label{from V4 back to M4 in deriv}
\end{equation}

(b) we replace $\phi_{(back)}(t)=\langle \phi(x)\rangle_{cosm.av.}$ with $\phi(x)$, that is,    $\Omega_{(back)}=1+\xi\frac{\phi_{(back)}^2}{M_P^2}$ is 
replaced by $\Omega=1+\xi\frac{\phi^2}{M_P^2}$;

(c) we replace  $\zeta_{(back)}$ with $\zeta=\frac{\Upsilon}{\sqrt{-g}}$;

(d) we return from the Einstein frame to the original frame with the help of
\begin{equation}
\tilde{g}^{\mu\nu}=\frac{1}{(1+\zeta)\Omega} g^{\mu\nu} \quad \text{and, therefore,} \quad \sqrt{-\tilde{g}}\tilde{g}^{\mu\nu}=(1+\zeta)\Omega\sqrt{-g}g^{\mu\nu};
\nonumber
\end{equation}

(e) we replace $R_{(back)}$ with $R(\Gamma, g)=g^{\mu\nu}R_{\mu\nu}(\Gamma)$, where $\Gamma$ stands for affine connection;

(f) we add the gravitational action in the Palatini formalism and the vacuum terms that are present in the primordial classical TMT action  (\ref{S-Higgs+gr class}).

Thus, the procedure of reconstruction of the 1-loop quantum effective primordial TMT action of the Higgs+gravity system is completed by replacing 
the classical primordial  TMT action  (\ref{S-Higgs+gr class}) with the action
\begin{eqnarray}
S_{H+gr}^{(eff)}&=&\int d^4x\Biggl[(\sqrt{-g}+\Upsilon)\left( -\frac{M_P^2}{2}\right)\left(1+\xi_{eff}\frac{2|H|^2}{M_P^2}\right)R(\Gamma,g)
\nonumber
\\
&+&(b_k\sqrt{-g}-\Upsilon) g^{\alpha\beta}\left(\partial_{\alpha}H\right)^{\dag} \partial_{\beta}H
 -(b_p\sqrt{-g}+\Upsilon)\lambda_{eff} |H|^4
\nonumber
\\
&+&\Bigl(\sqrt{-g}+\frac{\Upsilon^2}{\sqrt{-g}}\Bigr) q^4M_P^4+\frac{\sqrt{-g}}{(1+\zeta)^2\Omega^2}\mathcal{O}(1)\lambda^2M_P^4
\Biggr].
\label{S-Higgs+gr eff}
\end{eqnarray}
where the mass term $\propto m^2 |H|^2$ has been omitted since we intend to use  $S_{H+gr}^{(eff)}$ to study inflation, where it is negligible.
Our choice of vacuum parameters $V_1=V_2=-q^4M_P^4$ with $q^4=3\cdot 10^{-10}$ can also be taken into account: since $\lambda^2\sim 10^{-22}$, it is obvious that the effect of 1-loop quantum corrections to the vacuum energy is negligible compared to the contribution of $q^4M_P^4$.
This observation is reinforced by the fact that in the range of values of the Higgs field where the TMT effective potential becomes flat, $\Omega\gg 1$.

\section{On the integration constant when perfoming the TMT procedure
\\
 with the 1-loop effective primordal TMT action}
\label{About the integration constant}

It is necessary to recall one of the steps of the TMT procedure (briefly mentioned by means of eq.(\ref{M 2sqrt})) which is of fundamental importance for obtaining reasonable physical results. In the detailed description of the TMT procedure in ref. \cite{1-st paper}, applied to the classical primordial TMT action (\ref{S-Higgs+gr class}) of the Higgs+gravity sector of TMSM, it was shown that the zero or extremely small value of the cosmological constant is ensured by the choice of the integration constant $\mathcal{M}=2\sqrt{V_1V_2}(1+\delta)=2q^4M_P^4(1+\delta)$, where $\delta\sim\frac{\lambda\sigma^4}{q^4M_P^4}\lll 1$, and  $\sigma$ is the VEV in the CF, defined by the relation (\ref{s2}); its value was found using eq.(\ref{sigma via v}). To perform this necessary step in the TMT procedure with the quantum effective primordial TMT action $S_{H+gr}^{(eff)}$ we must ensure that the latter, in addition to the 1-loop quantum corrections in the up-copy (where $\phi >M_P$), is also capable of describing quantum corrections near the vacuum (where $\phi\sim\sigma$). The results concerning the 1-loop effective potential in the LPPF near the vacuum of GWS theory,  discussed briefly in the first paragraph of section \ref{Quantization procedures}, imply that the quantum corrections to $\lambda$ and $m^2$ at the electroweak energy scale are small (where $\xi$ is not relevant at all). The procedure for reconstructing the quantum effective primordial TMT action implemented in section \ref{Reconstruction of the TMT effective potential}, can be repeated (with appropriate modification) also to obtain the 1-loop quantum effective primordial TMT action applicable near the vacuum of GWS theory; it is easy to see that, except for the need to preserve the mass term $\propto m_{eff}^2\phi^2$, the form of the quantum effective primordial TMT action coincides with (\ref{S-Higgs+gr eff in main text}).  It is obvious that in LPPF quantum corrections do not have a significant effect on the value of $v$, and hence on the value of $\sigma$ (in CF). Therefore, the value of the integration constant $\mathcal{M}=2q^4M_P^4$ used  in section \ref{impact} remains the same as it was in the operations with the classical primordial action (\ref{S-Higgs+gr class}).
To summarize this discussion, it is worth noting that although the goal of this paper is the inflationary stage, there is a need to take into account the results arising from the quantum effective primordial action near the electroweak SM vacuum. This circumstance once again demonstrates that the applicability of the TMSM in the range from the electroweak SM vacuum to the inflationary energy scales is a direct consequence of the concept of the cosmological realization of the TMSM.

\end{document}